\documentclass[12pt]{nature} %
      \usepackage[round,authoryear]{natbib}

 \usepackage{ifpdf}
 \usepackage{epsfig}
 \usepackage{amsfonts}
 \usepackage{amsthm}
 \usepackage{amssymb, latexsym, amsmath}
 \usepackage{amsbsy}
 \usepackage{amstext}
 \usepackage{amsopn}
 \usepackage{amsgen}
\graphicspath{{Figures/}}
\usepackage{lineno}
 \usepackage{graphicx}
 \usepackage{color}
  \usepackage{verbatim}
  \usepackage[T1]{fontenc}

 \usepackage{amsbsy}

\usepackage{units}

 \usepackage[normalem]{ulem}
 \usepackage{wasysym}
 \usepackage{verbatim}
\usepackage[scale=0.8]{geometry}
\usepackage{fancyhdr}
\usepackage{bm}
 \newcommand{\s}{\nobreak\hspace{.11em}\nobreak}

 \newcommand{\be}{\begin{equation}}
 \newcommand{\ee}{\end{equation}}
 \newcommand{\ba}{\begin{eqnarray}}
 \newcommand{\ea}{\end{eqnarray}}
 \newcommand{\bs}{\begin{subequations}}
 \newcommand{\es}{\end{subequations}}

 \newcommand{\erbold}{\mbox{{\boldmath $
 r$}}}

  \newcommand{\eRbold}{\mbox{{\boldmath $
  {R}$}}}
  \newcommand{\Rbold}{\mbox{{\boldmath $
  {R}$}}}

\newcommand{\tnr}[1]{\textcolor{black}{#1}}

\begin{document}
  \title{
         ${{~~~~~~~~~~~~~~~~~~~~}^{^{^{
     %    To~be~
     %     Published~in~the~
     %              submitted~to
     %              ~~Celestial~Mechanics~and~Dynamical~Astronomy
     %              ~the~Astronomical~Journal~
     %               ~Icarus~(or~JGR~?)
     %   \,,~Vol.~764\,,~id.~26\,~(2013)
                  }}}}$\\
     %             ~
 {\Large{\textbf{{{
 Tidal insights into rocky and icy bodies: \\ An introduction and overview}
 ~\\
 ~\\}
            }}}}
 \author{Amirhossein~Bagheri$^{1}$, Michael~Efroimsky$^{2}$, Julie~Castillo-Rogez$^{3}$, Sander~Goossens$^{4}$,  Ana-Catalina~Plesa$^{5}$, Nicolas~Rambaux$^{6}$, Alyssa~Rhoden$^{7}$, Michaela~Walterov\'{a}$^{5}$, Amir~Khan$^{8,1}$, Domenico~Giardini$^{1}$\\
  \quad\\
 \quad\\
            %  {\Large{Author 1~,}}\\
                             $^{1}$ Institute of Geophysics, ETH Zürich, Zürich, Switzerland\\
 $^{2}$                             US Naval Observatory, Washington DC, USA\\
                             $^{3}$ Jet Propulsion Laboratory, California Institute of Technology, Pasadena CA, USA \\
 $^{4}$     NASA Goddard Space Flight Center, Greenbelt MD, USA                    \\
 $^{5}$   Institute of Planetary Research, DLR, Berlin, Germany  \\
   $^{6}$      IMCCE, CNRS, Observatoire de Paris, PSL Université, Sorbonne Université, Paris, France                    \\
      $^{7}$      Southwest Research Institute, Boulder, CO, USA           \\
  $^{8}$    Physik-Institut, University of Zürich, Zürich, Switzerland     \\
    \date{}}

 \maketitle

\clearpage

\section*{Abstract}

Solid body tides provide key information on the interior structure, evolution, and origin of the planetary bodies.
Our Solar system harbours a very diverse population of planetary bodies, including those composed of rock, ice, gas, or a mixture of all. While a rich arsenal of geophysical methods has been developed over several years to infer knowledge about the interior of the Earth, the inventory of tools to investigate the interiors of other Solar-system bodies remains limited. With seismic data only available for the Earth, the Moon, and Mars, geodetic measurements, including the observation of the tidal response, have become especially valuable and therefore, has played an important role in understanding the interior and history of several Solar system bodies. To use tidal response measurements as a means to obtain constraints on the interior structure of planetary bodies, appropriate understanding of the viscoelastic reaction of the materials from which the planets are formed is needed. Here, we review the fundamental aspects of the tidal modeling and the information on the present-day interior properties and evolution of several planets and moons based on studying their tidal response.
%It is necessary to understand the overall tidal response of celestial bodies, which is defined by interplay of their viscoelastic reaction and self-gravitation.
We begin with an outline of the theory of viscoelasticity and tidal response. Next, we proceed by discussing the information on the tidal response and the inferred structure of Mercury, Venus, Mars and its moons, the Moon, and the largest satellites of giant planets, obtained from the analysis of the data that has been provided by space missions. We also summarise the upcoming possibilities offered by the currently planned missions.

\section*{Keywords}
Tides, viscoelasticity, Rheology, Mercury, Venus, Moon, Mars, Icy worlds, Interior structure, Surface geology

\clearpage

\section{Introduction}

The Solar system harbours a diverse population of planetary bodies. These include objects formed of rock, ice, gas, as well as objects of a mixed composition. Over the past two decades, our understanding of these bodies' interior structure has been considerably improved owing to the valuable data provided by several successful space missions. A combined perception of celestial bodies' interiors and the mechanisms that govern their evolution can represent an efficient means to infer information about their past history and origins which help us to understand how the Solar system has formed and evolved. Given the scarcity of seismic data for the planetary bodies except the Earth, the Moon, and Mars, our studies have to rely on remote sensing and geodetic measurements. An efficient means to infer knowledge on the planets' and moons' interior structure is their tidal response to the gravitational attraction from other objects.

\begin{figure}[ht]
\includegraphics[width=.9\textwidth]{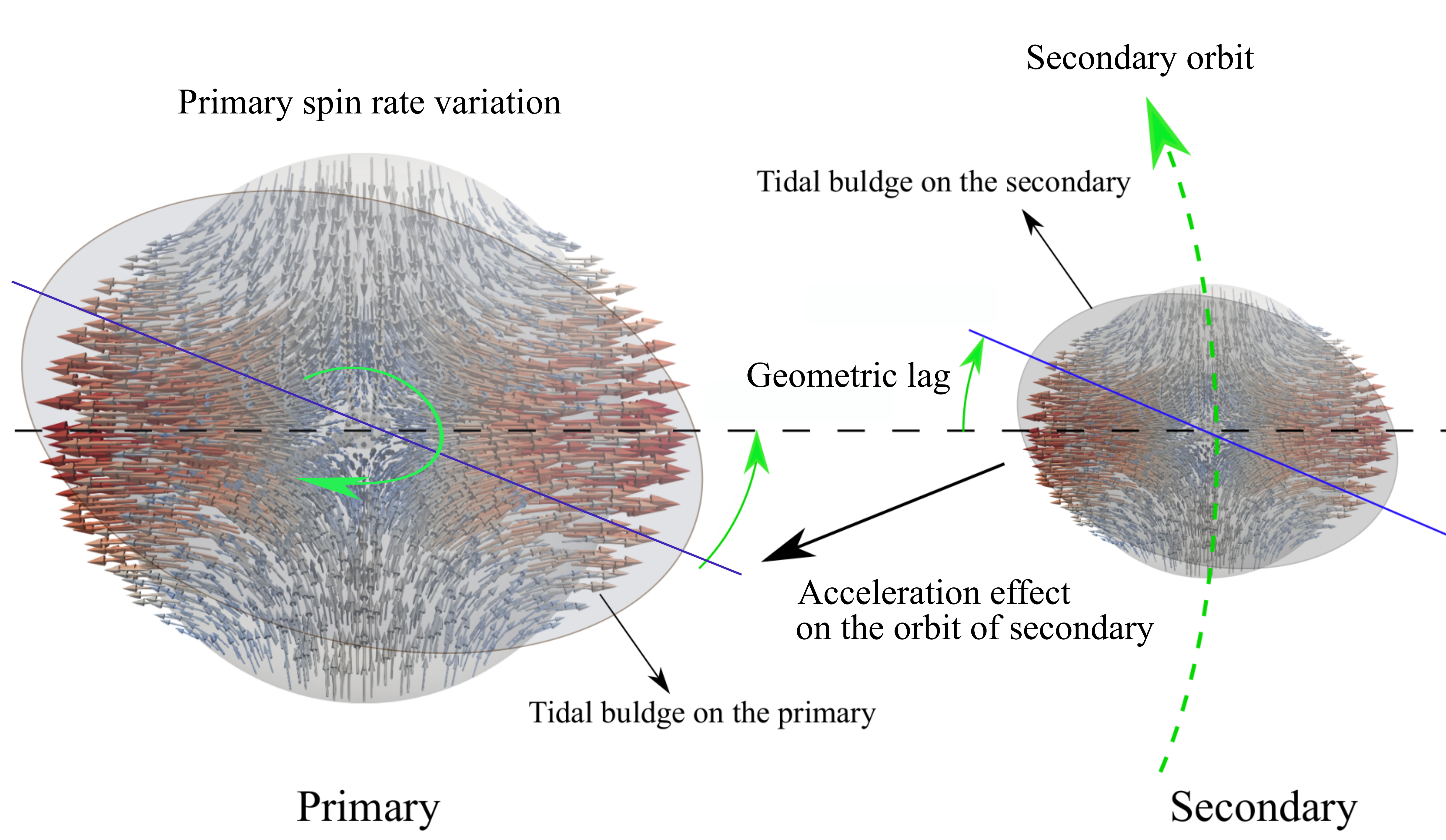}
\caption{Tidal force field and the geometric lag induced in a planet and its satellite resulting from the gravitational attraction of the two bodies towards each another. Tension in the body is shown with red and compression with blue arrows where the size of the arrows indicate the force magnitude. The force field is shown on the unperturbed bodies. The maximum tidal force in the perturbed body is induced in the side of the body that faces the perturber. The tidal evolution changes the orbit period of the satellite and spin rate of the rotation of two bodies, driving their orbital properties towards synchronism.}
\label{fig:Tidal_force_field}
\end{figure}

\noindent
A planet's side facing its satellite is attracted by the satellite stronger than the opposite side. Since the planet's rotation is, generally, not synchronised with the period of the orbiting satellite, the satellite exerts a periodically changing force field and a resulting deformation field in the hosting planet. These changes in the gravitational potential generated by the satellite are known as \textit{tides} (Figure~\ref{fig:Tidal_force_field}). Aside from the well-known ocean tides on the Earth, both the Earth and other planets demonstrate atmospheric tides and, most importantly, {\it bodily tides}, a phenomenon on which this review concentrates. This phenomenon is always reciprocal; so the satellite, in its turn, is experiencing periodic perturbation of the planet-generated potential, and is developing periodic deformation. The Sun can also play the same role and generate tides in the planets and itself experiences tides generated by them.

\noindent
Tidal deformation of a celestial body results in both vertical and horizontal displacement of its surface and in the ensuing perturbation of its gravitational field.
These variations are described by three \textit{Love numbers}: $h$, $l$, and $k$. The two former Love numbers ($h$, $l$) represent the deformations caused by the tidal force on the planet in the vertical and horizontal directions, respectively, whereas $k$ represents the induced perturbation in the gravitational field.

\noindent
The interior of a planetary body is not perfectly elastic and is affected by internal friction, as a result of which the tidal bulge does not exactly align with the position of the tide-raising body, but exhibits a phase lag. The \textit{quality factor} $Q$ is defined as the inverse of the sine of the absolute value of the phase angle between the tidal bulge and the direction towards the tide-raising body. It relates to the energy dissipated by friction per loading cycle, in such a way that a lower $Q$ implies higher dissipation. Both the tidal deformation magnitude and the phase lag are sensitive to the large-scale interior properties of the body, such that a larger deformation (higher Love numbers) would imply that the interior of the object is less \text{rigid} (contains fluid parts, highly porous material, etc.), while a larger phase lag would imply that the interior is composed of a material that is more viscous. Thus, the tidal response, represented in the form of the tidal Love numbers and tidal quality factor, was used to probe the structural properties of the planetary body. Owing to their long-wavelength nature, tides sample the large-scale interior properties of the body and can be used to probe its deep interior.

\noindent
Studying the planetary tidal response dates backs to \citet{Love11}, in the early 20th century, when the response of a compressible and homogeneous Earth model was first computed. With the increasing precision of modern geodesy, the tidal response of the Earth can be measured by satellite altimetry and the Global Positioning Systems \citep{yuan_etal13}; the semi-diurnal tidal Love numbers and the tidal quality factor of the Earth are $k_2 = 0.3531$, $h_2 = 0.6072$, $l_2 = 0.0843$, and $Q\sim10$, respectively  \citep{seitz_etal12,krasna_etal13} (For earlier measurement and analysis, one can refer to \citet{dehant87,mitrovica_etal94,smith_etal73,ryan_etal86,haasSchuh96,petrov2000}). Tidal dissipation in the Earth takes place predominantly in the oceans, especially in shallow seas. The dissipation in the oceans excluded, the solid-Earth tidal quality factor is found to be $Q\sim280\pm 70$ \citep{egbert_Ray03}. Based on such measurements, tidal tomography studies have been efficiently used to constrain the large-scale features of the Earth's deep interior, such as the nature of two large low-shear-velocity provinces (LLSVP) \citep{metivier_Conrad08, latychev_etal09, Lau_etal17}. %Also, the surface mass loading on the Earth caused by the ocean tides can be used to obtain knowledge of the shallow structure of our planet \citep[e.g.,][]{Martens_etal16, martens2016using}.

\noindent
\textcolor{black}{The purpose of this review is to provide an introduction to the existing knowledge on the interior of the planetary bodies based on studying their tidal activity. In this review, we cover the fundamental aspects and formulation of the tidal modeling and discuss the role played by tides in understanding the planetary bodies. First, we review the general aspects of viscoelasticity followed by several viscoelastic models used to mimic dissipation in materials, next we provide the fundamental information for modeling the tidal response of a planetary body followed by a brief introduction to tidal evolution and thermal coupling. Next, we discuss the constraints obtained by using the tidal response measurements in understanding the interior properties and evolution of several rocky and icy planetary bodies in the Solar system, i.e., Mercury, Venus, the Moon, Mars, and the largest moons of gas giants. We also discuss what can be learned from measuring the tidal response in the anticipated missions for the planetary bodies which currently lack such data, either completely or with sufficient precision. While this review mostly discusses the interior structure based on measurements of tidal response presented by tidal Love number and tidal quality factor, we also provide a short summary of the surface geological features as a result of tidal activity and its interpretations for the interior properties and evolution of the planetary bodies.}

%\subsection{Significance}\label{motivation}

\section{Viscoelasticity}\label{sec:Viscoelasticity}\

Tidal dissipation differs from seismic dissipation, because it is defined not only by the dissipative properties of the mantle but also by the interplay of these properties with the body's self-gravitation. Leaving the description of this interplay for Section \ref{sec:Tides}, in the current section, we address the rheological properties of the materials, from which the planetary bodies are formed.

\subsection{General aspects}

Dissipation in a material is, essentially, a relaxation process whose effectiveness is sensitive to the material composition, temperature, confining pressure, grain size and, importantly, to the frequency of forcing. The combined influence of these parameters on the energy damping rate is highly non-trivial, and is defined by complex micro-scale processes such as grain-boundary interactions, dislocation migration and the presence of voids and melt \citep[e.g.,][]{Jackson_etal02,JacksonFaul10,McCarthy_etal11, SundbergCooper10, Takei_etal14}. Hence, an appropriate modeling of the viscoelastic behaviour of a celestial body is crucial for the correct interpretation of its tidal measurements.

\noindent
Mars, as an example, appears to be an instructive case of necessity for such modeling. Compared to the solid-Earth quality factor at semi-diurnal tides ($Q = 280 \pm 70$, \cite{Ray_etal01}), the Martian quality factor at the period of Phobos's tides (T = 5.55 hr) is surprisingly low, i.e., $Q \approx 78-100$,  \citep[e.g.,][]{JacobsonLainey14}. Applying the simplistic Maxwell rheology to Mars results in a low quality factor implying an unreasonably low average viscosity ($\sim 10^{14}$~Pa.s) \citep{Bills_etal05}, in contrast to that estimated for the Earth's mantle ($\sim 10^{22-23}$~Pa.s) \citep{anderson_OConnell67,karato_Wu93}. This result is surprising, given that Mars is expected to have cooled faster on account of its smaller size, and therefore has a higher viscosity value than the solid Earth (See, e.g. \citet{Plesa_etal18,Khan_etal18}). The issue is resolved by employing a more realistic viscoelastic model attributing the high tidal dissipation to strong anelastic relaxation in Mars  \citep{CastilloRogezBanerdt12, NimmoFaul13,Khan_etal18,Bagheri_etal19}.

\noindent
Another example revealing the necessity for accurate rheological modeling is Venus. Application of a simplistic elastic model to the mantle would yield to an enticing but naive interpretation for the measured tidal Love number value \citep{Dumoulin_etal17,konopliv_yoder96}. This interpretation would suggest a fully liquid core \citep{konopliv_yoder96}. However, a fully solid iron core has been shown to be a plausible option, as well, when viscoelasticity is taken into account \citep{Dumoulin_etal17,Yodervenus}. Further insights about tides on Mars and Venus are provided in sections \ref{sec:Mars} and \ref{sec:venus}, respectively.
These examples reveal that accurate modeling of viscoelastic dissipation in planetary bodies is essential to understanding their interior structure.

%\subsection{Viscoelastic dissipation modeling}\label{sec:viscoelastic_modeling}
\noindent
Based on various friction mechanisms involved, several viscoelastic models have been proposed and constrained by means of laboratory experiments \citep[e.g.,][]{Gribbcooper98, Faul_etal04, jackson2005laboratory, JacksonFaul10, Takei_etal14}. Most of the experimental studies have focused on rocky materials such as olivine and orthopyroxene (see, e.g. \citet{Qu_etal21} and \citet{JacksonFaul10}) whereas few studies have considered the dissipation in ice \citep{MccarthyCooper16}. These models have been utilized to study the planetary data \citep{LauFaul19, Nimmo_etal12, NimmoFaul13, Khan_etal18}. Because of its ease of implementation, the Maxwell rheology has long been employed to model the viscoelastic behaviour of the planets and moons, and has been especially popular in the studies of very long time-scales such as glacial isostatic adjustment \citep{Al-AttarTromp13, Crawford_etal18, Ivins_etal21,lau_etal21}. This model, however, fails to accommodate the transient anelastic behaviour between the fully elastic and viscous regimes \citep[][]{Rambaux10, RenaudHenning18, CastilloRogezBanerdt12}, which makes this model inappropriate at seismic and oftentimes at tidal frequencies.

\noindent
Due to the shortcomings of the Maxwell model , in the later studies, it was suggested to rely on the combined Maxwell-Andrade model, often referred to as simply the Andrade model \citep[e.g.,][]{andrade1910, Rambaux10, Castillo, Efroimsky2012a}. %\citet{jackson2000laboratory}, \citet{Bills_etal05}, and \citet{SundbergCooper10} addressed
More sophisticated options, such as the Burgers model and the Sundberg-Cooper that incorporate anelasticity as a result of elastically-accommodated and dislocation- and diffusion-assisted grain boundary sliding processes \citep{burgers1935, SundbergCooper10,jackson2005laboratory,jackson2000laboratory,JacksonFaul10}. These models are corrected to take into account the effect of grain size, frequency, temperature, and pressure on the dissipation~-~a fundamental set of parameters that are needed to describe planetary interiors \citep[e.g.,][]{JacksonFaul10}.

\noindent
Geophysical analysis enables us to test the viscoelastic models against the attenuation data gleaned over a broad frequency range: from seismic wave periods ($\leq$1 s) over normal modes ($\sim$1 hr), bodily tides (hrs-days), and Chandler Wobble (months), to very long-period tides ($\sim$20 years). This frequency gamut is spanning five orders.
Equipped with this knowledge, we can then model the quality factors of planets and use the available measurements to predict the dissipation behaviour over a large range of periods. Figure~\ref{fig:quality_factor_earth} shows the measured quality factor of the solid Earth as a function of period, ranging from seconds to years, based on the extended Burgers viscoelastic model (described in Section \ref{sec: Burgers}) constrained by geophysical observations \citep{LauFaul19}. \textcolor{black}{ In this figure, dissipation in the measured surface waves, normal modes, semi-diurnal M2 tides, Chandler Wobble, and the 18.6~yr long-period tides are taken into account.}
As shown in the figure, studying the dissipation in the Earth, particularly its frequency dependence, has resulted in diverse interpretation, revealing the complexity of this process and the need for further considerations.

\noindent
Studying the tides is not limited to the rocky planets.
In the recent years, substantial attention has been devoted to the tidal dynamics of the icy systems such as Trans-Neptunian Objects, including both the Pluto-Charon binary and Kuiper belt objects \citep[e.g.,][]{Bierson_etal20, bagheri_etal22,Rhoden_etal20b,Saxena_etal18, arakawa_etal21, Renaud_etal21}, and the Galilean moons, Iapetus and Enceladus \citep[e.g.,][]{Kamata17, shoji_etal14,spencer_etal13, tyler2009,tyler2014,beuthe19,tyler11}. All these bodies are either presently tidally active or experienced significant tidal activity in their history. Understanding the tidal response of these bodies is essential to constrain their origin and in some cases, more importantly, to assess the potential habitability of their interior.
\noindent
Modeling their evolution in their past history requires information on dissipation in ice. Due to the much lower viscosity of ice in comparison to rocks \citep{MccarthyCooper16}, dissipation in ice can easily dominate that in the rock in the bodies containing both fractions. This applies also to binaries in the outer Solar system where the temperature is too low for the rocky fraction to dissipate energy, and most of the tidal dissipation is taking place in the ice layers. The viscoelastic behaviour of ice is studied using the same models as those used for rocks \citep{Renaud_etal21, SundbergCooper10,bagheri_etal22}.

\noindent
The mentioned viscoelastic models have been used to study dissipation in the planetary bodies (discussed in Section {\ref{interiors}), although the measurements in the case of other planetary bodies than the Earth are considerably more limited. In the next sections, we review the theoretical aspects of material viscoelastic properties followed by a summary of several laboratory-based rheological models. \textcolor{black}{Note that here we are mostly focusing on the solid body tides, not on the dissipation in fluid parts, such as in the molten core or in the surface or subsurface oceans. The effect of liquid parts is a complex process believed to be responsible for phenomena such as libration of Mercury \citep[e.g.,][]{rambaux2007, margot_etal07,peale05} or precession of the Moon \citep[e.g.,][]{yoder81, cebron_etal19, Williams_etal01}. Moreover, it has been argued that in some circumstances dissipation in the ocean can dominate that in the solid parts \citep{tyler08, tyler_etal15, tyler2009}. However, detailed discussion on the possibility of significant tidal dissipation in the ocean is not in the scope of this study }. 
\begin{center}
\begin{figure}[ht!]
\hspace{2.5cm}\includegraphics[width=0.68\textwidth]{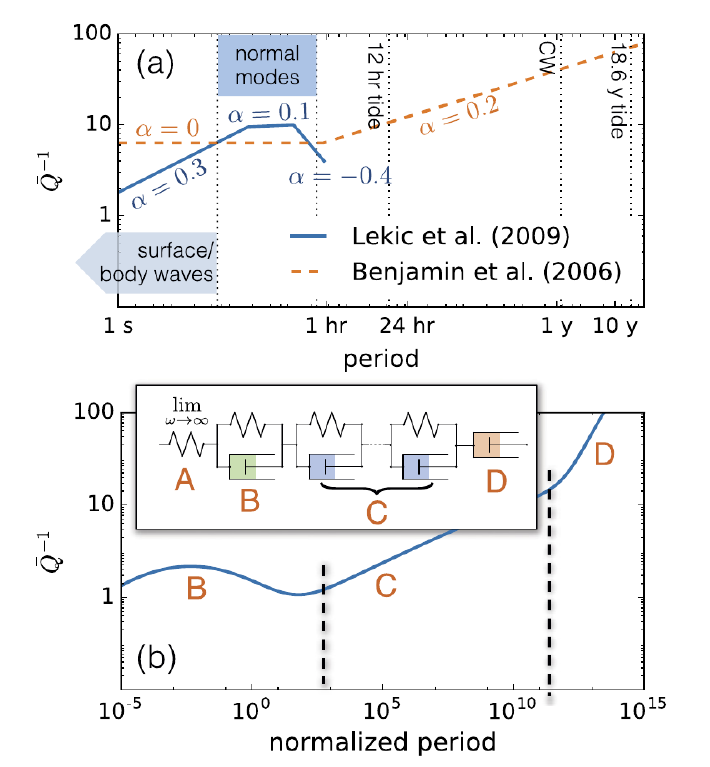}
    \caption{\small{(a) The frequency dependence of normalised attenuation, ${\bar{Q}}^{-1}$ for the solid Earth, displaying the contrasting absorption bands of \citet{Benjamin_etal06} and \citet{Lekic_etal09}  (orange–dashed and blue–solid lines, respectively). The former study used geodetic observations of the semi-diurnal and long period tides, and the Chandler Wobble (CW), whose periods are indicated by the vertical black dotted lines. The latter study used surface wave and normal mode data whose frequency bands are marked by blue boxes. (b) Main panel: schematic frequency dependence of attenuation of the Extended Burgers model \citep{FaulJAckson15}. Inset: the mechanical components of the Extended Burgers model in spring-dashpot representation. Figure~and caption modified from \citet{LauFaul19}.}}
    \label{fig:quality_factor_earth}
\end{figure}
\end{center}
  \clearpage

  \subsection{Constitutive equation}

 Rheological properties of a material are expressed by a constitutive equation linking the present-time deviatoric strain tensor $\,u_{\gamma\nu}(t)\,$ with the values assumed by the deviatoric stress $\,{\sigma}_{\gamma\nu}(t\,')\,$ over the time $\,t\,'\,\leq\,t\,$. When the rheological response is linear, i.e. when the resulting strain is linear in stress (which is usually the case for strains not exceeding $10^{-6}$ \citep{Karato08}\,),  the equation is a convolution, in the time domain:
 \begin{eqnarray}
 2\,u_{\gamma\nu}(t)
 %  \,=\,\hat{J}(t)~\sigma_{\gamma\nu}
 \,=\,\int^{t}_{-\infty}\stackrel{\;\centerdot}{J}(t-t\,')~
 {\sigma}_{\gamma\nu}(t\,')\,d t\,'~~,~~~
 \label{I12_4}
 \label{E1}
 \end{eqnarray}
 and is a product, in the frequency domain:
 \begin{eqnarray}
 2\;\bar{u}_{\gamma\nu}(\chi)\,=\;\bar{J}(\chi)\;\bar{\sigma}_{\gamma\nu}(\chi)\;\;.
 \label{LLJJKK}
 \label{E2}
 \end{eqnarray}
 In equation (\ref{E1}), the kernel $\stackrel{\;\centerdot}{J}(t-t\,')$ is a time derivative of the {\it{compliance function}}  ${J}(t-t\,')$, also called {\it{creep function}}, which carries all information about the (linear) rheological behaviour of the material.
 
 \noindent
 In equation (\ref{E2}), $\chi$ stands for the frequency, $\,\bar{u}_{\gamma\nu}(\chi)\,$ and $\,\bar{\sigma}_{\gamma\nu}(\chi)\,$ denote the Fourier images of the strain and stress tensors, while the {\it{complex compliance function}}, or simply {\it{complex compliance}} $\,\bar{J}(\chi)\,$
 is a Fourier image of $\,{J}(t-t\,')\,$. For details, see, e.g. \citet{Efroimsky2012a}.

\noindent
While elasticity is a result of bond-stretching along crystallographic planes in an ordered solid, viscosity and dissipation inside a polycrystalline material occur by motion of point, linear, and planar defects, facilitated by diffusion. Each of these mechanisms contributes to viscoelastic behaviour \citep[e.g.,][]{Karato08}.
Deformation of a viscoelastic solid depends on the time evolution of the applied load \citep{chawla_etal99}. For small deformations, the stress-strain relation is linear, and the response is described, in the time-domain, by the creep function $J(t)$. Its time derivative links the material properties and forcing stress (the input) with the resulting strain magnitude and phase lag  (the output), see equation (\ref{E1}). The response of the material to forcing comprises~ (a) an instantaneous elastic response,\, (b) a semi-recoverable transient flow regime where the strain rate changes with time, and (c) a steady-state creep. Accordingly, the creep function for a viscoelastic solid consists of three terms:
\begin{equation}
\underbrace{J(t)}_\text{Creep function} = \underbrace{J}_\text{Elastic} + \underbrace{f(t)}_\text{Transient strain-rate} + \underbrace{{t}{/}{\eta_{}}}_\text{Viscous},
\label{creepfunction}
\end{equation}
$t$ and $\eta$ being time and the shear viscosity. The Fourier image of $J(t)$ is the complex compliance
$\bar{J}(\chi) = \Re[\bar{J}(\chi)] + i \Im[\bar{J}(\chi)]$, where $\chi$ is the frequency.
The associated shear quality factor is given by
\bs 
\ba 
{Q_s(\chi)}^{-1}\,=\, \frac{|\,\Im[\bar{J}(\chi)] \,|}{\,\sqrt{\Re^{\textstyle{^2}}[\bar{J}(\chi)] + \Im^{\textstyle{^2}} [\bar {J}(\chi)]}}
\;\,.
\label{GandQmu1}
\ea
For solids far from the melting point, the instantaneous (elastic) part of deformation is usually sufficiently large: $\;|\s\Re[\bar{J}(\chi)]\s| \gg |\s\Im[\bar{J}(\chi)]\s|\,$, in which case we have
\ba 
{Q_s(\chi)}^{-1}\,\approx\,\frac{|\,\Im [\bar{J}(\chi)]\,|}{\Re [\bar J(\chi)]}\;\,.
\label{appro}
\ea 
\es
This quality factor is responsible for attenuation of seismic waves, and varies over depth and over geological basins\footnote{Note that seismic attenuation takes place as a result of three effects: intrinsic anelasticity, geometric spreading, and scattering attenuation. The viscoelastic models discussed here only represent the attenuation due to intrinsic anelasticity and not the other two effects, all of which have to be taken into account in the study of dissipation of seismic waves \citep[See, e.g.,][]{cormier89, Lognonne_etal20, bagheri_etal15,lissa_etal19,winkler_etal79,margerin_etal2000}}.\, An intrinsic material property, $Q_s(\chi)$ is different from the degree-$l$ tidal quality factors $Q_l(\chi)$ characterising the planet as a whole. As explained in \citet{Efroimsky2012a,Efroimsky15} and \cite{Lau_etal17}, the distinction comes from the fact that the tidal factors are defined by interplay of self-gravitation with the overall rheological behaviour (generally, heterogeneous). In simple words, self-gravitation pulls the tidal bulge down, thus acting as an effective addition to rigidity. While at sufficiently high frequencies this effect is negligible, it becomes noticeable at the lowest frequencies available to analysis.

\noindent
Below, we consider the Maxwell, extended Burgers, Andrade, Sundberg-Cooper rheologies, as well as a simplified power-law scheme. These models were derived from laboratory studies of various regimes of viscoelastic and anelastic relaxation. The applicability realm of each model depends on such parameters as the grain size, temperature, and pressure. 
 % For example, \cite{JacksonFaul10} explore the applicability of the Maxwell, extended Burgers, Andrade, and power-law models to melt-free polycrystalline olivine (grain sizes in the range 3--165~$\mu$m), at temperatures 800--1200$^\circ$C, subject to torsional forcing  oscillations with periods of 1--1000~s. A more elaborate model of \citet{SundbergCooper10} is based on torsional oscillation data obtained on a fine-grained (5~$\mu$m) peridotite (olivine+39 vol\% orthopyroxene) specimen (temperature range 1200--1300$^\circ$C and periods of 1 -- $\sim$200~s).
 % As shown in figure \ref{fig:springdashpot},
Each model can be represented as an arrangement of springs and dashpots connected in series, or in parallel, or in a combination of both connection methods as shown in Figure~\ref{fig:spring}. \citep{FindleyOnaran65,MozcoKristek05,NowickBerry72,cooper02,jackson07,mccarthyCastilloRogez13}. Instantaneous elastic response is mimicked by a spring, while a fully viscous behaviour is modeled with a dashpot. A series connection (a Maxwell module) implies a non-recoverable displacement, while a parallel connection (a Voigt module) ensures a fully recoverable deformation. Schematic diagrams of four viscoelastic models are shown in Figure~(\ref{fig:spring}).

\begin{figure}
\begin{center}
\includegraphics[width=0.765\textwidth]{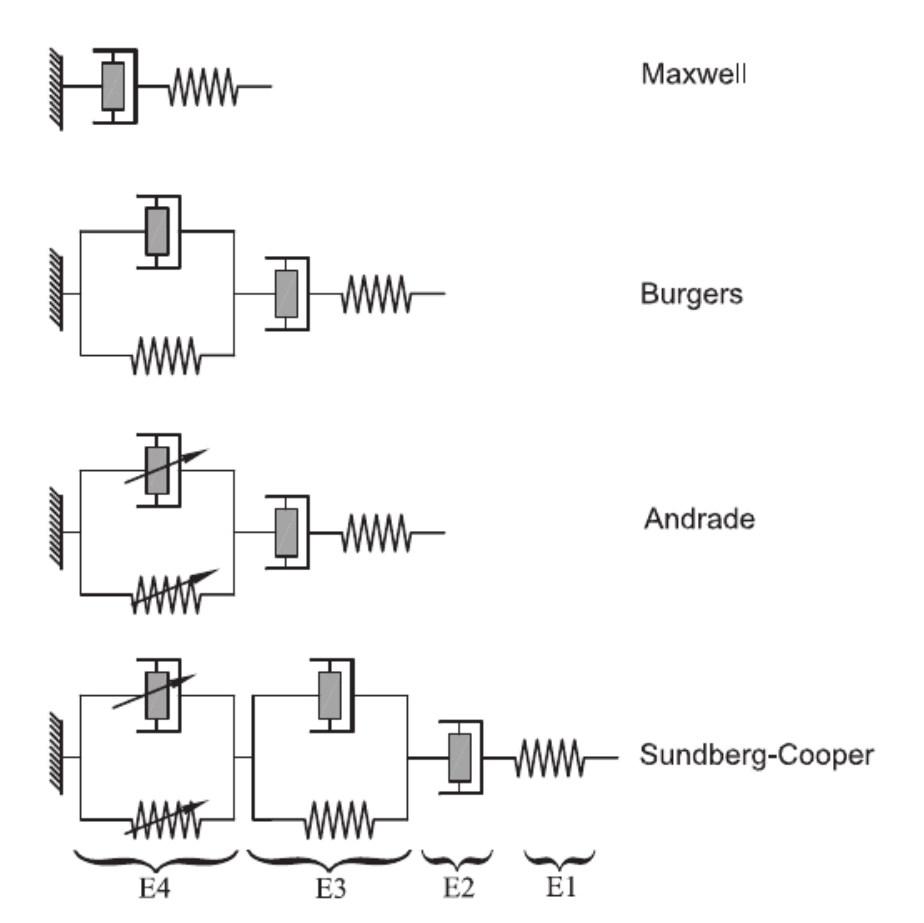}
\caption{\small{~Schematic representation of viscoelastic models in terms of springs and dashpots. A spring element (E1) represents a purely elastic response, while a dashpot element (E2) is representative of purely viscous damping. A series connection of elements 1 and 2 is representative of the response of a Maxwell model (irrecoverable), while a connection of elements 1 and 2 in parallel (element 3) results in an anelastic (recoverable) response with a discrete (single) spectrum of relaxation times. Arrows on spring and dashpot in element 4, conversely, indicate an element that incorporates a continuous distribution of anelastic relaxation times and results in a broadened response spectrum \citep{Bagheri_etal19}.}}
\label{fig:spring}
\end{center}
\end{figure}

\subsection{Maxwell}\label{maxwell}

Maxwell is the simplest model of  the viscoelastic behaviour and can be interpreted as a series connection of a spring and dashpot. In the time domain, the creep function for this model is: \begin{equation}
J(t) = J +  \frac{t}{\eta}\,\;,
\label{maxwellcreep}
\end{equation}
$\eta$ being the viscosity, and $J$ being the shear elastic compliance related to the shear elastic modulus $\mu$~by
\begin{eqnarray}
J\,=\,\frac{1}{\mu}\;\,.
\label{}
\end{eqnarray}
The real and imaginary parts of the complex compliance are
\begin{eqnarray}
 \Re[\bar J(\chi)] &=& J\;\,,
 \label{77}\\
 \Im[\bar J(\chi)] &=&-\,\frac{1}{\chi\,\eta} \,= \,-\,J\,\frac{1}{\chi\,\tau_M}\;\,,
 \label{88}
\end{eqnarray}
 %  wherefrom the real and imaginary parts of the complex shear modulus are
 %  \begin{equation}
 %  \Re[\bar G(\omega)] = \frac{{\tau_M}^2 \omega^2 }{ J_U({\tau_M}^2 \omega^2+ 1)},
 %  \end{equation}
 %  \begin{equation}
 %  \Im[\bar G(\omega)] = \frac{{\tau}_M \omega}{{J_U(\tau_M}^2 \omega^2 + 1)},
 %  \end{equation}
where $\chi$ is the frequency, and the Maxwell time is introduced as \begin{eqnarray}
\tau_M\,=\,\eta\s J\,=\,\frac{\eta}{\mu}~~.
\end{eqnarray}
\noindent
It is the timescale of relaxation of strain after the stress is abruptly turned off.

\noindent
Lacking a transient term, the Maxwell model implies an elastic regime at high frequencies and a viscous-fluid regime in the low-frequency limit. It provides a reasonable approximation for reaction to very-long-period loading such as glacial isostatic adjustments \citep{Peltier74}. On the other hand, the model is less adequate at tidal and especially seismic frequencies where transient processes are present.

\subsection{Burgers and Extended Burgers}\label{sec: Burgers}

The shortcoming of Maxwell's model in representing a transient response at the frequencies residing between the elastic and viscous bands can be rectified by introducing a time-dependent anelastic transition between these two regimes. This gives birth to the {\it{Burgers rheology}}:
 \begin{equation}
 \label{burgersdistribution}
 J(t) \s=\s J_U \s \left[\,1\,+ \,\Delta \s\left[ 1\s-\s \exp\left(\,-\;\frac{t}{\tau}\right)\right]\, +\,\frac{t}{\tau_{M}}\,\right]\;\,,
 \end{equation}
where $J_U$ is the elastic compliance, $\Delta $ is the so-called {\it{anelastic relaxation strength}}, and $\tau$ is a characteristic time of the development of anelastic response. While the Maxwell model does not discriminate between the unrelaxed and relaxed values of $J$, the Burgers model does. From the above expression for $J(t)$, we observe that while $J_U$ is the unrelaxed value, $J_U\s(1+\Delta)=J_R$ is its relaxed counterpart. This conclusion can be deduced also from the expression for the compliance in the complex domain:
\ba
 \bar{J}(\chi)\s=\s\Re\left[\bar{J}(\chi) \right]\,+\,i\,\Im\left[\bar{J}(\chi) \right]\;\;,
 \ea
 \ba
 \Re[\bar{J}(\chi)]\s=\s J_U\left(1\s+\s\frac{\Delta }{1\s+\s\chi^2\s\tau^2}\s\right)~~~,\qquad
 %  \ea  \ba
 \Im[\bar{J}(\chi)]\;=\;J_U\,\Delta \;\frac{\chi\s\tau}{1\s+\s\chi^2\s\tau^2}~~~.
 \label{}
 \ea
 By inserting these expressions for $\Re[\bar{J}(\chi)]$ and $\Im[\bar{J}(\chi)]$ into definition (\ref{GandQmu1}) for $\s Q^{-1}_s\s$, \,it is easy to demonstrate that for this model the inverse quality factor possesses a peak, which makes the model applicable to realistic situations where such a peak appears in experiments.

\noindent
More generally, the anelastic relaxation time $\tau$ can be replaced with a distribution $D(\tau)$ of relaxation times over an interval specified by an upper ($\tau_H$) and lower ($\tau_L$) bounds \citep{JacksonFaul10}. From a micromechanical point of view, this distribution is associated with diffusionally accommodated grain-boundary sliding. This is how the so-called {\it{extended Burgers model}} emerges:
 \begin{equation}
 \label{Eq:ExtBurgJ}
 J(t) = J_U \left[ 1 + \Delta \int_{\tau_L}^{\tau_H} D(\tau) \left[1 - \exp \left(- \;\frac{t}{\tau} \right)\right] d\tau + \frac{t}{\tau_M} \right]\,\;,
 \end{equation}
where $J_U\Delta$, as in the Burgers model, is the increase in compliance associated with complete anelastic relaxation.

\noindent
The corresponding components of the complex compliance are:
\ba
\Re[\bar J(\chi)] &=& J_U \bigg( 1 + \Delta \int_{\tau_L}^{\tau_H} \frac{D(\tau) }{1 + \chi^2 \tau^2}\;  d\tau \bigg)\,\;,\\
\nonumber\\
\Im[\bar J(\chi)] &=& -\,J_U \bigg( \chi\;\Delta \int_{\tau_L}^{\tau_H} \frac{\tau D(\tau) }{1 + \chi^2 \tau^2} \; d\tau\, +\, \frac{1}{\chi\s \tau_M} \bigg)\,\;.
\ea

\noindent
A commonly used distribution of anelastic relaxation times associated with the background dissipation is the absorption band model proposed by \citet{MinsterAnderson81}. Within that model, $D(\tau)$ is implemented by the function $D(\tau,\,\alpha)$ bearing a dependence on a fractional parameter $\alpha\,$:
\begin{eqnarray}
D_B (\tau,\,\alpha)\;=\;\left\{
\begin{array}{ll}
 0                                                                  &    \mbox{for}\;\;\tau\notin[\tau_L\,,\; \tau_H]\,\;,\\
 %  ~\\
 \frac{\textstyle \alpha \tau^{\alpha-1}}{\textstyle{{\tau_H}^\alpha - {\tau_L}^\alpha}}   &    \mbox{for}\;\;\tau_L < \tau < \tau_H \,\;,
\end{array}
\right.
\label{anderson}
\end{eqnarray}
where $0 < \alpha < 1\,$, while $\tau_L$ and $\tau_H$ denote the cut-offs of the absorption band where dissipation is frequency-dependent and scales, approximately,\,\footnote{~Using approximation (\ref{appro}),  estimating the real part of the compliance as its elastic term, and also neglecting the viscous term in the imaginary part, we observe that 
\ba 
Q^{-1}_s(\chi)\propto \chi\,\Delta\int_{\tau_L}^{\tau_H}\frac{\tau\,D(\tau)\,d\tau}{1+\chi^2\tau^2}\,=\,\frac{\alpha\,\Delta}{\tau_H^\alpha -\tau_L^\alpha}\int_{\chi\tau_L}^{\chi\tau_H}\frac{\chi^{-1}(\chi\tau)\;d(\chi\tau) }{1+\chi^2\tau^2}\;\chi^{1-\alpha}(\chi\tau)^{\alpha-1}\;=\;\frac{\alpha\,\Delta}{\tau_H^\alpha -\tau_L^\alpha}\;\chi^{-\alpha}\;A(\chi)~~,
\nonumber
\ea 
where 
$$
A(\chi)\equiv\int_{\chi\tau_L}^{\chi\tau_H}\frac{\textstyle z\;dz}{\textstyle 1+z^2}\;=\;\frac{\textstyle 1}{\textstyle 2} \ln(1+z^2)\Big{|}_{\chi\tau_L}^{\chi\tau_H}\,=\,\frac{1}{2}\;\ln\frac{1+(\chi\tau_H)^2}{1+(\chi\tau_L)^2}~~.
$$
Since the function $A(\chi)$ is slower that the power function $\chi^{-\alpha}$, we may say that within this crude approximation the quality factor scales as about $\chi^\alpha$.
} ~as $Q_s\propto \chi ^\alpha$. The lower bound of the absorption band ensures a finite shear modulus at high frequencies and restricts attenuation at those periods.

\noindent
 \cite{JacksonFaul10} found that their experimental data were better fit by including a dissipation peak in the distribution of anelastic relaxation times, which is superimposed upon the monotonic background along with the associated dispersion. This background peak is attributed mostly to sliding with elastic accommodation of grain-boundary incompatibilities \citep[see][for a different view]{Takei_etal14}. In this case, $D(\tau)$ writes as
\begin{equation}
D_P(\tau) = \frac{1}{\sigma \tau \sqrt{2\pi}} \exp \bigg( \frac{-\ln {({\tau}/{\tau_P})}}{2\, \sigma^2}        \bigg)
\label{tau}
\end{equation}
peaked around some $\tau_P\s$, a new timescale to be a part of the model.

\subsection{Andrade\label{Andrade}}

While the extended Burgers model incorporates a distribution of relaxation times within a restricted time-scale to account for the transient anelastic relaxation, the Andrade model
implies a distribution of relaxation times over the entire time domain:
\begin{equation}
J(t) =  J_U + \beta\s t^\alpha + \frac{t}{\eta}\,\;,
\end{equation}
%(represented by arrows on spring and dashpot). % This configuration of a Maxwell module and a ``modified" Voigt module results in a creep function of the form \citep{da1962validity}
where $\alpha$ defines the frequency-dependence of the compliance,\,\footnote{~Following a long-established convention, we are denoting the Andrade dimensionless parameter with $\alpha$. For the same reason, we denoted with $\alpha$ a parameter emerging in distribution (\ref{anderson}). It should however be kept in mind that these two $\alpha$`s are different parameters and assume different values.\label{differ}} \,while $\beta$ is qualitatively playing the same role as $\Delta$ in the extended Burgers model (and may, in principle, change with frequency).
\noindent
Having fractional dimensions, the parameter $\beta$ is somewhat unphysical. For this reason, it was suggested by \citet{Efroimsky2012a,Efroimsky15} to cast the compliance as
\begin{equation}
J(t) =  J_U\,\left[1\, +\, \left(\frac{t}{\tau_A}\right)^\alpha + \frac{t}{\tau_M}\right]\,\;,
\label{equation}
\end{equation}
with the parameter $\tau_A$ defined through
\ba
\beta\,=\,\tau_A^{-\alpha}\;J_U
\label{}
\ea
and named {\it{the Andrade time}}. A justification for this reformulation will be provided shortly.

\noindent
 With $\Gamma$ denoting the Gamma function, the complex compliance corresponding to (\ref{equation}) writes as:
 \ba
 \bar{J}(\chi)\,=\,J_U\,\left[
 1\,+\,(i\chi\tau_A)^{-\alpha}\,\Gamma(1+\alpha)\,-\,i(\chi\tau_M)^{-1}
 \right]\,\;,
 \label{}
 \ea
its real and imaginary parts being
\begin{equation}
\Re[\bar J(\chi)] = J_U\s \bigg [ 1 +  \Gamma(1+\alpha)\; (\chi\tau_A)^{-\alpha}\,\cos \left(\frac{\alpha\pi}{2} \right) \bigg]\,\;,
\label{28}
\end{equation}
\begin{equation}
\Im[\bar J(\chi)] =\,-\,J_U\s \bigg [ \Gamma(1+\alpha)\; (\chi\tau_A)^{-\alpha}\,\sin \left(\frac{\alpha\pi}{2} \right) \,+\, (\chi \tau_M)^{-1} \bigg]\,\;.
\label{29}
\end{equation}
\noindent
The absorption band in this model extends from 0 to $\infty$. In other words, anelastic relaxation effectively contributes across the entire frequency range from short-period seismic waves to geological time-scales. This generates two problems.

\noindent
First, in the situations where the anelastic behaviour is dominated by defect unjamming, it has a low-frequency cut-off, as explained by \citet[eqn 17]{KaratoSpetzler90}.\,\footnote{~According to Figure~3 in \citet{KaratoSpetzler90}, for the Earth mantle the threshold is located at about 1/yr. However, due to the sensitivity of this threshold to the temperature and pressure, for the mantle as a whole this threshold is smeared into a transition zone covering a decade or two.} The presence of this feature can be built into the Andrade model ``by hand,'' by assuming that the Andrade time quickly grows to infinity (or, equivalently, that the parameter $\beta$ quickly reduces to zero) as the frequency goes below some threshold value. So, for frequencies below this threshold, the response of the material is overwhelmingly viscoelastic, while above the threshold the response becomes predominantly anelastic. At this point, we can appreciate the convenience of employing $\tau_A$ instead of $\beta\,$: it turns out \citep[Fig 4]{Castillo} that for olivines $\tau_A$ and $\tau_M$  are similar over an appreciable band of frequencies:
\ba
\tau_A \approx \tau_M\,\;.
\label{}
\ea
%\textcolor{black}{JULIE,\\
%IS THIS APPROXIMATE EQUALITY VALID ALSO FOR ICES? IF YES, COULD YOU PLEASE ADD A SENTENCE WITH A REFERENCE?\\
%ALSO, DO YOU HAPPEN TO KNOW IF THE THRESHOLD DISCUSSED BY KARATO \& SPETZLER WAS ADDRESSED BY OTHER AUTHORS?\\}
\noindent
Second, within the Andrade model it is impossible, by distinction from the Burgers and extended Burgers models, to write down the relaxed value of $J$. This problem finds its resolution within the Sundberg-Cooper and extended Sundberg-Cooper models discussed below.

\subsection{Sundberg-Cooper}

Similarly to the Maxwell model, in the Andrade model it is conventional to identify the elastic compliance $J$ with its unrelaxed value $J_U$. Consequently, just as the Maxwell model is extended to Burgers, so the Andrade model can be extended to that of \cite{SundbergCooper10}, to account for the combined effects of diffusional background and elastically-accommodated grain-boundary sliding:
%
% Andrade's model does not feel the unrelaxed compliance in case elastically-accommodated grain-boundary sliding is sufficiently rapid.
%In order to reconcile an attenuation plateau observed in the measurements at high-temperatures (1200--1300$^\circ$ C) and low frequencies (0.0056--1~Hz),
%This model graphically consists of two Voigt modules and a Maxwell module (cf. Figure~\ref{figure}); One module is similar to that used in Andrade's model (E4), whereas the other module is equivalent to that of the extended Burgers model (E3).
\bs
\ba
J(t) = J_U +J_U\Delta  \,\left( 1 -   e^{-\;{t}/{\tau} }  \right) + \beta\, t^\alpha+  \frac{t}{\eta}\;\;.
\label{sc1}
\ea
In terms of the Maxwell and Andrade times, this creep function can be written down as
\ba
J(t) = J_U\s\left[\,1\,+\,\Delta  \,\left( 1 -   e^{-\;{t}/{\tau} }  \right) + \left(\frac{t}{\tau_A}\right)^\alpha+  \frac{t}{\tau_M}\,\right]\;\,.
\label{sc2}
\ea
\label{sc}
\es
 In the frequency domain, this compliance becomes
 \ba
\Re[\bar J(\chi)] = J_U \s\bigg [\s 1\, +\,\Gamma(1+\alpha)\; \left(\chi\s \tau_A\right)^{-\alpha} \cos \left(\frac{\alpha\pi}{2} \right) \,+\, \frac{\Delta}{1 + \chi^2 \tau^2}\,\bigg]\,\;,
\label{}
\ea
\ba
\Im[\bar J(\chi)] = \,-\,J_U \s\bigg [\s \Gamma(1+\alpha)\, (\chi\s \tau_A)^{-\alpha} \sin \left(\frac{\alpha\pi}{2}  \right) \s+\s \chi\; \frac{\tau\, \Delta }{1 + \chi^2 \tau^2}\;   \,+\, \left(\chi\s \tau_M\right)^{-1} \s\bigg]\,\;,
\label{}
\ea
 the corresponding $Q_s^{-1}(\chi)$ possessing a peak.

\noindent
 A further extension of the Sundberg-Cooper model can be performed in analogy with the extended Burgers model. The term containing the parameter $\tau$ can be replaced with an integral specifying a distribution of anelastic relaxation times $\tau$, as in equation (\ref{Eq:ExtBurgJ}).

\noindent
 % Similar to the case of extended Burgers model, modifications for the grain size, temperature, and pressure are allowed for through equation (\ref{gptmodification}).
% Equivalently, these modifications can be performed by introducing the pseudo-period master variable $X$ given by equation (\ref{XB}). 
The real and imaginary parts of the complex compliance for the extended Sundberg-Cooper model are:
\ba
\Re[\bar J(\chi)] = J_U \s\bigg [\s 1\, +\,\Gamma(1+\alpha)\; \left(\chi\s \tau_A\right)^{-\alpha} \cos \left(\frac{\alpha\pi}{2} \right) \,+\, \Delta \int_{\tau_L}^{\tau_H} \frac{D(\tau) }{1 + \chi^2 \tau^2} \;d\tau  \,\bigg]\,\;,
\label{37}
\ea
\ba
\Im[\bar J(\chi)] = \,-\,J_U \s\bigg [\s \Gamma(1+\alpha)\, (\chi\s \tau_A)^{-\alpha} \sin \left(\frac{\alpha\pi}{2}  \right) \s+\s \chi\;\Delta \int_{\tau_L}^{\tau_H} \frac{\tau\, D(\tau) }{1 + \chi^2 \tau^2}\;  d\tau  \,+\, \left(\chi\s \tau_M\right)^{-1} \s\bigg]\,\;,
\label{38}
\ea
where $D(\tau)$ is given either by expression (\ref{tau}) or by (\ref{anderson}).
In the latter case, it is important to mind the difference between the Andrade exponential $\alpha$ and the parameter entering distribution (\ref{anderson}).  While in equation (\ref{anderson}) we denoted the parameter with the same letter $\alpha$ and wrote the function as $D_B(\tau,\,\alpha)$, in equations (\ref{37} - \ref{38}) this function should appear as  $D_B(\tau,\,\alpha_1)$, with $\alpha_1$ generally different from the Andrade $\alpha$.

\subsection{Power-law Approximation}

Finally, we consider a power-law approximation sometimes used for fitting measurements \citep[e.g.,][]{Jackson_etal02}. As we shall now demonstrate, this description follows from the Andrade model (\ref{28} - \ref{29}) under the simplifying assumptions that the anelastic dissipation is weak and that the viscoelastic dissipation is even weaker: \footnote{~We indeed see from equation (\ref{28}) that assumption (\ref{39b}) 
 %  inequality $\,(\chi\s\tau_A)^{-\alpha}\s\ll\,1\,$ 
 implies the weakness of anelastic dissipation. It can also be observed from equation (\ref{29}) that assumption (\ref{39a}) % $\,(\chi\s\tau_M)^{-1}\,\ll\,(\chi\s\tau_A)^{-\alpha}\,$ 
implies the weakness of viscoelasticity as compared to anelasticity.} 
\bs 
 \ba 
  \mbox{Assumption~\,1}\s:& &
 \quad
%  (\chi\s\tau_M)^{-1}\,\ll\,
 (\chi\s\tau_A)^{-\alpha}\s\ll\,1\,\;,
 \label{39b}\\
    \mbox{Assumption~\,2}\s:& & \quad(\chi\s\tau_M)^{-1}\,\ll\,(\chi\s\tau_A)^{-\alpha}
 %  \s\ll\,1
 \,\;.\qquad\qquad
 \label{39a}
 \ea 
 \label{39}
 \es 
\noindent
Approximations (\ref{appro}) and (\ref{39a}) enable us to write the inverse shear quality factor as
\ba
Q_s^{-1}\,\equiv\;\frac{|\,\Im[\bar{J}(\chi)]\,|}{
|\bar{J}(\chi)|}\;\approx\;\frac{J_U}{|\bar{J}(\chi)|}\;\Gamma(1+\alpha)\;(\chi\s\tau_A)^{-\alpha}\;\sin\left(\frac{\alpha\s\pi}{2}\right)\,\;.
\label{40}
\ea
Equivalently,
\ba
Q_s^{-1}\,\cot\left(\frac{\alpha\s\pi}{2}\right)\approx\;\frac{J_U}{|\bar{J}(\chi)|}\;\Gamma(1+\alpha)\;(\chi\s\tau_A)^{-\alpha}\;\cos\left(\frac{\alpha\s\pi}{2}\right)\,\;,
\label{41}
\ea
 With aid of inequalities (\ref{39}), $|\bar{J}(\chi)|$ can be written down as 
 %  \ba
 %  \nonumber
 %  |\bar{J}(\chi)|  \approx  J_U\s\sqrt{1+2\s\Gamma(1+\alpha)\,(\chi\tau_A)^{-\alpha}\,\cos\left(\frac{\alpha\s\pi}{2}\right)\,+\,\left(\Gamma(1+\alpha)\,(\chi\s\tau_A)^{-\alpha}\right)^2\,}\;\,.
 %  \label{42}
 %  \ea
 % By inequality (\ref{39b}), it can be further simplified as
\ba
|\bar{J}(\chi)|\;\approx\;J_U\s\left[1\s+\s\Gamma(1+\alpha)\,(\chi\s\tau_{A})^{-\alpha}\,\cos\left(\frac{\alpha\s\pi}{2}\right)\s\right]~~.
\label{43}
\ea
%  or, the same:
%  \ba
%  1\;-\;\frac{J_U}{|\bar{J}(\chi)|}\;\approx\;
%  \frac{J_U}{|\bar{J}(\chi)|}\;\Gamma(1+\alpha)\,(\chi\s\tau_A)^{-\alpha}\s\cos\left(\frac{\alpha\s\pi}{2}\right)\;\,.
 %  \label{44}
 %  \ea
Combining equations (\ref{41}) and (\ref{43}), we arrive at
\ba
\frac{J_U}{|\bar{J}(\chi)|}\;\approx\;1\;-\;Q_s^{-1}\,\cot\left(\frac{\alpha\s\pi}{2}\right)\,\;.
\label{45}
\ea
Also, the approximate expression (\ref{40}) 
 % for the inverse shear quality factor 
 can be concisely reparameterised through the forcing period $2\pi/\chi\,$:
\ba
Q_s^{-1}\s\approx\,A\,\left(\frac{2\s\pi}{\chi}\right)^{\alpha}~~~, \qquad A\s=\s\Gamma(1+\alpha)\,\sin\left(\frac{\alpha\pi}{2}\right)\,(2\pi\tau)^{-\alpha}
~~~.
\label{46}
\ea
 Together, equations (\ref{45} - \ref{46}) constitute a simplified version of the Andrade model, a version that is valid when both assumptions (\ref{39}) are fulfilled, which is often the case at seismic frequencies. 
 \noindent
\textcolor{black}{Figure~\ref{fig:visc_models_shear} compares the frequency-dependent shear modulus and the quality factors rendered by different rheological models \citep{Bagheri_etal19}. This figure readily reveals that Maxwell model and power-law fail to provide appropriate results in a wide range of periods. The other three models, i.e. the extended Burgers, Andrade, and Sundberg-Cooper ones, sometimes render similar dependencies, though in Section~\ref{interiors} we shall see that the choice of the right rheology may become very important in modeling of tides.}
 
 \begin{center}
\begin{figure}[ht]
\hspace{-10mm}\includegraphics[width=1.1\textwidth]{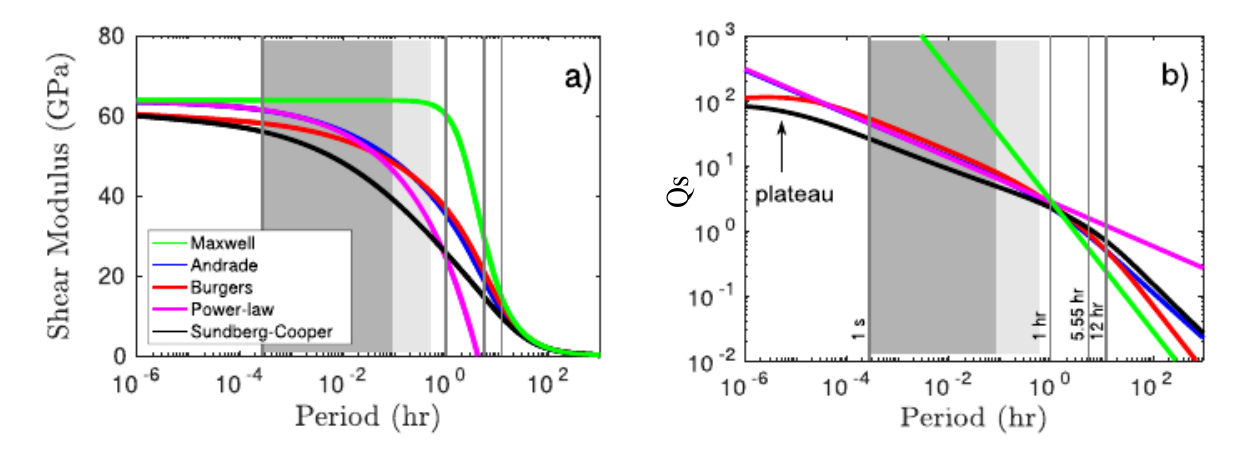}
\caption{Computed variations of relaxed shear modulus ($G_R$) and shear attenuation ($Q_s$) with period for the ﬁve rheological models considered in this study. (a, b) ($G_R$) and ($Q_s$) as a function of period at constant temperature and grain size; the vertical lines show periods of interest: seismic body waves (1 s), normal modes (1 hr), main tidal excitation of Phobos on Mars (5.55 hr), and main tidal excitation of the Sun (12.32 hr). Figure~is modified from \citet{Bagheri_etal19}}
\label{fig:visc_models_shear}
\end{figure}
 \end{center}
% Also, dDifferent viscoelastic models applied to Mars are discussed and compared in Section \ref{sec:Mars}. 

\subsection{Rescaling for different values of the temperature, pressure, and grain size}

 \noindent
 To make a rheological model practical, it is necessary to endow the timescales and other parameters of the model with a dependency on the grain size $d$, temperature $T$, and pressure $P$.
 For the Maxwell time, \citet{MorrisJackson09,JacksonFaul10,McCarthy_etal11} suggested the following rescaling prescription:
\begin{equation}
\label{eq:1}
\tau_M (T,P,d) = \tau_{M0} \bigg(\frac{d_g}{d_{0}}\bigg)^{m_{gv}} \exp \bigg[ \bigg(\frac{E^{*}}{R}\bigg)\bigg(\frac{1}{T} - \frac{1}{T_{0}}\bigg)\bigg] \exp \bigg[  \bigg(\frac{V^{*}}{R}\bigg) \bigg(\frac{P}{T} - \frac{P_{0}}{T_{0}}  \bigg) \bigg],
\end{equation}
where $R$ is the gas constant, $E^{*}$ is the activation energy, $V^{*}$ is the activation volume, $m_{gv}$ is the grain size exponent for viscous relaxation, $P$ is the pressure, $T$ is the temperature, and $\tau_{M0}$ is a normalised value of the Maxwell time at a particular set of reference conditions ($d_0$, $P_0$, and $T_0$).
In this expression, both exponential functions come from the rescaling of viscosity $\eta$, under the assumption that the rigidity $\mu$ stays unchanged under the variations of
both $T$ and $P$. This assumption is acceptable for temperatures not exceeding about 3/4 of the melting temperature, which in its turn, is a function of pressure.

 \noindent
 While the above formula is sufficient to rescale the Maxwell model, it is more complicated in the case of the Burgers model with its additional timescale $\tau$,
 and even more involved for the extended Burgers model with its three additional timescales $\tau_L$, $\tau_H$, $\tau_P$.
 \citet{JacksonFaul10} suggested to extend  (\ref{eq:1}) to all these timescales: \begin{equation}
\label{gptmodification}
\tau_i (T,P,d) = \tau_{i0} \bigg(\frac{d_g}{d_{0}}\bigg)^{m_g} \exp \bigg[ \bigg(\frac{E^{*}}{R}\bigg)\bigg(\frac{1}{T} - \frac{1}{T_{0}}\bigg)\bigg] \exp \bigg[  \bigg(\frac{V^{*}}{R}\bigg) \bigg(\frac{P}{T} - \frac{P_{0}}{T_{0}}  \bigg) \bigg],
\end{equation}
 where all parameters are as in equation (\ref{eq:1}), while $i = M,\s L,\s H, \s P$. The grain size exponential $m_g$ can be different for the anelastic ($m_{ga}$ for $i = L, H, P$) and viscous ($m_{gv}$ for $i = M$) relaxation regimes.
 % To more realistically account for variations of the unrelaxed shear modulus with temperature and pressure, \cite{JacksonFaul10} suggest the following modification
 % \begin{equation}\label{Eq:dGdPdT}
 % J_U (T,P) = \bigg[ G_U (T_{0},P_{0}) + (T-T_{0}) \frac{\partial G_U}{\partial T} + (P - P_{0})\frac{\partial G_U}{\partial P} \bigg]^{-1}.
 % \end{equation}
 % Values for the temperature and pressure derivatives are given in Table~\ref{table:visco_params}.
 %
 % It is recommended by \cite{JF10} that extended Burgers model is more realistic than other models as it better describes the transition from elasticity to grain-size sensitive viscoelastic behaviour.

 \noindent
 In the case of the Andrade model, \cite{JacksonFaul10} adjust for the variations of $d$, $T$, and $P$ by replacing the actual period with the following pseudo-period:
 \begin{equation}
 X = \chi^{-1}  \bigg(\frac{d_g}{d_{0}}\bigg)^{-m_g} \exp \bigg[ \bigg(\frac{-E^{*}}{R}\bigg)\bigg(\frac{1}{T} - \frac{1}{T_{0}}\bigg)\bigg] \exp \bigg[ \bigg(\frac{-V^{*}}{R}\bigg) \bigg(\frac{P}{T} - \frac{P_{0}}{T_{0}}  \bigg) \bigg]\,\;.
 \label{XB}
 \end{equation}
This is equivalent to a simultaneous rescaling of both the Maxwell and Andrade times by formula (\ref{eq:1}). To apply it to the Andrade time, one may hypothesise that $\tau_A$ and $\tau_M$ either are equal or are staying proportional within the considered realm of temperatures, pressures and grain sizes. For $\tau_A$, the power $m_g$ should assume its anelastic value $m_{ga}$, like in the case of the Burgers model. While this rescaling is natural for the Maxwell time, its applicability to the Andrade time still requires justification. 

\noindent
Similarly to the complete Andrade model, its simplified version (\ref{45} - \ref{46}) becomes subject to prescription (\ref{XB}). In equation (\ref{46}), a replacement of the actual forcing period $2\pi/\chi$ with the pseudo-period $X$ given by equation (\ref{XB}) yields:
 \ba
 Q_s^{-1}\approx A\s X^{\alpha}\;\,.
 \ea
 Just as in the case of the full Andrade model, the applicability of this rule to the power law needs further study.

\noindent
\textcolor{black}{The parameters involved in the viscoelastic models often need to be better constrained. Extensive experimental work has been carried out in this direction. \citep{jackson2000laboratory, jackson2005laboratory, JacksonFaul10, jackson07, Jackson_etal02, SundbergCooper10, Qu_etal21, McCarthy_etal11}  ~In addition, understanding of the effects of porosity or the presence of partial melt needs further research. Constraints on the effect of the involved parameters will help to obtain more detailed insights into the interiors of the planetary bodies and to develop interpretation of the existing and upcoming measurements. In the literature on tides, each of the afore-mentioned viscoelastic models has been used. In Sections~\ref{sec:Tidal_Thermal_evolution} and~\ref{sec:Tides}, we address other aspects required for modeling the tidal response of planetary bodies.  Those aspects are used in combination with the viscoelastic models to constrain interior properties of the planets and moons, as discussed in Section~\ref{interiors}.
}

\section{Tidal and thermal evolution in planetary systems}\label{sec:Tidal_Thermal_evolution}

\subsection{Tidal evolution}\label{tidal_evolution_modeling}

Viscoelastic tidal dissipation in gravitationally interacting planetary bodies results in an angular momentum exchange between the spin and orbit of the bodies. This process results in an evolution of the spin and orbital rates  towards low spin-orbit resonances, e.g. 1:1 for most moons or 3:2 for Mercury, \textcolor{black}{
and damping of the eccentricity and inclination of the orbit. }
%(Another option is the so-called pseudo-synchronous end-state \citep{makarov_efroimsky13}. While this rotation mode is not unequivocally proven for any Solar-system body, it was hypothesised that the tectonic pattern of Europa might be a consequence of mildly super-synchronous (i.e., faster than synchronous) rotation \citep{greenberg2003,greenberg2002,hoppa2001}. It is also possible that planet b in the TRAPPIST-1 system is pseudo-synchronous \citep{2018ApJ...857..142M})
The conservation of the angular momentum of the two-body system implies an evolution of the separation, eccentricity, and inclination, while the dissipation of the rotational kinetic energy leads to heat deposition in the orbital partners.
When the perturbed body rotates faster than the perturber is orbiting above its surface, the tidal bulge leads the perturber and exerts such a torque on the perturber that the semimajor axis of the orbit expands. On the other hand, if the tidally perturbed body rotates slower than the perturber's orbital rate, the bulge lags behind and exerts on the perturber a torque contracting the semimajor axis, provided the orbital eccentricity is not very large. The boundary between these cases is commonly defined by a distance between the two bodies, known as \textit{synchronous radius}, at which the perturber's mean motion equals the hosting body's spin rate. This popular belief, however, was shown by \citet{Bagheri_etal21} to be only applicable for the low-eccentricity orbits, i.e., in the case of highly eccentric orbits, even a satellite above the synchronous radius can migrate towards its hosting planet.

\noindent
Studying of the orbital evolution of celestial bodies helps to constrain the history of the planetary systems and their origin. For example, numerous studies have targeted the dynamical evolution of the Earth-Moon system, exploiting the Moon's present-day separation rate of $38.08\pm 0.19$~mm/yr observed from Lunar Laser Ranging (LLR), employing various tidal evolution models \citep{Webb82,touma_wisdom98,Williams_etal14,rufu_canup20,wisdomTian15,cuk_etal16, canup_Asphaug01,zahnle_etal15}. The discovery of the unexpectedly rapid migration of the Saturnian moon Titan ($\sim11$~cm/yr) \citep{lainey_etal20} has been used to explain the large obliquity of Saturn. 
This measurement invalidates the traditional belief \citep{hamiltonWard04} that the presently observed obliquity of the rotation axis of Saturn is a result of the crossing of a resonance between the its spin-axis precession and the nodal orbital precession mode of Neptun that has happened during the late planetary migration more than 4 Gyrs ago. Instead, \citet{saillenfest_etal21} proposed that the resonance
was encountered more recently, about 1 Gyr ago, and forced Saturn's obliquity to increase from a small value to its current state.
Another example is the measurement of the Martian moon, Phobos's migration rate towards its host at a rate of $\sim1.8$~cm/yr. This observation has been used to constrain the Martian moons' origin and interior properties \citep{Yoder82, Bagheri_etal21, Singer1968, Samuel_etal19} as discussed in Section \ref{sec:Mars}. Based on this observation, it has been also shown that Phobos will collide with Mars's surface in $\sim$30-50 Myrs \citep[e.g.,][]{Bills_etal05}}.

%, the corresponding orbit being referred to as the {\it synchronous orbit}. 

\noindent
The rate at which tidal evolution takes place depends on the orbital parameters such as the distance between the two bodies, spin and orbital periods, eccentricity of the orbit, and on the physical properties of these bodies' interiors, that affect viscoelastic dissipation.
Modeling tidal evolution comprises the following major steps:

(1) ~Decomposition of the tidal potential into Fourier harmonic modes.\

(2) ~Assigning to each Fourier mode a specific phase delay and magnitude decrease.\

\noindent
The first step can be carried out by means of a development by \citet{Kaula64} who explicitly wrote down Fourier expansions for both the perturbing potential and the additional tidal potential of the perturbed body. To perform the second step, simplified tidal models such as  constant phase lag model (CPL) \citep{Macdonald64,Goldreich66,MurrayDermott99} and  constant time lag model (CTL) \citep{Singer1968,Mignard79,Mignard80,Mignard81,Hut81,Heller_etal11} were  introduced for analytical treatment and applied to rocky moons and planets, as well as gas giants.

\noindent
Despite the common use of the CTL and CPL models, they have been shown to suffer problems of both physical and mathematical nature \citep{EfroimskyMakarov13, EfroimskyMakarov14}; The CTL model implies that all the tidal strain modes experience the same temporal delay relative to the corresponding stress modes \citep{EfroimskyMakarov13, makarov_efroimsky13}. The CPL model, on the other hand, is not supported by physical principles because it assumes a constant tidal response independent of the excitation frequency which is incompatible with geophysical and laboratory data as shown in various studies \citep{JacksonFaul10,jackson2005laboratory, Khan_etal18,Bagheri_etal19,LauFaul19,Nimmo_etal12,NimmoFaul13}.
\noindent
These shortcomings of the CTL and CPL models are resolved by assigning separate phase lag and amplitude decrease to each Fourier mode. This assignment is defined by the rheology of the body and is also influenced by its self-gravitation. A combination of the properly performed steps (1) and (2) provides a means for calculating spin-orbit evolution of planets and moons \citep{BoueEfroimsky19}, including modeling their capture into spin-orbit resonances \citep{Noyelles_etal14}. Further extensions to the formulation presented by \citet{BoueEfroimsky19} were developed to incorporate highly eccentric orbits and included the effect of dissipation as a result of libration in longitude \citep{Bagheri_etal21,Renaud_etal21,bagheri_etal22}.

\subsection{Tidal-thermal evolution coupling}\label{sec:coupling}

Tidal dissipation is  not only responsible for the observed orbits and spin states of celestial bodies, but also can affect these bodies' thermal evolution. Thermal evolution is responsible for the planetary bodies' differentiation, melting, and volcanism.
The principal heat sources in a binary system are (see e.g., \citet[][]{Hussmann_etal06}): ~(a) the heating associated with accretion during planet formation, (b) the gravitational energy released during planetary differentiation, (c) radiogenic heating in the silicate component due to the decay of long-lived radioactive isotopes (U, Th, and K), and (d) tidal heating due to viscoelastic dissipation. % \citep{EfroimskyMakarov13,Efroimsky18, peale_cassen78,RenaudHenning18,Saxena_etal18}.
%\end{itemize}
Of these sources, only (c) and (d) are of relevance for the long-term evolution of the planet while the two first sources are mostly linked to the early stages of planetary accretion.

\noindent
Thermal evolution implies heat being transported to the surface of the body either by conduction or convection, and as a consequence, the interior temperature varies with time. This results in substantial changes in the interior structure and physical properties of the body, such as viscosity and rigidity, which in turn, can considerably change the tidal response of the body and affect its orbital evolution and spin. \textcolor{black}{In the solar system, we can find several examples where tidal and thermal evolution affect one another}. For example, in Europa, tidal heating can be intense enough to maintain the presence of a liquid surface or subsurface ocean\textcolor{black}{, though there is no consensus on whether the tidal dissipation is taking place predominantly in the ocean or the ice shell} \citep[e.g.,][]{tyler_etal15,choblet_etal17, HussmannSpohn04, tobie_etal03,rhodenWalker22, sotin_etal09}.
Moreover, tidal heating can result in volcanism, like in Io, considerably exceeding the heating by long-term radiogenic isotopes \citep[e.g.,][]{peale_etal79,VanHoolst_etal20,kervazo2022,foley_etal20,dekleer_etal2019,deKleer_etal19}. When considerable portion of ice or liquid water is present inside the body, tidal heating can entail cryovolcanism. This is believed to be the case in several icy moons such as Enceladus, Titan, Europa, and Triton \citep[e.g.,][]{spencer_etal09,sohl_etal14,vilella_etal20,hansen_etal21,hay_etal20}. In all of the mentioned examples, the tidal and thermal evolution have co-modulating effects on each other. This,  necessitating their joint consideration in evaluating the evolution and present-day state of planetary systems.  
%  Importantly, intense tidal heating can change the rheology, thus preventing thermal runaways. 
  % and also enabling the planet to leave  a higher spin-orbit resonance 
 % \citep{Makarov15}.
 \textcolor{black}{Since in this paper we mostly focus on tides with referring to thermal evolution only in cases where tidal dissipation plays an important role, we do not provide more details on thermal evolution modeling.}

\section{Tidal potential, Love numbers, and tidal response}\label{sec:Tides}

\subsection{Static tides}

 Consider a spherical body of radius $\s R\s$. An external perturber of mass $\s M^{\,*}\s$, located at a point $\,{\erbold}^{\;*} = (r^*,\,\lambda^*,\,\phi^*)\,$, in the body frame, generates the following disturbing potential at a point
   $\,\Rbold = (R,\phi,\lambda)\,$ 
 on the surface of the body:
 \ba
 W(\eRbold\,,\,\erbold^{~*})~=~\sum_{
 n=2}^{\infty}~W_{n}(\eRbold\,,~\erbold^{~*})~=~-~\frac{G\;M^*}{r^{
 \,*}}~\sum_{{\it{n}}=2}^{\infty}\,\left(\,\frac{R}{r^{~*}}\,\right)^{\textstyle{^{\it{n}}}}\,P_{\it{n}}(\cos \gamma)~~,
 \label{1}
 \ea
 where $\s R\,<\,r^*\,$, the letter $\s G
 % =6.7\times10^{-11}\,\mbox{m}^3\,\mbox{kg}^{-1}\mbox{s}^{-2}
 $ denotes the Newton gravity constant,
 %   $\,R\,$ stands for the radius of the tidally perturbed body ($\,r^*\geq R\,$),
 $\phi$ is the latitude reckoned from the spherical body's equator, $\lambda$ is the longitude reckoned from a fixed meridian, $P_n(\cos\gamma)$ are the Legendre polynomials, while $\gamma$ is the angular
 separation between the vectors $\,{\erbold}^{\;*}\,$ and $\,\Rbold\,$ pointing from the center of the perturbed body.
 
 \noindent
 In a static picture, the additional tidal potential arising from the deformation of the perturbed body is a sum of terms $U_n$ each of which is equal to $\,W_n\,$ multiplied by a mitigating factor $\,k_{\textstyle{_n}}\left({R}/{r}\right)^{\,n+1}\,$, where $\,k_{\textstyle{_n}}\,$ is an $\,n$-degree Love number. With the perturber residing in $\,\erbold^{\,*}\,$, the additional potential at a point $\,\erbold = (r,\phi, \lambda)\,$ is
  \ba
 U(\erbold\,,\;\erbold^{\;*})&=&\sum_{{\it n}=2}^{\infty}~U_{\it{n}}(\erbold)~=~\sum_{{\it n}=2}^{\infty}~k_{\it
 n}\;\left(\,\frac{R}{r}\,\right)^{{\it n}+1}\;W_{\it{n}}(\eRbold\,,\;\erbold^{\;*})~~.~~~~~~~~
 ~~~~~~~~~~~~~~~~
  \label{2}
 \ea
Note that in this equation, it is implied that while the surface point has the coordinates $\,\Rbold = (R,\,\phi,\,\lambda)\,$, the coordinates of the exterior points are $\,\erbold = (r,\phi,\lambda)\,$, with $r\geq R$. In simple words, the point $\erbold$ is located right above $\Rbold$.
 \noindent
 Along with the potential Love numbers $k_n$, the vertical displacement Love numbers $h_n$ and the horizontal displacement Shida numbers $l_n$ \citep{shida12} are in use. They appear in the expressions for degree-$n$ vertical displacement $H_n$ and horizontal displacement $L_n$ of a surface point:
 \begin{equation}
    H_n\,=\,\frac{h_n}{g}\;W_n(\pmb R,\,\pmb r^{*})\;\,,
    \end{equation}
\begin{equation}
   L_n\,=\,\frac{l_n}{g}\;\nabla W_n(\pmb R,\,\pmb r^{*})\;\,,
\end{equation}
\textcolor{black}{where $g$ is the surface gravity. While static tides imply permanent deformation of the planet or moon, in the non-synchronous orbits time-varying tides are raised in the bodies.}

 \subsection{Actual situation: time-dependent tides}

 For time-dependent tides, the above formalism acquires an important additional detail: the reaction lags, as compared to the action. Within a simplistic approach, we might simply take each $W_n$ at an earlier moment of time. In reality, this simplification is too crude, because lagging depends on frequency; so each $W_n$ must be first decomposed into a Fourier series over tidal modes, and then each term of the series should be endowed with its own lag. The magnitude of the tidal reaction is also frequency dependent, as a result of which each term of the Fourier series should now be multiplied by a
 dynamical Love number of its own. Symbolically, this may be cast in a form similar to the static-case expression:
 \ba
 U(\erbold\,,\;\erbold^{\;*})&=&\sum_{{\it n}=2}^{\infty}~U_{\it{n}}(\erbold\,,\;\erbold^{\;*})~=~\sum_{n=2}^{\infty}~\left(\,\frac{R}{r}\,\right)^{ n+1}\;\bar{k}_{n}\;W_{n}(\eRbold\,,\;\erbold^{\;*})~~.
 \label{3}
 \ea
 The hat in $\,\hat{k}_{l}\,$ serves to remind us that this is not a multiplier but a linear operator that mitigates and delays differently each Fourier mode of $W_n$.

\noindent
A degree-$n$ component of potential (\ref{3}) can be found by means of a convolution-type  {\it{Love operator}} \citep{Efroimsky2012a}:
 \ba
 U_{n}(\erbold,\,t)\;=\;\left(\frac{R}{r}
 \right)^{{\it n}+1}\int_{-\infty}^{t} {\bf\dot{\it{k}}}_{\textstyle{_n}}(t-t\,')~W_{n}
 (\eRbold\,,\;\erbold^{\;*},\;t\,')\,dt\,'~.
 \label{chuk}
 \label{6}
 \ea
 This is not surprising, because the linearity of tides implies that, at a time $\,t\,$, the magnitude of reaction depends linearly on the perturbation magnitudes at all the preceding moments of time, $\,t\,'\leq t\,$. The inputs from the actions at earlier times emerge owing to the inertia (delayed reaction) of the material. A perturbation applied at a moment $\,t\,'\,$ enters the integral for $\,U_{n}(\erbold,\,t)\,$ with a weight $\,{\bf\dot{\it{k}}}_{\textstyle{_n}}(t-t\,')\,$ whose value depends on the time elapsed. Here, overdot denotes time derivative, so the weights are time derivatives of some other functions $k_n(t-t^{\,\prime})$.  Following \citet{churkin}, who gave to this machinery its current form, we term the weights as {\it{Love functions}}.

\noindent
 In the frequency domain, the convolution operator becomes a product:
 \ba
 \bar{U}_{\textstyle{_{n}}}(\pmb r,\,\omega)\;=\;\left(\,\frac{R}{r}\,\right)^{n+1}\bar{k}_{\textstyle{_{n}}}(\omega)\;\,\bar{W}_{\textstyle{_{n}}}(\pmb R,\,\pmb r^{*}, \,\omega)\;\;,
 \label{gek}
 \ea
 where $\,\omega=\omega_{\textstyle{_{nmpq}}}\,$ is the tidal mode; $\,\bar{U}_{\textstyle{_{n}}}(\omega)\,$ and $\,\bar{W}_{\textstyle{_{n}}}(\omega)\,$ are the Fourier images of the potentials $\,{U}_{\textstyle{_{n}}}(\erbold,\,t)\,$ and $\,{W}_{\textstyle{_{n}}}(\Rbold,\,\erbold^*,\,t)\,$;
 while the complex Love numbers
 \ba
 \bar{k}_{\textstyle{_n}}(\omega)
 % \;=\;{\cal{R}}{\it{e}}\left[\bar{k}_{\textstyle{_l}}(\omega)\right]\;+\;\inc\; {\cal{I}}{\it{m}}\left[\bar{k}_{\textstyle{_l}}(\omega)\right]
 \;=\;|\bar{k}_{\textstyle{_n}}(\omega)|\;e^{\textstyle{^{-i\,\epsilon_{{_n}}(\omega)}}}\;=\;{k}_{\textstyle{_n}}(\omega)\;e^{\textstyle{^{-i\,\epsilon_{{_n}}(\omega)}}}~~,
 \label{number}
 \label{8}
 \ea
 are the Fourier components of the Love functions $\,{\bf\dot{\it{k}}}_{\textstyle{_n}}(t-t\,')\,$.
 \noindent
A pioneer work devoted to  development of the functions $\,\bar{U}_{\textstyle{_{n}}}(\omega)\,$ and $\,\bar{W}_{\textstyle{_{n}}}(\omega)\,$ into Fourier series was presented by \citet{1879RSPS...30..255D} who derived several leading terms of this expansion. A full expansion was later provided in a monumental work by \citet{Kaula61,Kaula64}. A reader-friendly explanation of this machinery can be found in \citet{EfroimskyMakarov13}. The tidal Fourier modes $\omega=\omega_{nmpq}$ over which these functions are decomposed are parameterised with four integers $nmpq$ and can be approximated as
\ba
\omega_{nmpq}\,\approx\,(n-2p+q)\,\dot{\cal{M}}\,-\,m\,\dot{\theta}\;\,,
\label{}
\ea
where ${\cal M}$ and $\dot{\cal{M}}$ are the mean anomaly and mean motion of the perturber, while $\theta$ and $\dot{\theta}$ are the rotation angle and rotation rate of the tidally perturbed body.\,\footnote{~An accurate expression for $\omega_{nmpq}$ includes also terms proportional to the apsidal and nodal precession rates of the perturber. Usually, these terms are small.} The actual forcing frequencies in the body are \citep{EfroimskyMakarov13}
  \ba
  \chi_{nmpq}\,=\;|\s\omega_{nmpq}\s|\;\,.
  \label{frequency}
  \ea
  Below, whenever this promises no confusion, we drop the subscript and simplify the notation as
  \ba
  \omega\equiv\omega_{nmpq}~~~,\quad
  \chi\s\equiv\s\chi_{\textstyle{_{nmpq}}}\;\,.
  \label{}
  \ea
  The Darwin-Kaula theory of tides has to be re-worked considerably for bodies experiencing physical libration \citep{FrouardEfroimsky17}. Negligible for planets and large satellites, the impact of physical libration on tidal evolution becomes strong for middle-sized satellites, and very strong for some of the small moons. For example, in Phobos, it more than doubles the tidal dissipation rate, while in Epimetheus it increases the dissipation rate by more than 25 times \citep{Efroimsky18, Bagheri_etal21}.

\noindent
\textcolor{black}{Similarly to the potential tidal Love numbers $k_n$, the time-dependant displacement Love numbers, $h_n$ and $l_n$ can be derived. Except for the Earth and the Moon, no robust measurements of the displacement Love numbers for other bodies have been made. Such measurements would require delivery of precise geophysical instruments on the surface of a planetary body. Thus, most of the studies focused on tides have to rely on the measured potential Love number as discussed in Section~\ref{interiors}}.

\subsection{Complex Love numbers}

Expressing the degree-$n$ Love number as
 \ba
 \bar{k}_{n}(\omega)\;=\;
 %  {\cal{R}}{\it{e}}
 \Re
 \left[\bar{k}_{n}(\omega)\right]\;+\;i\;
 %  {\cal{I}}{\it{m}}
 \Im
 \left[\bar{k}_{n}(\omega)\right]\;=\;|\bar{k}_{n}(\omega)|\;
 e^{\textstyle{^{-i\epsilon_{n}(\omega)}}},
 \label{L36}
 \ea
 we introduce the dynamical Love number
 \ba
 {k}_{n}(\omega)~=~|\bar{k}_{n}(\omega)|\,\;.
 \label{}
 \ea
 We also define the phase as $\,-\,\epsilon_n\,$, with a ``minus" sign, thus endowing $\,\epsilon_n\,$ with the meaning of phase lag.
 It can also be shown \citep{EfroimskyMakarov13} that  ~Sign$\,\epsilon_l(\omega)$ = ~Sign$(\,\omega\,)$. The so-called {\it quality function}
 \bs
 \ba
 {K}_{n}(\omega)\,\equiv\,\;-\;
 %{\cal{I}}{\it{m}}
 \Im
 \left[\,\bar{k}_{n}(\omega)\,
 \right]\;=\;
 {k}_{n}(\omega)\;\sin\epsilon_{n}(\omega)
 \label{14a}
 \ea
 can be written down also as
 \ba
 {K}_{n}(\omega)\,\equiv\,\;-\;
 %{\cal{I}}{\it{m}}
 \Im
 \left[\,\bar{k}_{n}(\omega)\,
 \right]\,=\,
 \frac{{k}_{n}(\omega)}{Q_{n}(\omega)}\;\,\mbox{Sign}({\,\omega})~~,
 \label{14b}
 \ea
 \label{tidalqualityfactor}
 \es
 where $Q_n(\omega)$ is the tidal quality factor defined through
 \ba
 Q_n^{-1}(\omega)\,=\,|\s\sin\epsilon_{n}(\omega) \s|\;\,.
 \label{15}
 \ea
 The quality function $K_n(\omega)$ appears in the expressions for tidal forces, tidal torques, tidal heating \citep{EfroimskyMakarov14}, and tidal evolution of orbits \citep{BoueEfroimsky19}.

\noindent
 While $\sin\epsilon_n(\omega)$ is an odd function, $Q_n(\omega)$ is even~-- and so is $k_n(\omega)$. Hence, no matter what the sign of $\omega$ and $\epsilon_n$, we can always regard both  $Q_n(\omega)$ and $k_n(\omega)$ as functions of the frequency $\chi\equiv|\omega|\;$:
 \ba
 Q_n(\omega)\,=\,Q_n(\chi)~~,\quad k_n(\omega)\,=\,k_n(\chi)~~~.
 \label{}
 \ea
  The mode-dependency $\,\bar{k}_{\textstyle{_{n}}}(\omega)\,$ and, consequently, the dependencies $\,{k}_{\textstyle{_n}}(\omega)
   %  =k_n(\chi)
  \,$, $\,{\epsilon}_{\textstyle{_n}}(\omega)\,$,  $\,Q_n(\omega)
  %  =Q_n(\chi)
  \,$ can be derived from the expression for the complex compliance $\,\bar{J}(\chi)\,$ or the complex rigidity $\,\bar{\mu}(\chi)=1/\bar{J}(\chi)\,$, functions containing the information about the rheology of a body.

\noindent
Overall, tidal dissipation is a very complex process wherein
self-gravitation \footnote{~Self-gravitation is pulling the tidal bulge down, effectively acting as additional rigidity. Negligible over the frequencies much higher than the inverse Maxwell time, gravity becomes an important factor at lower frequencies.} and rheology are intertwined. Its quantification necessitates elaborate viscoelastic modeling, to appropriately interpret observation of tides, and to make these observations an effective tool to constrain the deep interior.

%\section{Tidal response of a planetary body}\label{tidal_response}
  \subsection{Quality function of a homogeneous celestial body}

  By a theorem known as the {\it{correspondence principle}} or the {\it{elastic-viscoelastic analogy}}  \citep{1879RSPS...30..255D, biot_etal54},
 the complex Love number of a spherical uniform viscoelastic body, $\,\bar{k}_{n}(\chi)\,$, is related to the complex compliance $\,\bar{J}(\chi)\,$ by the same algebraic expression through which the static Love number $\,{k}_{n}\,$ of that body is related to the relaxed compliance $\,{J}_R\,$:
  \ba
 \bar{k}_{n}(\chi)
 ~=~\frac{3}{2\,(n\,-\,1)}\;\,\frac{\textstyle 1}{\textstyle 1\;+\;{\cal{B}}_{n}/\bar{J}(\chi)}
 ~~~,~\quad~
  \label{k2bar}
 \ea
 where
  \ba
 {\cal{B}}_{n}\,\equiv~\frac{\textstyle{(2\,n^{\,2}\,+\,4\,n\,+\,3)}}{\textstyle{n\,\mbox{g}\,
 \rho\,R}}~=\;\frac{\textstyle{3\;(2\,n^{\,2}\,+\,4\,n\,+\,3)}}{\textstyle{4\;n\,\pi\,
 G\,\rho^2\,R^2}}~\;.
   \label{B}
   \ea
In this expression, $G$ denotes Newton's gravitational constant, while $g$, $\rho$, and $R$ are the surface gravity, density, and radius of the body.
 From equation (\ref{k2bar}), we find how the quality function $ K_n(\omega_{lmpq})$
 entering an $\,nmpq\,$ term of the expansions for the tidal torque and tidal dissipation rate is expressed through the rheological law $\,\bar{J}(\chi)\;$:
 \ba
 K_{n}(\chi)\;\equiv\;
 {k}_n(\chi)\;\sin\epsilon_{n}(\chi)~=~-~\frac{3}{2(n-1)}~\frac{{\cal{B}}_{\textstyle{_n}}\;\Im\left[\bar{J}(\chi)\right]
 }{\left(\Re\left[\bar{J}(\chi)\right]+{\cal{B}}_{\textstyle{_n}}\right)^2+\left(\Im
 \left[\bar{J}(\chi)\right]\right)^2}\;\;.
 \label{E10a}
 \label{22}
 \ea
\begin{figure}
\begin{center}
\includegraphics[width=0.765\textwidth]{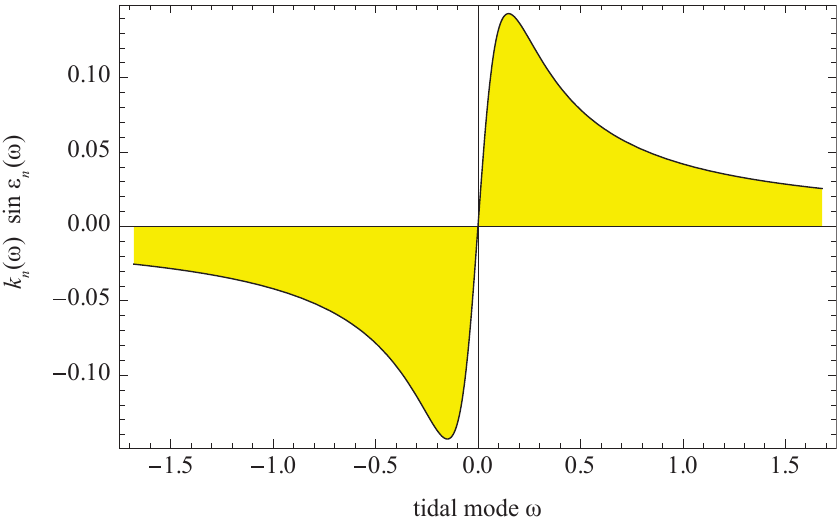}
\caption{\small{~A typical shape of the quality function $\,k_n(\omega)\,\sin\epsilon_n(\omega)\,$, ~where $\,\omega\,$ is a shortened notation for the tidal Fourier mode
 $\,\omega_{\textstyle{_{nmpq}}}\,$. ~~(From Noyelles et al. 2014.)}}
\label{figure}
\end{center}
\end{figure}
 For the Maxwell or Andrade model, the dependence of $~K_n\s\equiv\s k_{\textstyle{_n}}~\sin\epsilon_{\textstyle{_n}}\,$ on a tidal mode has the shape of a kink, depicted in Figure~\ref{figure}.   It can be demonstrated \citep[eqn 45]{Efroimsky15} that the  frequency-dependence of the inverse quality factor
 $\s Q_n(\chi)^{-1}\;\equiv\; |\s\sin\epsilon_n(\chi)\s|\s$ 
  % \ba 
  % \label{23}
  %  \sin\epsilon_l(\chi)\,=\,-\;\frac{{\cal{B}}_{\textstyle{_l}}\;\; \Im  \left[\bar{J}(\chi)\right] }{ \sqrt{ \left(\, \Re  \left[\bar{J}(\chi)\right]\,\right)^2+\,\left(\, \Im   \left[\bar{J}(\chi)\right]\,\right)^2\,}~ \sqrt{\left(\,  \Re \left[\bar{J}(\chi)\right]+{\cal{B}}_{\textstyle{_l}}\right)^2+\,\left(  \Im  \left[\bar{J}(\chi)\right]\right)^2}\, }\quad.\quad
 %   \ea
has a similar shape, with a similarly positioned peak. 

\noindent
 For a Maxwell body, the extrema of the kink  $\,K_n(\omega)\,$ are located at
 \ba
 {\omega_{peak}}_{\textstyle{_n}}\,=\;\pm\;
 \;\frac{\tau_M^{-1}}{1\,+\,{\cal{B}}_n\,\mu}\;\approx
 \;\pm\;\frac{1}{{\cal{B}}_n\,\eta}\,\;,
 \label{wh}
 \ea
 which can be checked by insertion of formulae (\ref{77} - \ref{88}) into equation (\ref{22}).\,\footnote{~The approximation in equation (\ref{wh}) is hinging on the inequality $\s{\cal B}_n\s\mu\gg 1\s$. Barely valid for a Maxwell Earth ($\s{\cal B}_2\,\mu\approx 2.2\s$), \,it fulfils well for Maxwell bodies of Mars's size and smaller.}

 \noindent
 %  Unless a binary is extremely close,  only the quadrupole ($l=2$) terms matter. For them,
 %  \ba
 %  \omega_{peak}\,\approx\;\pm\;\frac{1}{{\cal{B}}_{2}\;\eta}\;=\;\pm\;\frac{\textstyle{8\,\pi\,G\,\rho^2\,R^2}}{\textstyle{57\;\eta}}\quad,
 %  \label{peak}
 %  \ea
  Between the peaks, the quality function $\,K_n(\omega)=k_n(\omega)\,\sin\epsilon_n(\omega)\,$ is about linear in frequency:$\,$\footnote{~This is the reason why the Constant Time Lag (CTL) tidal is applicable solely for $\,|\s\omega\s|<|\s{\omega_{peak}}_n\s|\,$, and renders incorrect results for higher frequencies.}
 \ba
 \label{444}
 |\s\omega\s|\;<\;|\s{\omega_{peak}}_n\s|\quad\Longrightarrow\quad K_n(\omega)\;
 % \simeq\;3.4\;\frac{\eta\;\omega}{G\,\rho^2\,R^{\,2}}\;
 \approx\;\frac{3}{2\s(n-1)}\;\frac{\omega}{|\s{\omega_{peak}}_n\s|}\,\;.
 \ea
 Outside the inter-peak interval, the function $\,K_n(\omega)\,$ falls off as the inverse frequency:
 \ba
 |\s\omega\s|\;>\;|\s{\omega_{peak}}_n\s|\quad\Longrightarrow\quad
 K_n(\omega)
 %  \;\simeq\;\frac{2}{3}\;\frac{G\,R^{\,2}\,\rho^2}{\eta\,\omega}
 \;\approx\;\frac{3}{2\s (n-1)}\;\frac{|\s{\omega_{peak}}_n\s|}{\omega}\;\;.
 \label{do}
 \label{555}
 \ea
  Naturally, the insertion of $\,\omega={\omega_{peak}}_n\,$ in any of these expressions renders the same value for the peak amplitude:
\ba
K_n^{\rm{(peak)}}\s\approx\;\pm\;\frac{3}{2\s(n-1)}\,\;.
\label{}
\ea

\noindent
While the peaks' amplitude is insensitive to a choice of the viscosity value $\s\eta\s$, the spread between the extrema depends on $\s\eta\s$.
Expression (\ref{wh}) indicates that for a higher viscosity the peaks are residing close to zero, i.e. to the point of resonance $\omega\equiv\omega_{nmpq}=0$. If the viscosity evolves and assumes lower values (which happens when a body is getting warmer), the peak frequency grows, eventually superseding the orbital frequency $\s\dot{\cal M}\s$. In realistic situations, this requires very low viscosities and happens for bodies at high temperatures or, possibly, for bodies close to rubble.
Outside the inter-peak interval, the quality function $\,K_n(\omega)\,$ behaves as $\,\sin\epsilon_n(\omega)=Q^{-1}_n(\omega)\,\mbox{Sign}(\s\omega\,)$, and its values change slowly with frequency.

\noindent
Owing to the near-linear mode-dependence of $K_n$ in the inter-peak interval, the tidal torque value transcends spin-orbit resonances continuously \citep{makarov_efroimsky13,Noyelles_etal14}.\,\footnote{~The linearity of $\,k_n\,\sin\epsilon_n\,$ in $\,\omega\,$ is equivalent to the frequency-independence of the time lag: $\,\Delta t_n(\omega_{\textstyle{_{nmpq}}})\,=\,\Delta t\,$, see
 \citet{EfroimskyMakarov13}. This is why the tidal response of a terrestrial body can be described with the constant-$\Delta t\,$  model $\,${\it{only when all considered tidal frequencies are lower than}} $\,|\,\omega_{peak}\,|\;$
 -- $\,$or, equivalently, when all mean motions and spin rates are lower than $\,|\,\omega_{peak}\,|\,$.
 This usually requires a very low viscosity. We now see why the application of the CTL (constant time lag) tidal model to solid or semi-molten silicate planets is seldom possible (while for liquified planets this entire formalism is not intended anyway).}
 From expression (45) in \citet{Efroimsky15}, it can be derived that for a Maxwell body with $\s{\cal B}_n\s\mu\gg 1\s$, the locations of extrema of the kink function $\,\sin\epsilon_n(\omega)=Q^{-1}_n(\omega)\,\mbox{Sign}(\s\omega\,)$ virtually coincide with the locations of the extrema (\ref{wh}) for $\s K_n\s$.
Each of these two functions has only one peak for a positive tidal mode, when the regular Maxwell or Andrade models are used.
 This changes if we insert into formula (\ref{22}), and into its counterpart for $\,\sin\epsilon_n(\omega)=Q^{-1}_n(\omega)\,\mbox{Sign}(\s\omega\,)$, a complex compliance corresponding to a more elaborate rheology, such as the Sundberg-Cooper one. In that situation, an additional peak will appear.

\subsection{Layered bodies}

Analytical solutions for the tidal response of a homogeneous planetary body using other viscoelastic models can be derived \citep{RenaudHenning18}. However, in most geophysical applications, more sophisticated modeling is required for precise interpretations. This is due to the fact that the material properties of the planetary bodies vary with depth. This results in variation of the tidal response of the planetary body compared to a homogeneous planet. The variation of temperature, pressure, and grain size within the planetary bodies can be taken into account using the viscoelastic models discussed in Section~\ref{sec:Viscoelasticity}. Such an approach has been followed by, e.g., \citet{bagheri_etal22,Bagheri_etal19, Khan_etal18, Nimmo_etal12,NimmoFaul13,Padovan_etal14,steinbrugge2021,Plesa_etal18}.
To model a layered planetary body with depth-dependent properties, numerical methods have been used in such studies \citep{tobie2008, robertsNimmo08,behounkova2015}, while another widely used class of methods is based on the propagator matrix technique, derived in the scope of the normal mode theory \citep[e.g.,][]{alterman1959,takeuchi1962,wu1982,vermeersen1996,sabadini2004}. Similar approach is used to calculate the tidal response by, e.g., \citet{Plesa_etal18, MooreSchubert00, Padovan_etal14} to obtain the interior structure models further discussed in Section \ref{interiors}. \citet{Martens_etal16} developed a Python toolbox to compute the tidal and load Love numbers in an elastic regime and exploited it \citep{martens_etal19} to study the Earth's tides. \citet{Bagheri_etal19} used a numerical code based on the spectral-element-method to compute the tidal response using several viscoelastic models. \citet{dmitrovskii_etal21} used the same technique in 3D to model tides in an irregularly shaped body (Phobos); but only modeled the elastic response instead of a general viscoelastic behaviour.

\noindent
The overall tidal response of layered bodies depends on the interplay between the individual layers. For example, an ocean or a global molten layer below the crust or lithosphere might effectively decouple the interior of the planet from the surface and diminish the tidal deformation of the lower layers.
\begin{figure}
    \centering
    \includegraphics[width=\textwidth]{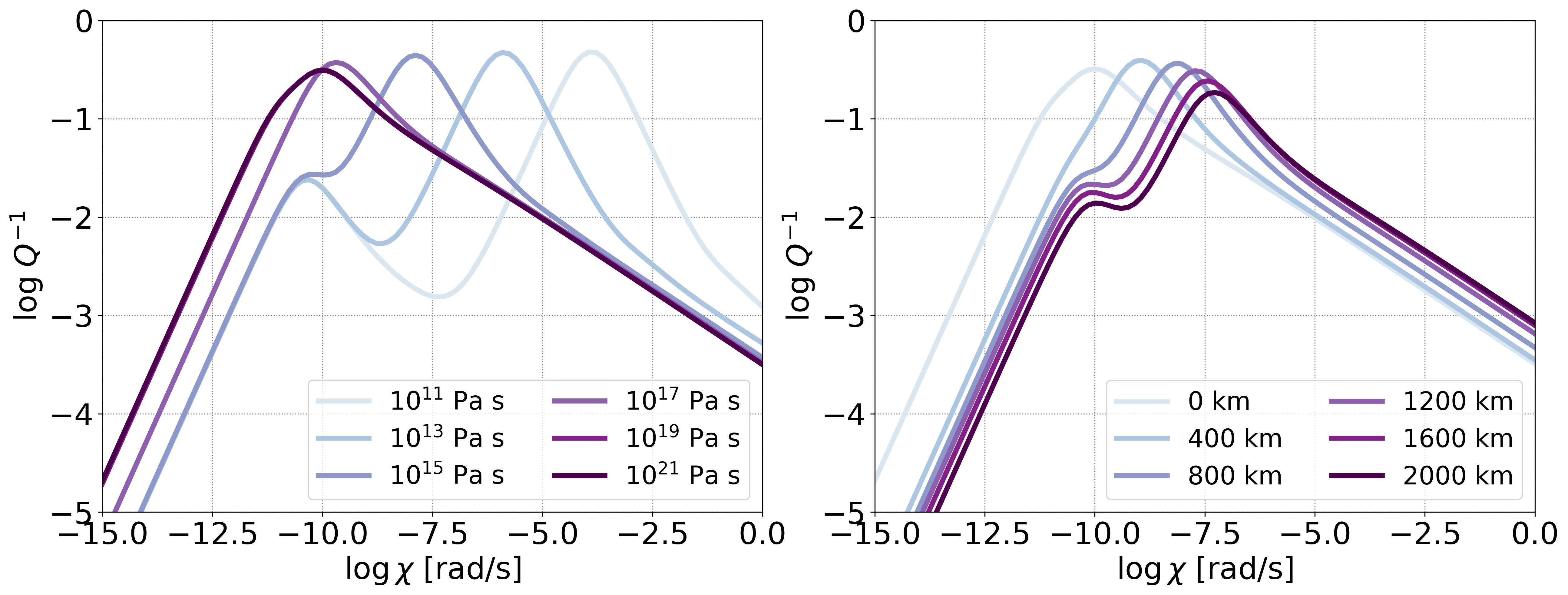}
    \caption{The inverse tidal quality factor of a model Earth-sized planet with a liquid core, a mantle of viscosity $\unit[10^{21}]{Pa\;s}$ and rigidity $\mu=\unit[200]{GPa}$, and a weak surface layer (e.g., a global icy crust). The left panel illustrates the effect of the upper layer's viscosity, keeping its thickness fixed to $\unit[200]{km}$ and its rigidity to the mantle value. On the right panel, the layer's viscosity is held constant at $\unit[10^{17}]{Pa\; s}$ and the individual lines correspond to different thicknesses.}
    \label{fig:two_layers}
\end{figure}
\noindent
%\textcolor{black}{Most of the planetary bodies in the Solar system considerably depart from a homogeneous object, which results in important differences in interpreting the measurements compared to a homogeneous body.
As an instructive example, a comparison between the tidal response of a planetary body incorporating layers of different physical properties is shown in   Figure~\ref{fig:two_layers}. This figure depicts the tidal quality factor of a three-layered Earth-sized planet consisting of a liquid core, a solid mantle governed by the Andrade rheology, and a weak surface layer with the same properties as the mantle, but a lower viscosity. The presence of two dissipative layers leads to the emergence of two peaks in the $Q^{-1}$ spectrum. In this sense, the response of a layered body might be mimicked by an advanced rheological model, for example the Sundberg-Cooper rheology \citep{gevorgyan2021} or by its extensions. Moreover the peak corresponding to the planetary mantle is shifted to higher frequencies and it becomes weaker with decreased upper-layer viscosity (or increased upper-layer thickness).

\noindent
\textcolor{black}{Appropriately modeling the tidal response on the layered bodies can also help in understanding the } the tidal dissipation pattern \citep[e.g.,][]{beuthe2013}, \textcolor{black}{ and interpret them to infer knowledge about the interior properties}. As shown by \citet{segatz1988} in the case of Jovian moon Io, the presence of a semi-molten layer (an astenosphere) between the solid mantle and lithosphere can considerably affect the pattern of surface tidal heat flow. The same is true for the tidal dissipation within icy moons with a subsurface ocean \citep{tobie_etal05} and for hypothetical exoplanets with icy crust overlying a silicate mantle \citep{henning2014}.

%\textcolor{black}{MICHAELA, ~COULD YOU PLEASE ADD SOMETHING HERE ?}

\section{Tides as a probe of the deep interior}\label{interiors}

\textcolor{black}{Having introduced the theoretical aspects of modeling tides in previous sections, here we address particular planetary bodies, focusing on how information about their interiors was obtained by studying their tidal response.
We summarise the constraints on the interior properties of Mercury, Venus, the Moon, Mars and its moons, and the largest moons of giant planets. We also mention the expected future improvements in measuring these bodies' tidal response.}

\subsection{Mercury}\label{sec:mercury}
%Investigations based on the tidal response of Mercury by \citet{vermaMargot16} and \citet{Steinbrugge18} using MESSENGER's radio tracking data has provided estimations of its tidal potential Love number. The BepiColombo mission has recently reached Mercury and will be able to provide geodetic measurements using its Laser Altimeter that can significantly improve the estimations of the deep interior, particularly the inner core size, as prospected by \citet{Thor_etal20}.
%
%....

% Part on Mercury by Sander Goossens
\noindent
Studies of Mercury's interior have long focused on its magnetic and gravitational field, and only recently the first measurements of its tidal response were obtained. The Mariner 10 flybys in 1974 and 1975~\citep{dunne1974} provided us with the first clues of Mercury's interior, by detecting its magnetic field~\citep{ness1974}, and with the first measurements of its gravitational field~\citep{anderson87}. A much more detailed view of Mercury and its environment was provided by NASA's MErcury Surface, Space ENvironment, GEochemistry, and Ranging (MESSENGER) spacecraft, the first to orbit Mercury~\citep{solomon2007}. Pre-MESSENGER studies of Mercury often focused on a combination of rotation and tides, together with its spin-orbit resonance, providing predictions that could later be tested against MESSENGER data~\citep{vanhoolst2003b,vanhoolst2007,rambaux2007,rivoldini2009,dumberry2011, matsuyamanimmo09}.

\noindent
One of MESSENGER's many goals was to map Mercury's gravity field, which could then be used to determine the state of Mercury's core. \citet{peale1976} and \citet{peale2002} showed that, because Mercury is in a Cassini state (where its spin axis, its orbit normal, and the normal to the invariable plane are co-planar), its polar moment of inertia and the moment of inertia of the solid outer shell (mantle and crust) can be determined from 3 quantities: Mercury's obliquity, the amplitude of its longitudinal librations, and its second degree gravitational harmonic coefficients. The first two were determined from Earth-based radar data~\citep{margot_etal07}, and MESSENGER finally provided the first precise measurement of Mercury's second degree harmonics~\citep{smith2012}. During the MESSENGER mission, estimates of its gravity field were updated as more data were collected. Mercury's gravitational tidal response as expressed in its degree two Love number $k_2$ was also determined.

\begin{figure}[!ht]
\begin{centering}
\noindent\includegraphics[width=0.6\linewidth]{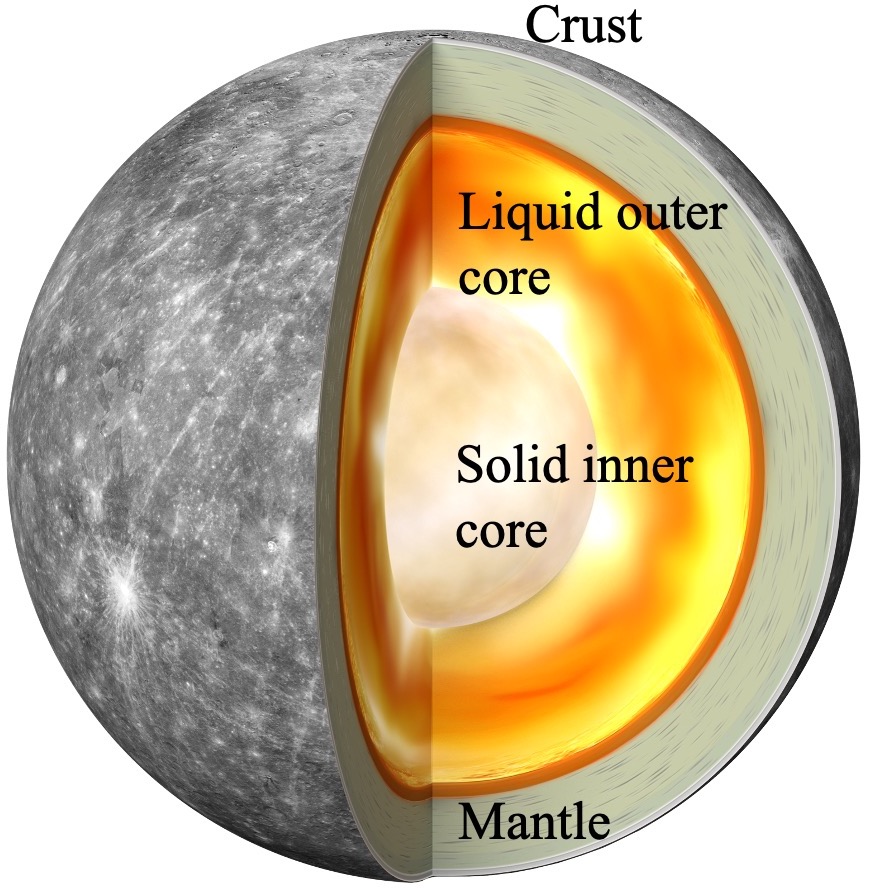}
\caption{The view of Mercury's interior based on recent measurements of its rotational state. Adapted from \protect\citet{genova2018core}.}
\label{fig:mercury_interior}
\end{centering}
\end{figure}

\noindent
The first estimate of $k_2 = 0.451 \pm 0.014$ was reported by~\citet{mazarico2014b} using three years of MESSENGER radio tracking data. The Love number was co-estimated along with gravity field parameters and rotational parameters, but they found that the radio data was not sensitive to parameters describing the forced librations. The value of $k_2$ was consistent with pre-MESSENGER analyses of Mercury's tidal response, which indicated a range of 0.4 -- 0.6~\citep{vanhoolst2003b,rivoldini2009}. Using the newly determined gravitational parameters, \citet{rivoldini2013} and \citet{Hauck_etal13} investigated Mercury's interior structure as constrained by its moments of inertia, but without considering the tides.

\noindent
Initial results using MESSENGER data proposed the existence of an FeS layer on top of the core, to account for the higher mantle density that was a result of a larger-than-expected value for the moment of inertia of the outer shell~\citep{smith2012}. Results using MESSENGER's X-Ray Spectrometer measurements of the ratio of Ti and Si also argue against the FeS layer~\citep{cartier2020}. Mercury's core itself is mostly considered to be metallic, with light elements of S and/or Si~\citep{rivoldini2009,Hauck_etal13,chabot2014core,knibbe2015,knibbe2018}.

\noindent
\citet{Padovan_etal14} were the first to comprehensively consider Mercury's tidal response in the light of MESSENGER's results. They considered several end-member models such as a hot or cold mantle, and found that the results presented in~\citet{mazarico2014b} fell in their range, and would be mostly consistent with their cold mantle, without a layer of FeS on the top of the core. The latter was initially considered by~\citet{smith2012} to account for the relatively large moment of inertia of the outer shell. Updates to the estimate of Mercury's obliquity by \citet{margot2012} reduced this value, and an FeS layer was no longer necessary~\citep{Hauck_etal13,rivoldini2013,knibbe2015}. The $k_2$ value was also later confirmed by an independent analysis by~\citet{vermaMargot16}, with a value of $k_2 = 0.464 \pm 0.023$. \citet{steinbrugge_etal18} further investigated Mercury's tidal response, and computed models consistent with MESSENGER measurements of mean density, mean moment of inertia, moment of inertia of mantle and crust, and $k_2$. They showed that the ratio of $h_2$ (the radial displacement Love number) and $k_2$ can provide better constraints on the size of a possible solid inner core than the geodetic measurements such as moments of inertia can.

\noindent
The MESSENGER mission had several extensions, where the altitude and location of its periapsis changed, with lower and lower altitudes obtained in the northern hemisphere, down to 25 km above surface. This increased the sensitivity of the tracking data with respect to smaller scale gravity features, and to Mercury's tidal response. Using the entire set of tracking data, \citet{genova2018core} presented a gravity model that included estimation of Mercury's rotational parameters and tidal Love number. Their estimate of Mercury's obliquity unambiguously satisfies the Cassini state. Their obliquity value results in a lower normalised polar moment of inertia of $0.333 \pm 0.005$, whereas earlier results yielded normalised polar moment of inertia values around 0.346~\citep{margot2012,Hauck_etal13,mazarico2014b}. Using this updated value and smaller error, they modeled Mercury's interior with a Markov Chain Monte Carlo (MCMC) \citep{MosegaardTarantola95} approach and found evidence for the existence of a solid inner core, with the most likely core size being between 0.3 and 0.7 times the size of the liquid core. Their updated Love number, $k_2 = 0.569 \pm 0.025$, was also higher than the previous estimate.

\noindent
An analysis by \citet{bertone2021} also finds Mercury's rotational parameters unambiguously satisfying the Cassini state, yet with a different obliquity that results in a normalised polar moment of inertia value of $0.343 \pm 0.006$. Their analysis is based on laser altimetry data from the Mercury Laser Altimeter (MLA, \citet{cavanaugh2007}), using crossovers (where two laser tracks intersect, the difference in measured altitude can be used to infer rotation and tidal parameters, for example). This discrepancy could point to differences in the rotation state of the entire planet as measured by gravity and the rotation state of the outer shell as measured by laser altimetry. \citet{bertone2021} did not estimate $k_2$ but they did provide the first estimate of the radial displacement Love number, $h_2 = 1.55 \pm 0.65$. Due to the sparsity of crossovers, this parameter is difficult to measure. Finally, an analysis by \citet{konopliv2020}, using the entire MESSENGER tracking data set, determined Mercury's Love number in close agreement with that of \citet{genova2018core}, with a value of $k_2 = 0.53 \pm 0.03$.

\noindent
The differences in moment of inertia values and newly determined Love numbers have implications for our knowledge of Mercury's interior structure, especially for the size of the liquid core. \citet{steinbrugge2021} performed an analysis of the lower normalised polar moment of inertia value of 0.333 and the higher Love number of 0.569, and found several challenges in determining interior structure models that fit these parameters: they find a relatively large inner core (> 1000 km), a relatively high temperature at the core-mantle boundary (CMB; above 2000 K), low viscosities at this boundary (below $10^{13}$ Pa s), and a low mantle density (markedly below 3200 kg m$^{-3}$). They also indicate that the low viscosities required to match $k_2$ imply a significantly weaker mantle. They indicate that such challenges do not exist for the higher normalised polar moment of inertia value of $\sim 0.346$. It should be noted that they focused their analysis on models that matched the central values of parameters such as the moments of inertia and $k_2$. If they take into account the quoted errors, they indicate some of the challenges are alleviated.

\begin{figure}[!ht]
\begin{centering}
\noindent\includegraphics[width=\linewidth]{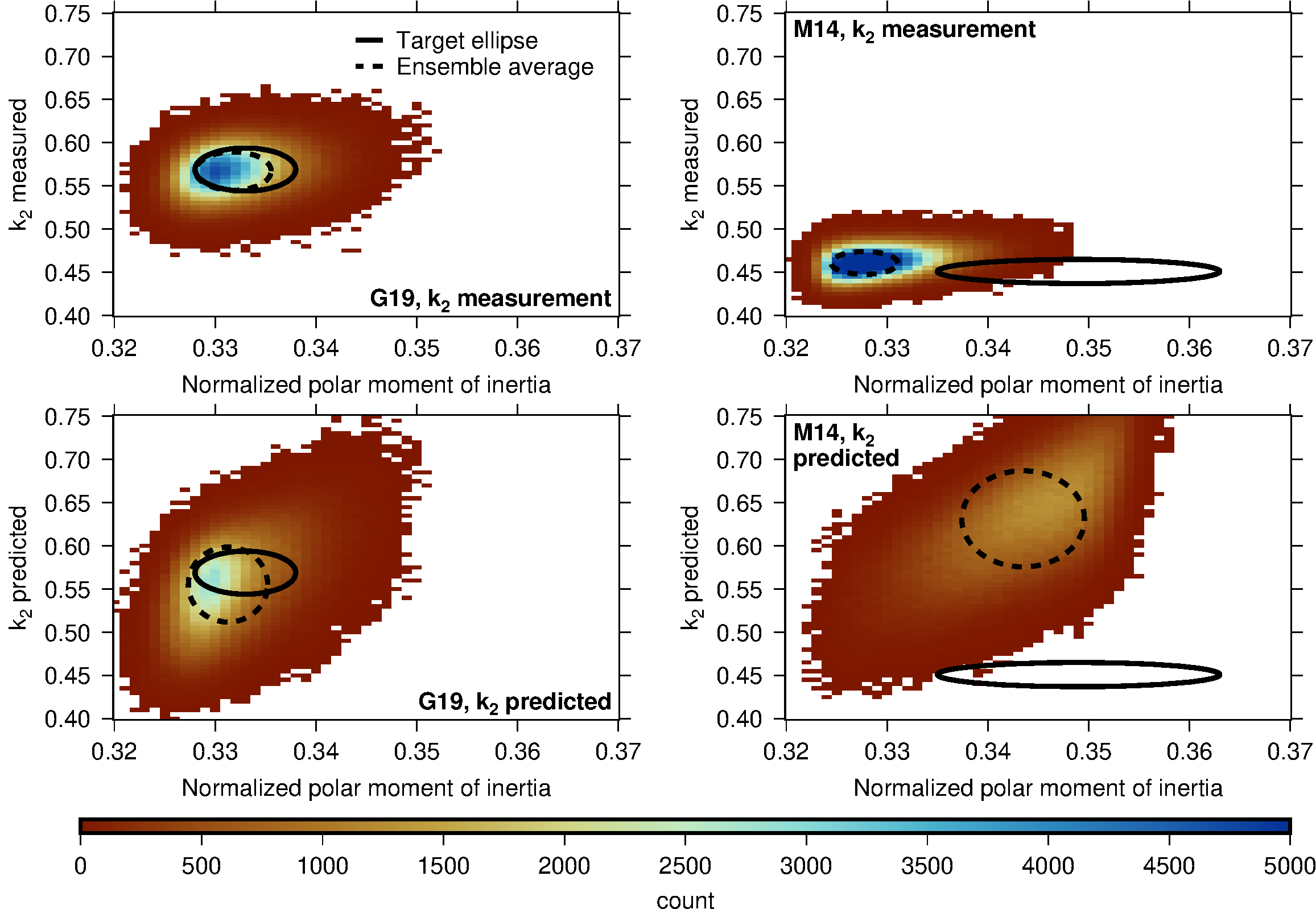}
\caption{Results of the MCMC analysis by \protect\citet{goossens2022_tides}, which determined models of Mercury's interior structure using different measurements for the normalised polar moment of inertia and tidal Love number $k_2$. The label ``G19" refers to \protect\citet{genova2018core} and ``M14" to \protect\citet{mazarico2014b}. The results are shown as heatmaps of the mapped quantities for all accepted models in the MCMC analysis. For the top panels, $k_2$ was used as a measurement in the MCMC analysis, whereas it was predicted for the results in the bottom panels. These results indicate that the set of measurements of G19 s the most consistent. Higher values of the normalised polar moment of inertia, as provided by M14, cannot simultaneously fit the measured $k_2$, and would indeed predict an even higher value.}
\label{fig:heatmap_mercury}
\end{centering}
\end{figure}

\noindent
A recent analysis by \citet{goossens2022_tides} also investigated the different values for moments of inertia and $k_2$, using an MCMC method to map out models of Mercury's interior that satisfy the measurements and their quoted errors. They find that models that match the lower normalised polar moment of inertia value of 0.333 \citep{genova2018core} also match or predict the Love number value of $k_2 = 0.569$. Models that match the higher normalised polar moment of inertia of $\sim 0.346$ indicate even higher Love numbers, larger than 0.6, with a wide spread. Their study thus indicates that the higher normalised polar moment of inertia values are not consistent with the current measurements of the Love number. In addition, they also find lower CMB temperatures than \citet{steinbrugge2021} indicated, in the range of 1600 -- 2200 K but with a peak at 1800 K. While their study does indicate low viscosity values at the CMB, models with a constant mantle temperature, mimicking a convecting mantle rather than a conducting one, predict lower temperatures and higher viscosities. Models that satisfy the lower normalised polar moment of inertia and the updated moment of inertia for the outer shell do indicate mantle densities that are lower than previously assumed ($3089 \pm 135$ kg m$^{-3}$). A study by \citet{lark2022a} indicates that the presences of sulfides in the mantle can explain this lower density. \citet{goossens2022_tides} also provide a prediction for the radial displacement Love number $h_2 = 1.02 \pm 0.04$ for models that satisfy the measurements from \citet{genova2018core}.

\noindent
The next spacecraft that will orbit Mercury is the European Space Agency's BepiColombo mission \citep{benkhoff2010}. This spacecraft will provide precise gravity measurements \citep{genova2021_ssr} as well as laser altimetry \citep{thomas2021_ssr}, both with a more global coverage than was possible with MESSENGER, due to the latter's elliptical orbit around Mercury. BepiColombo data will provide updated measurements of the moments of inertia and the Love numbers $k_2$ and $h_2$ \citep{steinbrugge2018_pss,thor_etal20,genova2021_ssr}, as well as for its rotational state and gravity, which will help resolve the current challenges in understanding Mercury's interior structure.

\subsection{Venus}\label{sec:venus}
\begin{figure}
    \centering
    \includegraphics[width=0.45\textwidth]{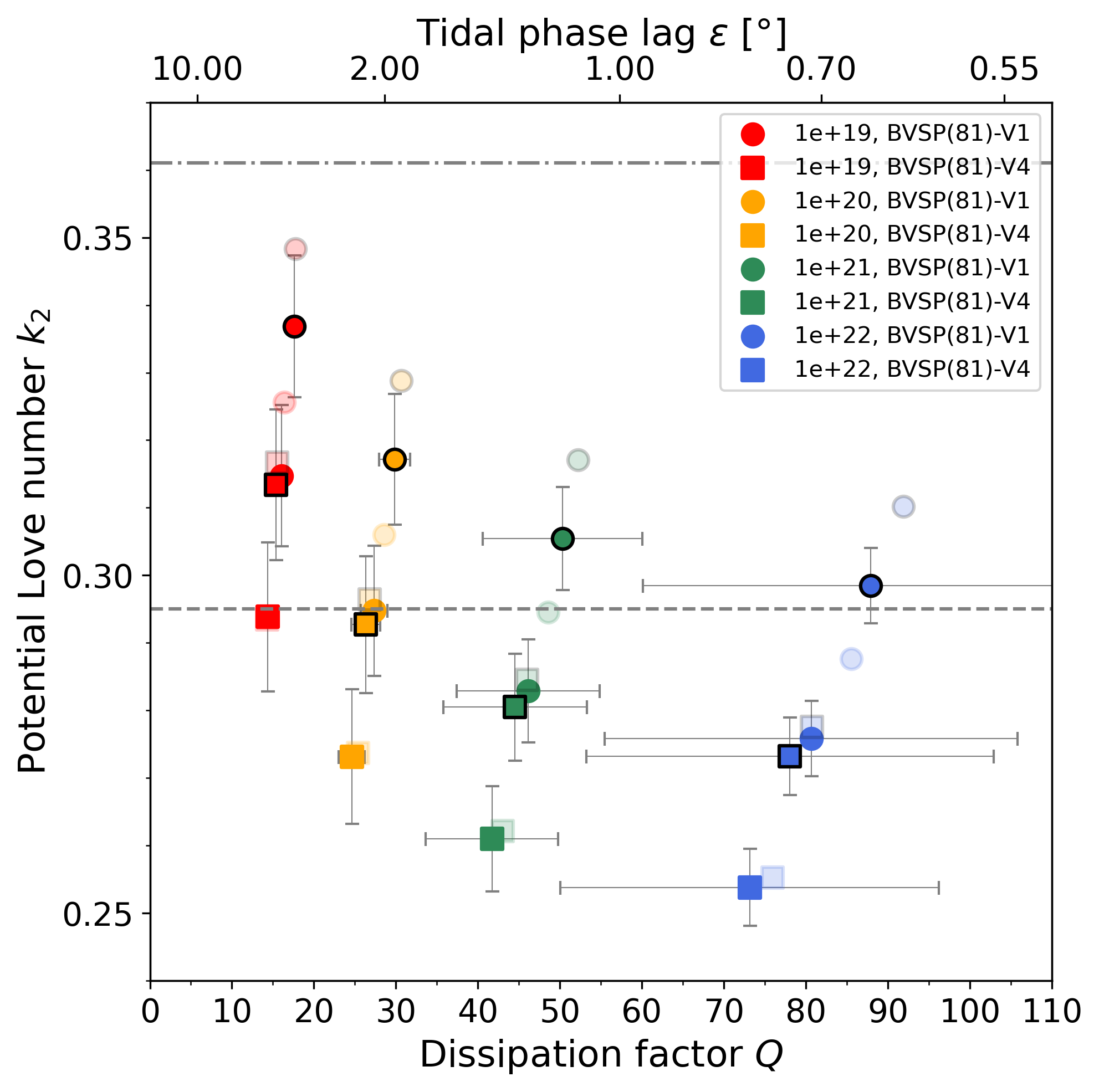}
   \vspace{1mm}     \includegraphics[width=0.5\textwidth]{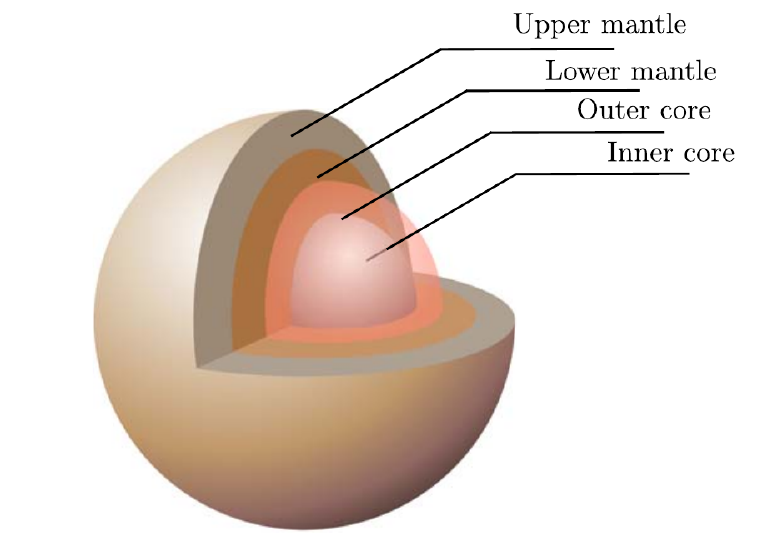}
    \caption{Left: Overview of Venusian tidal parameters for two mantle mineralogies and two mantle temperature profiles, assuming an incompressible Andrade model, reproduced after \citet{Dumoulin_etal17}. Circles symbolise an iron-poor ($0.24$ wt$\%$ FeO) mantle model and squares an iron-rich ($18.7$ wt$\%$ FeO) model according to \citet{bvsp1981}. Different colours stand for different mantle viscosities; the markers with black edges correspond to a hot interior \citep{armann12} and the markers without contour correspond to a cold interior \citep{steinberger10}. The dashed and dot-dashed lines indicate the mean value and the upper bound of the $2$-$\sigma$ interval for the tidal Love number obtained by \citet{konopliv_yoder96} using Magellan and Pioneer data: $k_2=0.295\pm0.066$. Finally, the error bars illustrate the span of results calculated for different values of the Andrade parameter $\alpha\in(0.2,0.3)$ (see Section~\ref{Andrade}), and the transparent markers, taken from \citet{Dumoulin_etal17}, indicate the effect of compressibility. Right: Schematic interior structure of Venus from \citet{shah_etal22}}
    \label{fig:Venus_k2_Q}
\end{figure}

The tidal response of Venus to semi-diurnal Solar tides was measured more than a quarter century ago using Magellan and Pioneer tracking data. \citet{konopliv_yoder96} estimated the $2$-$\sigma$ interval for the potential Love number as $k_2=0.295\pm0.066$ and concluded, following the predictions of \citet{Yodervenus}, that Venus has a fully liquid core. However, this conclusion was based on a purely elastic model of the tidal response. A new reassessment of the problem with a compressible Andrade model \citep{Dumoulin_etal17} indicated that the question of size and state of the Venusian core cannot be resolved with the data available. The wide range of admissible Love numbers, combined with the absence of $Q$ measurements and with the large uncertainties on the planet's moment of inertia \citep{margot2021} constrains neither the core state, nor the mantle mineralogy and temperature profile. \citet{Dumoulin_etal17} illustrated that only a future measurement of $k_2$ below $0.26$ and a large phase lag \footnote{~\citet{Dumoulin_etal17} define their ``tidal phase lag" as $\,\frac{\textstyle 1}{\textstyle 2}\s\arcsin{Q^{-1}}$.
For the semidiurnal tide (and, more generally, for the $nmpq$ tidal components with $m=2$), this quantity coincides with the \textit{geometric lag}, a quantity not to be confused with the {\it phase lag}. See \citet[eqn 26]{EfroimskyMakarov13} for details.} ($\epsilon_2>4^{\circ}$) would indicate a fully solid core. At higher Love numbers, the core can be interpreted as at least partially liquid and a precise measurement of the phase lag would further help to discern between different mantle viscosities and thermal states (see Figure~\ref{fig:Venus_k2_Q}).

\noindent
In addition to solid-body tides, the deformation of Venus is also affected by atmospheric tides consisting of two components: one due to the gravitational loading by the Sun and the other resulting from thermal forcing \citep{auclair2017,correia2003,ingersoll1978,gold1969}. The interplay between the gravitational and thermal tides leads to the instability of synchronous rotation and it has probably driven the planet to its present-day retrograde spin configuration. Close to its nonsynchronous, yet stationary rotation state, Venus might also be influenced by the weak gravitational pull of the Earth, as hypothesised by \citet{dobrovolskis1980} and \citet{gold1969}. Besides the contribution to the planet's rotational dynamics, the atmosphere acts as a global surface load and diminishes the tidal deformation of the Venusian surface by approximately $0.3\%$ \citep[see][]{remus2012anelastic,dermott1979}.

\noindent
\textcolor{black}{The tidal phase lag and thus the tidal quality factor $Q$ can also be used to constrain the thermal state of the interior, given the temperature dependence of the mantle viscosity. In addition to the present-day values, knowing the past thermal state and $Q$ of Venus can further help us to understand its tidally-induced rotational evolution \citep{bolmont2020}. Since Venus and the Earth are often referred to as twin planets, due to their similar size and mass, the tidal quality factor $Q$ may be similar between the two bodies. However, while on the Earth plate tectonics represent an efficient way to cool the interior, on Venus large scale subduction may be absent today. In fact, a recent study by \citet{Rolf_etal2018} concluded that if Venus may have experienced one or multiple episodes of plate tectonics in the past, the last of those episodes likely ended 300-450 Myr ago, otherwise thermal evolution models cannot match the observed surface gravity spectrum. If Venus has been in a stagnant lid regime, with an immobile surface over the past 500 Myr, its interior would likely have a much higher temperature than that of the present-day Earth. This may indicate a much more dissipative interior, i.e. lower tidal quality factor $Q$, than that of the solid Earth, for which $Q$ is around 280 \citep{Ray_etal01}. Even if Venus currently experiences some sort of surface mobilisation on a much smaller scale than tectonic plates provide on the Earth, with small patches of the surface being recycled in the interior (i.e., the so-called plutonic squishy lid regime \citet{Lourenco_etal2020}), this might still lead to a hotter interior than on present-day Earth. This type of surface mobilisation is thought to have operated on the Earth in the past during the Archean, when the interior temperature were higher than they are today \citep{Rozel_etal17, Lourenco_etal2020}.}

% text added by Ina
\noindent

\noindent
That the interior of Venus may be characterised by high mantle temperatures, which in turn may indicate a higher dissipation, is also supported by the small elastic lithosphere thickness that are indicators for a thin and hot lithosphere \citep{Smrekar_etal2018}. Small elastic thicknesses are estimated at coronae \citep{Orourke_Smrekar2018}, steep side domes \citep{Borrelli_etal2021}, and crustal plateaus \citep{Maia_Wieczorek2022}, and likely 20~km or less may be representative for a significant part of the planet \citep{Anderson_Smrekar2006}. Moreover, the temperatures inside the Venusian mantle may allow for volcanic activity at present-day. This has been suggested by several observations. For example, the presence of recently active hot-spots in the interior of Venus has been inferred based on their thermal signature \citep{Shalygin_etal2015} and emissivity data of Venus Express, which allow to distinguish fresh from weathered basaltic material \citep{Helbert_etal2008,Smrekar_etal2010, Dincecco_etal2017}. In addition, gravity, topography and surface deformation structures at the locations where recent volcanic activity has been suggested are consistent with the presence of mantle plumes in the interior \citep{Kiefer_etal1991,Smrekar_etal1991}. Furthermore, SO$_2$ variations in the atmosphere of Venus recorded by Pioneer Venus Orbiter \citep{Esposito_1984,Esposito_1988} and later by Venus Express \citep{Marcq_etal2013} provide additional hints at recent volcanic activity. All this evidence, although indirect, may indicate that the present-day interior of Venus is characterised by high mantle temperatures and hence low mantle viscosities that would lead to a lower dissipation factor than that of the Earth. Whether the above conclusions hold or not, it needs to be tested by future data that would allow to place tighter constraints on the thermal state of the Venusian interior.

\noindent
Owing to the lack of recent and accurate tidal measurements, Venus still remains the most enigmatic of terrestrial worlds. Recently revived interest in its interior and atmospheric conditions, nurtured by the putative detection of biosignatures (\citealp{greaves2021}; cf. \citealp{villanueva2021}), foreshadowed the selection of two geophysical mission concepts for a launch in late 2020s or early 2030s. The VERITAS mission (Venus Emissivity, Radio Science, InSAR, Topography, and Spectroscopy; \citealp{Smrekar_etal_2020}) of NASA's Discovery Program will address the present geological and volcanic activity, the link between interior and atmospheric evolution, and the global accurate mapping of Venusian gravity field. In order to reduce the uncertainties of the gravity field (and tidal response) measurements, introduced by the planet's rotation, the mission will combine standard Earth-based Doppler tracking with the systematic observation of surface features by the onboard instrument VISAR (Venus Interferometric Synthetic Aperture Radar). With this approach, the expected $3$-$\sigma$ accuracy of the Love number $k_2$ is $4.6\times10^{-4}$ and of the tidal phase lag $0.05^{\circ}$ \citep{cascioli2021}.

% Whether the above conclusions hold or not, it needs to be tested by future data of VERITAS and EnVision that in addition to the tidal deformation measurements will also search for signs of volcanic and tectonic activity that will allow to place tighter constraints on the thermal state of Venus' interior.
\noindent
The medium-sized ESA's Cosmic Vision Programme mission \textit{EnVision} \citep{Ghail_etal16} will focus on the geological structures of Venus that are of interest for understanding its past thermal evolution. As in the case of VERITAS, the primary objectives of the mission also include the determination of a uniform high-resolution gravity field, with a spatial resolution better than $\sim\unit[170]{km}$. The expected $1$-$\sigma$ error of both the real and the imaginary parts of the Love number $k_2$ is $10^{-3}$, implying a $0.1^{\circ}$ uncertainty for the tidal phase lag \citep{Rosenblatt_etal21}. Despite the similarity of the two geophysical missions, VERITAS and \textit{EnVision} are expected to be synergistic. First, the launch of \textit{EnVision} is planned for a later date than the launch of VERITAS, and the two orbiters will only operate simultaneously for a part of the respective mission durations. Second, while VERITAS aims at providing a global geophysical survey, \textit{EnVision} is designed for more targeted, repeated observations of the regions of interest identified by the former \citep{ghail2021}. With the new data, a refined estimate of the Venusian core size will be possible and the first measurements of phase lag will enable constraining the average mantle viscosity within an order of magnitude \citep{Dumoulin_etal17,Rosenblatt_etal21}.

%The tidal response of Venus to Solar tides was measured using Magellan and Pioneer Venus Orbiter tracking data \citep{Konopliv_yoder96} and used to constrain the size and state of its core \citep{Dumoulin_etal17}. Further improvements are envisaged to constrain the Venusian interior based on geodetic data in the context of the upcoming EnVision mission \citep{Rosenblatt_etal21}.

\subsection{The Moon}

\tnr{The Moon is one of the best studied bodies in the Solar system because of its proximity to the Earth. It is only 60 Earth radii away and its surface undergoes a monthly tidal deformation of $\pm$~0.1~m generated by the Earth's gravitational field. This tidal potential impacts also the Moon's gravity field and its orientation periodically. In addition, the Moon does not have an ocean as on the Earth and the tidal variations are only incorporated in solid tides. These variations are detected by space missions orbiting the Moon and by Lunar-Laser Ranging measurements performed from the stations on the Earth. These accurate regular measurements over the past 50 years have made the Moon the best place to test tidal theory and dissipation mechanisms. The interpretation of these variations provides information about the lunar interior, but many questions remain about its internal structure and dissipation mechanisms.}

\noindent
\tnr{The determination of the Love number $k_2$ at the monthly period of the Moon (its orbital period) has most recently been obtained by the Gravity Recovery and Interior Laboratory (GRAIL) space mission radio science experiment, which consists of precise measurements of two satellites in orbit around the Moon \citep{zuber_etal13}. The link between the satellites is in Ka-band and the link to the ground stations in S-band. This experiment is similar to the GRACE mission around the Earth and GRAIL has determined the lunar gravity field up to degree 900 \citep{konopliv_etal14} and provided the monthly Love $k_2$ number of the Moon at 0.02416 $\pm$ 0.00022  \citep{konopliv_etal13,konopliv_etal14, lemoine_etal13, Williams_etal14}. The Love number $h_2$ is determined on the Moon using the Lunar Reconnaissance Orbiter mission which carries a laser altimeter, Lunar Orbiter Laser Altimeter LOLA which measures displacements of the Moon's surface to an accuracy of 10 cm \citep{smith_etal17}. The Moon's Love number $h_2$ is determined to be 0.0371 $\pm$ 0.0033 \citep{mazarico_etal14} and 0.0387 $\pm$ 0.0025 \citep{thor_etal21}. Finally, the dissipation of the Moon is extracted from LLR ranging measurements by analysing the lunar orientation and libration \citep{park_etal21,pavlov_etal16,viswanathan_etal18} and see reference therein). The Lunar-Laser Ranging (LLR) consists of precise measurements of the round-trip travel time of a photon emitted from a laser on an Earth-ground station and lunar retroreflectors deployed by US astronauts and the Russian Lunakhod robotic missions \citep{murphy_etal13,chabe_etal20}. These measurements allow to determine the Earth-Moon distance centimeter level and the lunar rotation to the milli-arcsecond level \citep{viswanathan_etal18,park_etal21}. Lunar orientation and libration analysis allows for  extraction of the lunar dissipation. Currently, the monthly lunar tidal bulge delay time is estimated to be 0.1 days which corresponds to a monthly dissipation factor Q = 38 \citep{Williams_etal01, WilliamsBoggs15}.}

\noindent
Beyond the monthly Love number $k_2$ and dissipation factor $Q$, precise analyses of the LLR data have determined their frequency dependence. Indeed, the Moon's orbit is strongly perturbed by the presence of the Sun and its orbital motion results from a three-body motion. The Moon's orbit is described according to the Delaunay arguments $\ell$, $\ell'$, $F$, $D$ which represent the mean anomaly of the Moon, the mean anomaly of the Earth-Moon barycentre, the latitude anomaly of the Moon and the elongation of the Moon of periods 27.55 days, 365.25 days, 27.21, 29.53 days, to which must be added the precession of the lunar orbit of 18.6 years (for a description of the lunar orbit see \citet{ chapront_etal88,Chapront-Touze91}. Thus, librations and tidal shifts are mixed and tidal frequencies appear as mixed combinations of these frequencies with a period spectrum ranging from 2 weeks to 18.6 years and frequencies at 1 month, 7 months and 1 year \citep{Williams_etal01,rambauxWilliams11, WilliamsBoggs15}. The extraction of dissipation factors from LLR fits and the orientation of the Moon is challenging, but the terms at 27.2 days and 365 days stand out, while two others at 1095 and 2190 days provide upper bounds on lunar dissipation, whereas the term at 206 days is very complicated \citep{Williams_etal01,WilliamsBoggs15}. The $k_2/Q$ curve as a function of period has a kinked shape as shown in Section~\ref{sec:Viscoelasticity}, and the rheological models can then be compared with the observations to identify plausible internal structure patterns \citep{WilliamsBoggs15}.

\begin{figure}[ht!]

\hspace{3cm}\includegraphics[width=.73\textwidth]{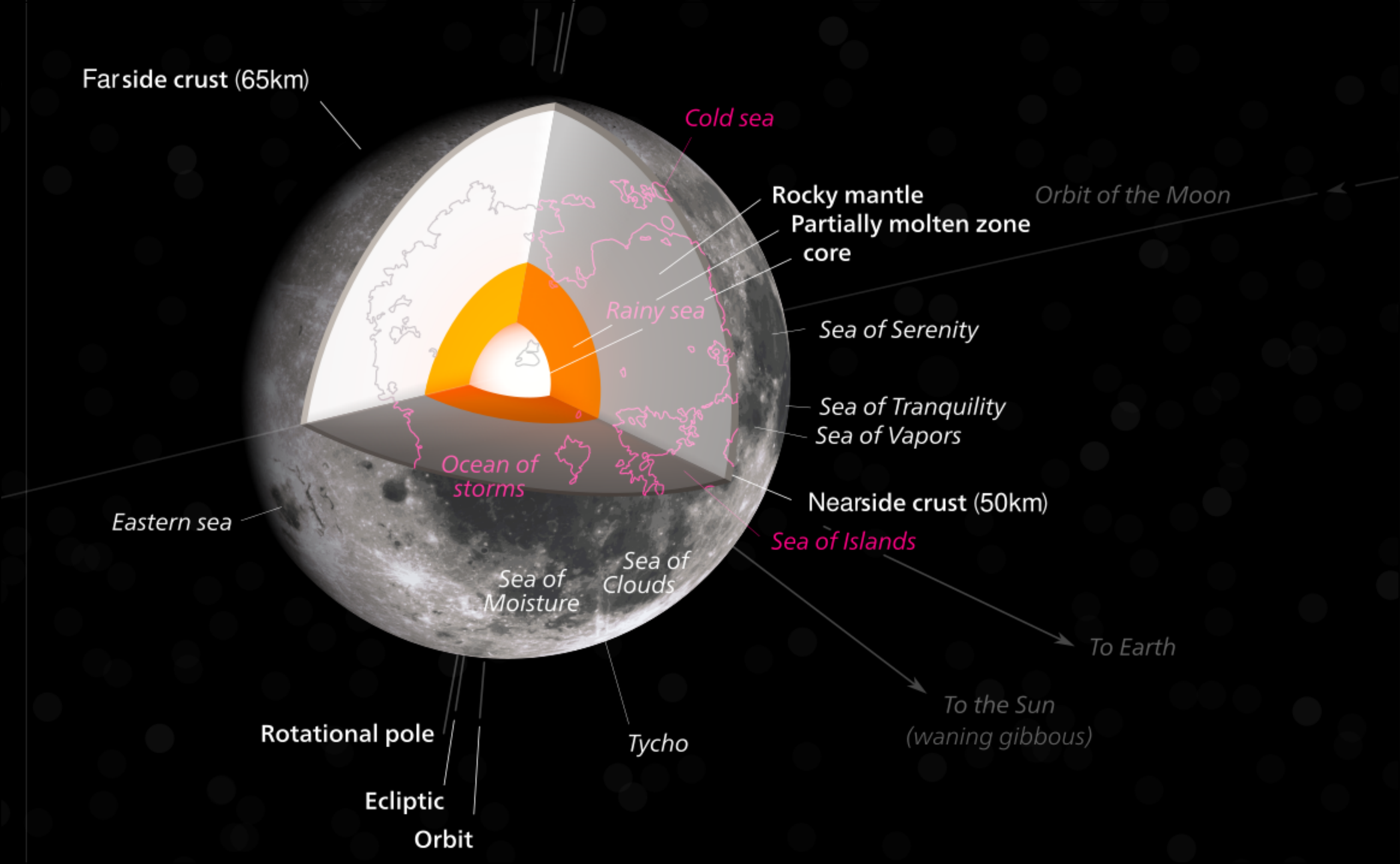}\\
\quad
\vspace{1cm}
\hspace{1.9cm}\includegraphics[width=.8\textwidth]{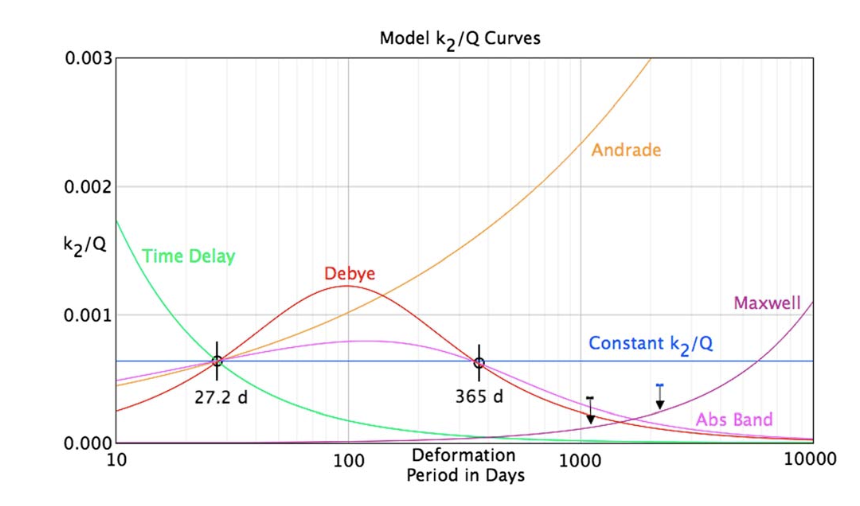}
\caption{Top: Artistic view of the interior structure and surface features of the Moon;
Bottom: Frequency dependence of k2/Q for various lunar rheological models (in colors) and LLR measurements points and limits (in black). The absorption band and Debye model presents a kink reproducing the measurements (figure from \citet{WilliamsBoggs15}"}
\label{fig:moon}
\end{figure}

\noindent
Several rheological models, such as Maxwell, Andrade, absorption band (See Figure~\ref{fig:moon}), were employed to explain the relatively low monthly dissipation factor Q=38 and the bell-shaped behaviour of the $k_2/Q$ law \citep[see e.g.,][]{WilliamsBoggs15}. Three physical mechanisms to explain this dissipation have been suggested: (i) the properties of the lunar material, (ii) the presence of water at depth and (iii) the presence of a partially molten zone. 

\noindent
(i) By analogy with the Earth, \citet{Nimmo_etal12} examined the influence of polycrystalline olivine melt-free to explain the observed dissipation. They were able to reproduce the dissipation at seismic and monthly tidal frequencies but not the slope behaviour associated at longer tidal periods. \citet{matsuyama_etal16} developed a Bayesian approach to test the presence or not of a transient layer above the core from the observed mass, solid moment of inertia, and elastic Love numbers $k_2$ and $ h_2$ to constrain the density and rigidity profile of the Moon. Their model favors an internal structure without a transition zone. However, it should be noted that this observation is based on a purely elastic approach.
\textcolor{black}{\citet{matsumoto_etal15} adopted a similar approach to \citet{matsuyama_etal16} to study the  interior of the Moon using an MCMC method. Their results are slightly different than \citet{matsuyama_etal16} in that, for example,  \citet{matsuyama_etal16} obtain a unimodal probability distribution for the liquid outer core radius, while \citet{matsumoto_etal15} obtained a bimodal distribution.  \citet{matsuyama_etal16} attribute this difference to the ways of computing the Love numbers, i.e., \citet{matsuyama_etal16} apply a simple anelastic correction to the observed Love numbers, while \citet{matsumoto_etal15} use a viscoelastic model to compute the Love numbers at the tidal forcing period, or, alternatively, to the use of additional tidal quality and seismic constraints by \citet{matsumoto_etal15}.}

\noindent
(ii) \citet{Karato13} proposed that water concentration would be the mechanism behind dissipation, also by analogy with the Earth's asthenosphere. Although this model reproduces the low value of the dissipation factor it does not describe the frequency dependence. 

\noindent
(iii) Finally, the molten layer model proposed by Nakamura as early as 1973 from seismic data seems to be the most robust. A Bayesian approach developed by \citet{Khan_etal14} used Love number, mass distribution, composition and electromagnetic data to explore the parameter space and conclude that the melt layer is present. Later, an attempt to fit the LLR data to the power scaling law $\,Q\sim\chi^{p}\,$
 resulted in a small {\it{negative}} value of the exponential: $\,p\,=\,-\,0.19~$ \citep{Williams_etal01}. A subsequent
 reprocessing of the data in \citep{WilliamsBoggs2009} rendered the value $\,p\,=\,-\,0.07~$.
\citet{Efroimsky2012a} proposed that since the frequency-dependence of $Q^{-1}$ has a kink form, as in Figure~\ref{figure}, the negative slope found by the LLR measurements may be located slightly to the left of the pick of the kink. For a Maxwell Moon, this necessitates the mean viscosity as low as 
$\sim\s 3\times 10^{15}\s$ Pa s, 
implying the presence of a considerable amount of partial melt. More recently, various models were used to explain by a partially molten layer the LLR data on the frequency dependence of dissipation \citep[e.g.,][]{Harada_etal14, Harada16, Rambaux14c, WilliamsBoggs15, Tan21, Briaud_etal22}.

\noindent
\tnr{We are at the beginning of the study of lunar dissipation, and our knowledge is still fragmentary. Several directions of improvement are now possible. Firstly, from an observational point of view, the installation of new, more compact and optimised single lunar Cube Corner Retroreflectors \citep{turyshev_etal13,dellAgnello_etal15}, better distributed spatially on the Moon during the next Artemis missions, will make it possible to improve the return of the photon number and the precision of the measurements \citep{dellAgnello_etal11, dehant_etal12}. 
The accuracy of these measurements is about 1-2 centimeter \citep{viswanathan_etal18, park_etal21} while theoretically the accuracy is expected to be around one millimeter \citep{Samain98}. This means that data analysis and dynamic models need to be improved to reach the theoretical accuracy. Finally, a better knowledge of long-period lunar tides will be possible by combining other measurements, such as altimetry, active retroreflectors, differential lunar laser ranging or from the zenith telescope \citep{petrova_etal13,zhang_etal20,dehant_etal12,thor_etal20}. Lunar seismic measurements will also help to better constrain the Moon's interior and to better understand tidal effects \citep{garcia_etal19}. Finally, laboratory measurements of the dissipation processes of lunar and supposedly lunar materials will help constrain the measurements and models currently used.}

\begin{comment}

 An attempt to fit the LLR data to the power scaling law $\,Q\sim\chi^{p}\,$
 resulted in a small {\it{negative}} value of the exponential: $\,p\,=\,-\,0.19~$ \citep{2001JGR...10627933W}. Later
 reprocessing of the data provided the value $\,p\,=\,-\,0.07~$. $\,$According to \citet{WilliamsBoggs2009}.\\
 ~\\
 ``{\it{There is a weak dependence of tidal specific dissipation $\,Q\,$ on period. The $\,Q\,$ increases from $\,\sim 30\,$ at a month to $\,\sim 35\,$ at one year. $~Q\,$ for rock is expected to have a weak dependence on tidal period, but it is expected to decrease with period rather than increase. The frequency dependence of $\,Q\,$ deserves further attention and should be improved.}}"
 \vspace{3mm}

\noindent
\citet{Efroimsky2012a, Efroimsky2012b} proposed that since the frequency-dependence of $k_2/Q$ has a kink form, as in Figure~\ref{figure}, the negative slope found by the LLR measurements may be located slightly to the left of the pick of the kink. For a Maxwell Moon, this necessitates the mean viscosity as low as $\sim\s 3\times 10^{15}\s$ Pa s, implying the presence of a considerable amount of partial melt.

\end{comment}

\subsection{Mars}\label{sec:Mars}

\noindent
Owing to numerous successful missions, particularly in the past two decades, Mars has become the most geophysically-explored planet after the Earth. As the only terrestrial planet except the Earth which has natural satellites, body tides in Mars are generated mainly by the Sun and Phobos, whereas the tides caused by Deimos are much weaker because of its small size and larger separation.
A vast volume of information about Mars has been obtained from orbiting and landed spacecraft (Mars Pathfinder, Mars Global Surveyor, Mars Odyssey, Mars Reconnaissance Orbiter and Mars Express), and from the geodetic Doppler ranging data
\citep{Yoder_etal03,Bills_etal05,Lainey_etal07, Konopliv_etal06, Konopliv_etal11,JacobsonLainey14, Konopliv_etal16, konopliv_etal20, Genova_etal16,lainey_etal20}.
The Mars InSight lander touched down on the planet on November 2018, carrying on board a suite of geophysical instruments, including a seismometer, a magnetometer, a radio science experiment, and a heat flow probe to explore the planet's interior \citep{Smrekar_etal19,Banerdt_etal20, Giardini_etal20,Lognonne_etal20}. Before the recent improvements in our knowledge of the interior structure of Mars owing to the seismic measurements provided by InSight \citep{Banerdt_etal20,Lognonne_etal20,Stahler_etal21,Khan_etal21,duran_etal22,Knapmeyer_etal21}, geodetic measurements used to be the sole constraint on the Martian interior.

\begin{figure}[ht!]
\hspace{-2mm} \includegraphics[width=.659\textwidth]{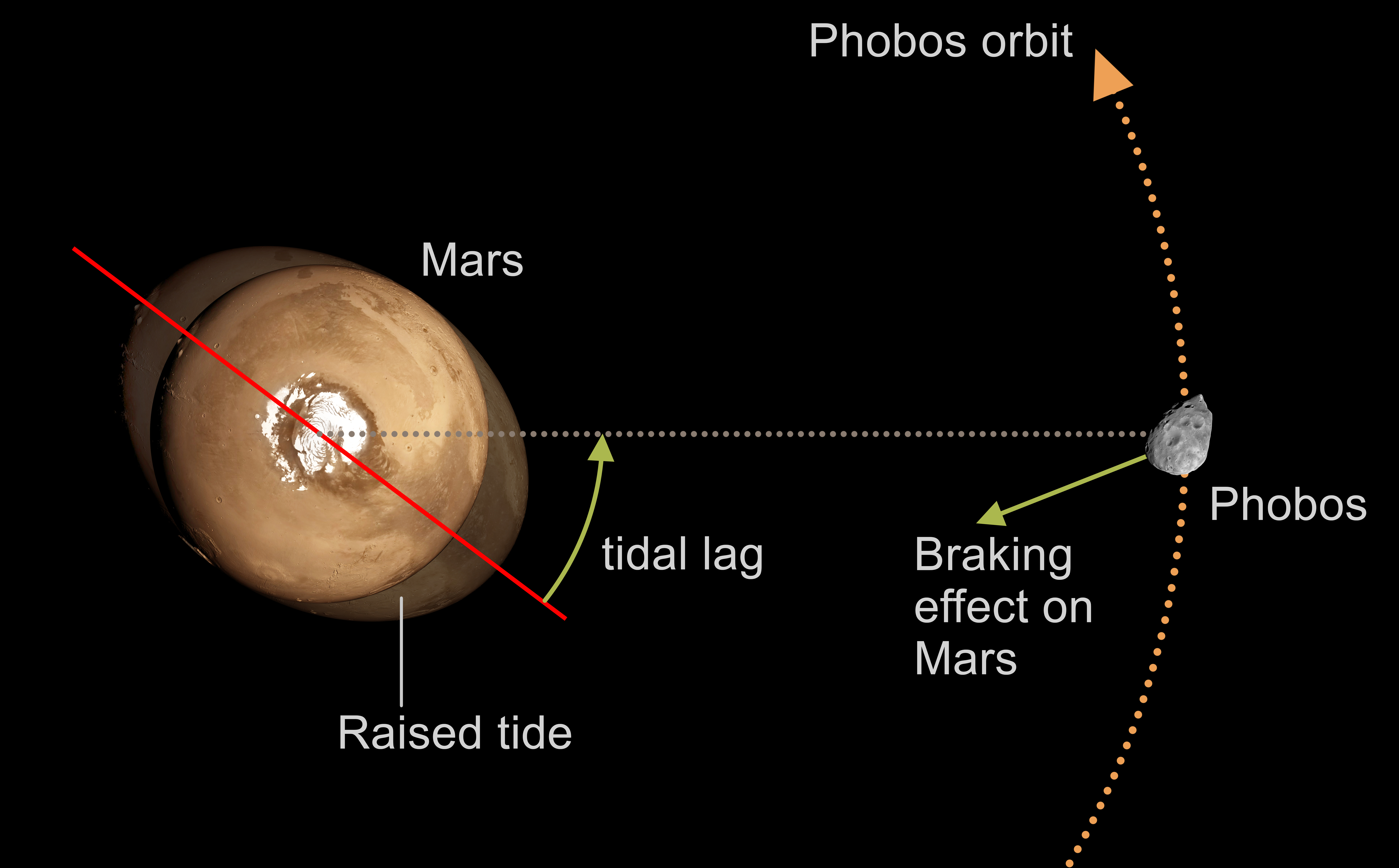}
\includegraphics[width=.361\textwidth]{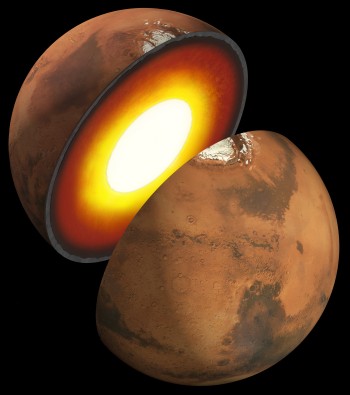}
\caption{Artistic view of tides raised by Phobos on Mars (left) and interior of structure Mars with crust, mantle and core (right). Image credit: Left:David Ducros/IPGP, Right: NASA/JPL-CalTech}
\label{fig:lander}
\end{figure}
\noindent
Constraining the tidal Love number of Mars can considerably help to understand the size and state of the Martian core. Along with that, the Martian tidal quality factor and its frequency-dependence can impose effective constraints on the mantle's temperature and rheological properties. Several studies based on different data sets have aimed at constraining Mars's tidal Love number, a summary of which can be found in Figure~6 of Chapter 3 of this volume. Basing their analysis on the tracking data collected by the Mars Orbiter Laser  Altimeter on the Mars Global Surveyor (MGS), \citet{smith_etal01}, obtained a Love number as low as 0.055$\pm$0.008, as the first measurement of the Martian Love number. This value is very close to the $k_2$ of a Mars-size hypothetically homogeneous solid planet, which implies that the Martian core has solidified. Yet, a  later analysis of the MGS Doppler and range data by  \citet{Yoder_etal03}, combined with the information from Mars Pathfinder and Viking Lander, suggested a much larger tidal Love number of 0.153 $\pm$ 0.017 at the period of the Solar tides (11 hr 19 min). Such a large value is incompatible with the existence of a solid core, and instead indicates the presence of a large liquid core. \citet{Yoder_etal03} obtained a liquid iron core of 1520-1840~km in radius based on their measured $k_2$. Using the measurement of the secular acceleration of Phobos, they proposed for the tidal quality factor a value of Q = 92$\pm$11. A caveat in computing the tidal quality factor is that the secular acceleration deals with the Phobos-raised tides at the synodic period (T~=~5.55~hr), whereas the tidal Love number is measured at the semi-diurnal tides period (T~=~24.62~day), and it is not clear whether this difference is taken into account. By removing the anelastic effect, in analogy to the Earth's tidal response to the Moon,
\citet{Yoder_etal03} also suggested that the elastic tidal Love number of Mars is in the range of 0.145$\pm$0.017.

\noindent
Several later studies dealt with constraining the tidal Love number and quality factor of Mars by analysing the accumulated data from different Mars orbiters. The deduced values for $k_2$ and $Q_2$ were not dramatically different from those obtained by \citet{Yoder_etal03}.
Having collected six years of MGS tracking data and three years of Mars Odyssey tracking data,
\citet{Konopliv_etal06} derived a global solution for Mars's gravity field, including the Solar-tide Love number of $k_2$ = 0.152$\pm$0.009 and the elastic tidal Love number of $k_2$ = 0.148$\pm$0.009. From the computed values of the Love numbers, these authors inferred the presence of a fluid core of a 1600-1800~km radius. \citet{lemoine_etal06} employed six years of MGS data, starting from 1999, to obtain $k_2$ = 0.176$\pm$ 0.041. Including tracking data from the MRO spacecraft, \citet{Konopliv_etal11} improved their previous finding and achieved a noticeable progress in determination of the high frequency portion of the spherical harmonic Mars gravity ﬁeld. This
resulted in a slightly larger value of the tidal Love number ($k_2$ = 0.164$\pm$0.009). Using the Deep Space Network tracking data of the NASA Mars missions, Mars Global Surveyor, Mars Odyssey, and the Mars Reconnaissance Orbiter, \citet{Genova_etal16} presented a higher-order and high-degree spherical harmonic solution of the Martian gravity field,  finding $k_2$ = 0.1697$\pm$0.0027. Note that \citet{Genova_etal16} directly take into account the atmospheric effects in their estimations, whereas all other studies need to correct for this effect after computing the Love number. Most recently, \citet{konopliv_etal20} analysed several years of data to detect the Chandler wobble of the Martian pole, for the first time for any Solar system body other than the Earth. In line with several previous measurements, this study suggested  $k_2$ = 0.169$\pm$0.006 at the Solar tides period without taking into account the atmospheric effects, and  $k_2$ = 0.174$\pm$0.008 after correcting for this effect.

\noindent
\textcolor{black}{\citet{pou_etal21} modeled the response of Mars to Phobos tides using tidal potential deduced from JPL Horizons ephemerides and predicted how the Very Broad Band seismometer (VBB) on the InSight lander can be used as a gravimeter to constrain the tidal response of Mars; however, after 3 years, extracting the tidal signal has proven to be more difficult and to require further data accumulation to reduce the signal-to-noise ratio to the necessary level.}

\noindent
The tidal quality factor of Mars has been constrained by measuring the secular acceleration of Phobos, as the most feasible method. Both Martian moons are very small, faint, and rapidly orbiting the planet, and are difficult to observe from the Earth-based instruments. Nevertheless, before the spacecraft era, several astrometric measurements had been carried out to constrain the orbital properties of the moons. The precision of most of those observations was, however, limited. A list of such observations can be found in \citet{JacobsonLainey14}.  \citet{Bills_etal05} used the observed transit of the shadow of Phobos from MOLA, to estimate the tidal quality factor of the planet, and found the secular acceleration of Phobos s = (1.367$\pm$0.006)$\times 10^{-3}~\rm deg/yr^2$. Therefrom, these authors deduced the tidal quality factor to be Q = 85$\pm$0.37. Further analysis was performed by \citet{raineyAharonson06, Lainey_etal07,Jacobson10, lainey_etal20}, with the aid of the Martian moons' ephemerides. In sumary, the quality factor derived by these studies resides in the range 78$\lesssim Q \lesssim$ 105.

\noindent
The observed tidal response, i.e. the values of the Martian tidal Love number and quality factor, can significantly improve our knowledge of its interior. \citet{Rivoldini_etal11} used the tidal Love number and the moment of inertia of Mars to constrain the core radius and sulfur content. They considered two main end-members of cold and hot interior models and found a core radius of 1794$\pm$65~km and a sulfur concentration of 16$\pm$2~\rm wt\%. These authors concluded that the geodetic data alone are not capable of constraining the mineralogy of the mantle and the crust. Note that \citet{Rivoldini_etal11} only considered an elastic Mars without taking into account the tidal  dissipation in the planet.

\begin{figure}[ht]
\begin{center}
\includegraphics[width=.7\textwidth]{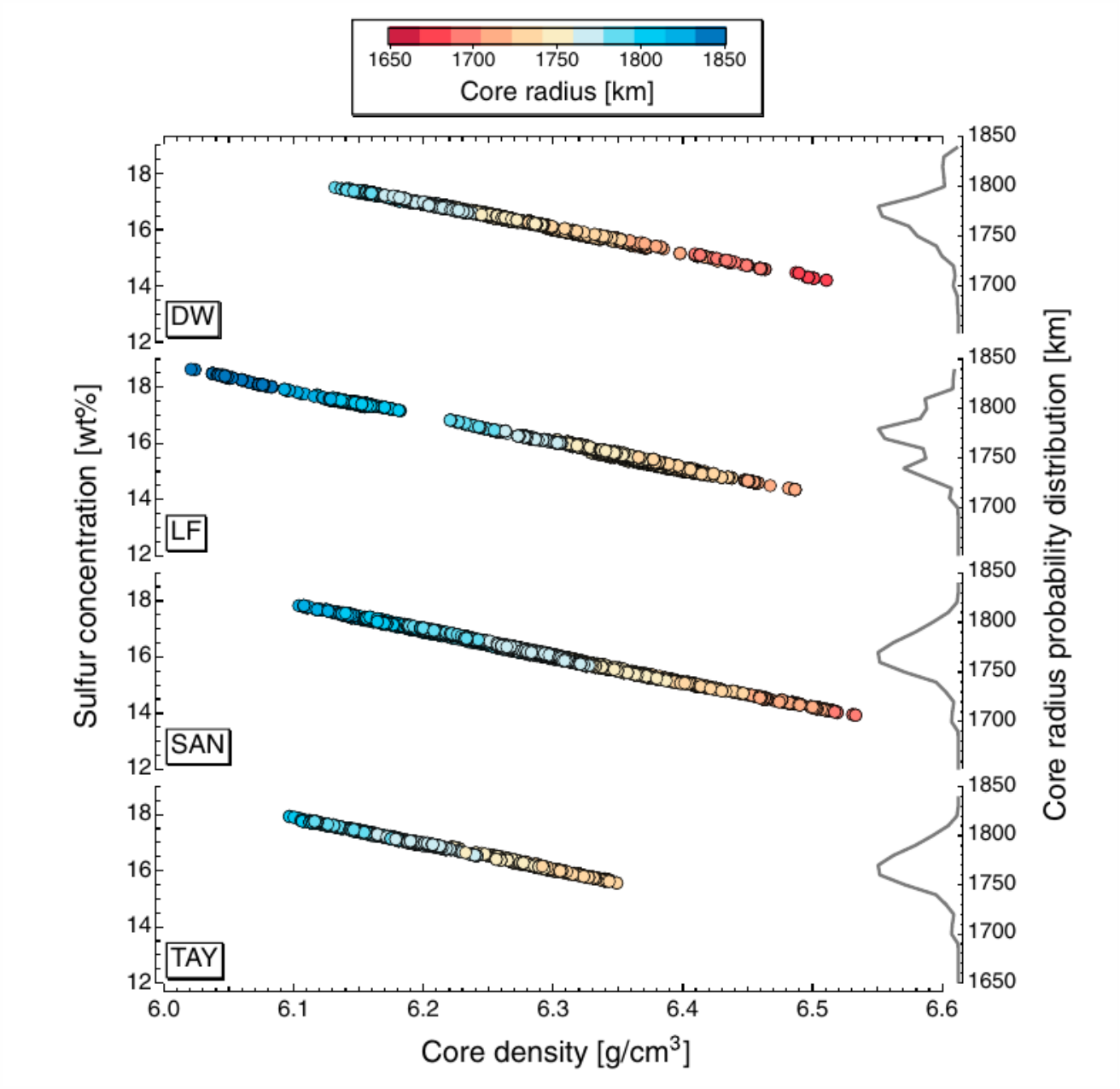}
\caption{ Sampled core properties in a probabilistic inversion method used by \citet{Khan_etal18}: radius, sulfur content, and density for the four main compositions (deﬁned in
Table 1). The histograms on the right axis show the probability distributions of core radius for each of the main
compositions.  }
\label{fig:mars_core}
\end{center}
\end{figure}

\noindent
\citet{Khan_etal18} utilised the knowledge on the Martian tidal response, along with mass and moment of inertia, to constrain, in the context of a probabilistic inversion method, the planet's major interior properties such as the size of the core, mantle, and crust, the composition of each part, the dissipative properties of the mantle, and the temperature profile. They employed a laboratory-based viscoelastic dissipation rheology (extended Burgers, \textcolor{black}{as discussed in Section~\ref{sec: Burgers}}) to model the anelasticity of the Martian mantle, and used the known tidal response in inversion. They examined different crust and mantle compositions based on the analysis of Martian meteorites \citep[i.e.,][]{DreibusWanke87, MorganAnders79,LoddersFegley97,Sanloup_etal99,Taylor13}. Figure~\ref{fig:mars_core} from \citet{Khan_etal18} demonstrates the inversion results for the properties of the core. As shown in this figure, they obtained a core radius of 1730-1840~km and 15-18.5$~\rm wt \%$ sulfur, similar to \citet{Rivoldini_etal11}. The colour code in the figure also shows the trade-off between the core radius and core density: the larger the radius, the lighter the core~-~and the more light elements in it, such as sulfur. Their results show that, except for the composition model of \citet{MorganAnders79}, all models were able to fit the data, and hence are not distinguishable based on the available geodetic measurements. \citet{Khan_etal18}  also determined that the parameter $\alpha$ from the extended Burgers model (see equation (\ref{anderson}) above) takes values in the range 0.24-0.38. They also found the temperature at the lithosphere to be in the range 1400-1460$~^\circ \rm C$.
\noindent
\clearpage
% Text added by Ina
\noindent
\textcolor{black}{A recent study by \citet{Plesa_etal18} used global thermal evolution models and calculated the tidal deformation of Mars using an Andrade pseudo period model \citep[Section \ref{Andrade} in this paper,][]{JacksonFaul10,jackson2005laboratory}. Model results indicate that a cold mantle that may result from concentrating nearly all radioactive heat producing elements in the crust would have a smaller quality factor $Q$ (i.e, the mantle would not be dissipative enough) than the available estimates. Conversely, models with a hot interior, either due to the presence of a large amount of heat sources in the mantle or due to an inefficient heat transport caused by a high mantle viscosity, would be too dissipative to satisfy the constraint given by the tidal quality factor estimates. In order to match the values of the tidal Love number $k_2$ of \citet{Konopliv_etal16} and \citet{Genova_etal16}, the combined thermal evolution and tidal deformation models of \citet{Plesa_etal18} indicated a core size larger than 1800 km with a preferred value of 1850 km, a value range consistent to the core size estimates of \citet{Rivoldini_etal11} and \citet{Khan_etal18}.}

\noindent
\citet{Bagheri_etal19} used an approach similar to \citet{Khan_etal18}, but employed several viscoelastic models \textcolor{black}{mentioned in Section~\ref{sec:Viscoelasticity}} to constrain the interior properties of the planet, particularly the dissipation. They examined the extended Burgers, Andrade, power-law, Sundberg-Cooper, and Maxwell models to fit the observations. Their results show that all these models, except Maxwell, are capable of fitting the tidal response measurements for Mars. The Maxwell model, however, can only fit the observed tidal response only at the expense of unrealistically low mantle viscosities. This is consistent with the conclusions from the work by \citet{Bills_etal05}, in which a very low average viscosity of the mantle was found and was attributed to the possible presence of partial melt within the planet \citep{kiefer03}, or a more volatile-rich mantle, or a tidally forced flow of a fluid within a porous solid \citep{nield_etal04}.
However, the results of \citet{Khan_etal18} and \citet{Bagheri_etal19} demonstrate that the tidal dissipation rate in Mars can be explained by the anelastic strength of the the mantle, without the need for the said phenomena. This observation reveals the necessity of utilising more elaborate viscoelastic models for the planetary bodies, especially for tidal frequencies, because at these frequencies the anelastic relaxation is important.
\citet{Bagheri_etal19} determined that the parameter $\alpha$ that appears in the Andrade and Sundberg-Cooper models, and in the power-law approximation should assume values in the interval of 0.22-0.42, which values are higher than the Earth's value of 0.15 \citep{IERS}. This serves as another indication that the Martian interior is colder, because generally $\alpha$ tends to decrease with the increase of temperature, especially on approach to melting point \citep{Fontaine,Bagdassarov}, and it is possible that the value of $\alpha$ for the Earth is so low due to the presence of partial melt.

\begin{center}
\begin{figure}[ht!]
\hspace{3cm}\includegraphics[width=0.65\textwidth]{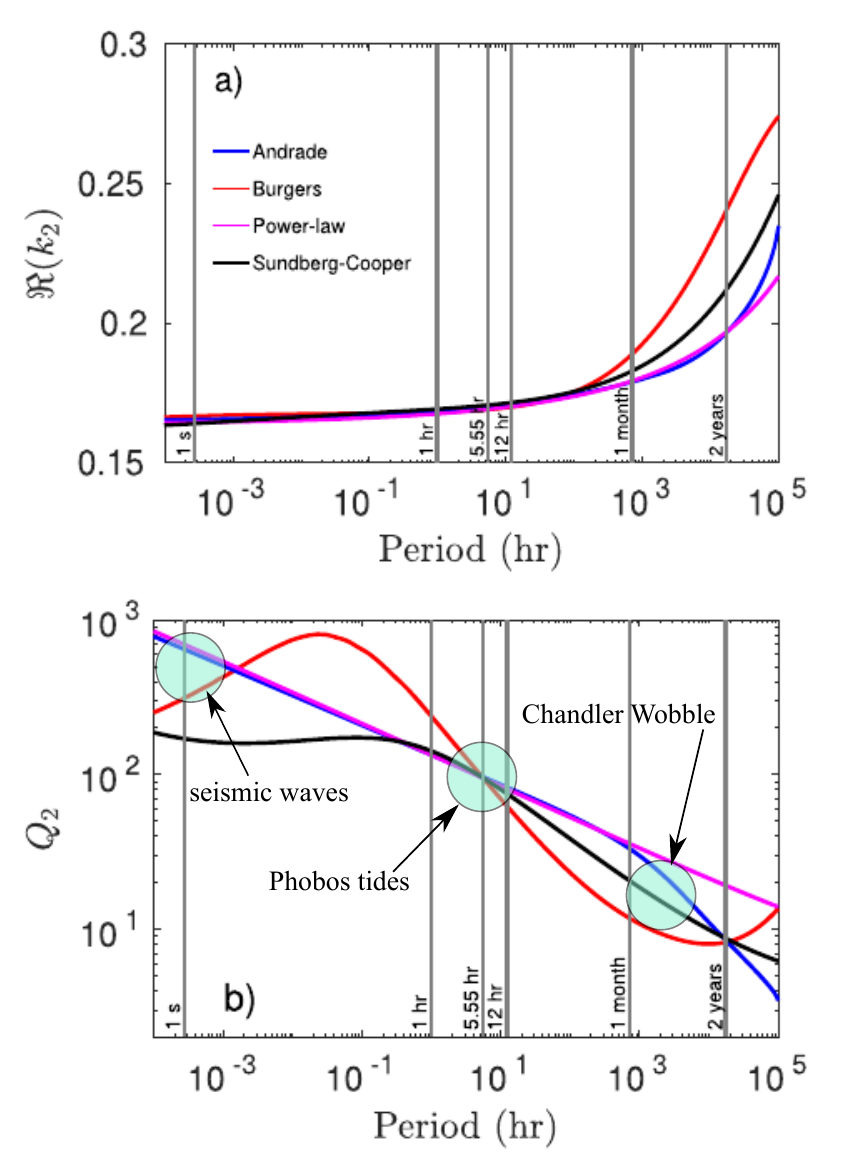}\\
    \caption{\small{The potential Love number and quality factor of Mars as a function of period, for four rheological models: Andrade, Burgers, power-law, and Sundberg-Cooper. Green areas indicate the periods at which measurements currently exist. These are the period of tides raised by Phobos ($\sim$5.55 hrs) and the periods of seismic waves (seconds) \citep{lainey_etal20,Lognonne_etal20}.}}
    \label{fig:quality_factor_mars}
\end{figure}
\end{center}

\noindent
\citet{Bagheri_etal19} also found the tidal displacement Love numbers $h_2$ (0.22-0.24) and $l_2$ (0.037-0.040) for Mars. These authors provided estimations of dissipation ($Q$) at different periods, such as seismic waves, normal modes, and Solar tides, based on each viscoelastic model, and showed that while all models are capable of fitting the only available data point for dissipation (the Phobos tides), they do not predict similar dissipative behaviour for other periods \citep{Bagheri_etal19} (See Figure~\ref{fig:quality_factor_mars}). The recent findings of seismic attenuation and anelasticity at the Chandler-wobble period can further improve their finding.
%\clearpage
\noindent
\textcolor{black}{\citet{Pou_etal22} considered modeling the tidal response of Mars with the aim of linking between the tidal and seismic attenuation using different viscoelastic models. Due to the proximity of Phobos to Mars, they take into account higher degrees of tidal response, up to and including  degree 5 tides. They constrain the core radius as 1805$\pm$75~km and the average shear attenuation of the mantle in the range of 100 and 4000. The large uncertainty implies that the current available tidal observations are not precise enough to constrain the attenuation properties of the Martian mantle.}
\noindent
In the recent studies based on analysing the seismic data from Marsquakes detected by InSight's Very Broad Band Seismic Experiment for Interior Structure (SEIS), the radius of the Martian core was precisely constrained \citep{Stahler_etal21}. The results show a liquid core of a 1830$\pm$40~km radius and the core density of 5700-6300$~\rm kg/m^3$. These findings were refined with further analysis of the seismic data;  \citet{duran_etal22} used the analysis of the direct, reflected, and converted seismic waves to obtain the radius of the core as 1820-1870~km and its mean density (6000–6200~$\rm kg/m^3$). Moreover, \citet{khan_etal22} used a joint geophysical and cosmochemical approach to constrain the radius and density of the core as 1840$\pm$10~km and 6150$\rm  \pm $46$~\rm kg/m^3$, respectively. These findings from precise seismic analysis compare very well with the conclusions based on the tides. This agreement illustrates the strength of the geodetic measurements in probing the large-scale planetary interior structure.

\noindent
Studying tides in the Mars-Phobos-Deimos system does not only provide information about the interior of the planet, but also can be used to infer knowledge on the interior structure and origin of its two companion moons. Both moons are presently tidally locked in 1:1 spin-orbit resonance to Mars and given their small eccentricity and radius, the dissipation in the system is dominated by that within Mars. In a recent study, \citet{Bagheri_etal21} used the available information on the tidal response of Mars, in combination with a laboratory-based viscoelastic model and an advanced tidal theory, to integrate back in time the current orbital configuration of the Martian moons and obtain information on the origin of the two moons. Both the tides produced in Mars by its moons and the tides produced in the moons by Mars contribute to these moons' orbital evolution. The results demonstrated that with a high probability, two moons are remnants of a disintegrated progenitor which was disrupted between 1--2.7 Gyrs ago. Moreover, based on the constraints obtained from the tidal evolution of the moons, they concluded that Phobos and Deimos are highly dissipative bodies and have a highly fractured and porous interior.  In an earlier study of tidal evolution of Phobos and Deimos, \citet{Yoder82} used a simplified tidal model and, assuming that the tidal friction in Phobos is negligible, suggested that Phobos might have a substantial internal strength. \citet{Samuel_etal19} used the same assumption, i.e. a monolithic Phobos, to jointly study the orbital evolution of Phobos, along with thermal evolution of Mars. However, this conclusion is not compatible with the constraints obtained by modeling the impact that have created the large Stickney crater on Phobos \citep{Asphaug_etal98}. 

\noindent
A question about Phobos and Deimos is their different tidal quality function ($k_2/Q_2$) obtained from studying their orbital evolution. All such studies which consider an orbital evolution time of the moons longer than $\sim $1 Gyrs remain to justify the large difference between the tidal quality function ($k_2/Q_2$) of the moons, as a measure of the interior structure. This difference results from the difference in the moons' present-day eccentricities, which is as large as two orders of magnitude and considering the eccentricity jumps associate with their crossing several resonance periods in their history. The puzzling origin of Phobos and Deimos is a long-standing problem and is out of the scope of this article. A complete review of the scenarios on the origin of the moons is provided by \citet{Rosenblatt11}.

\begin{center}
\begin{figure}[ht!]
\includegraphics[width=0.76\textwidth]{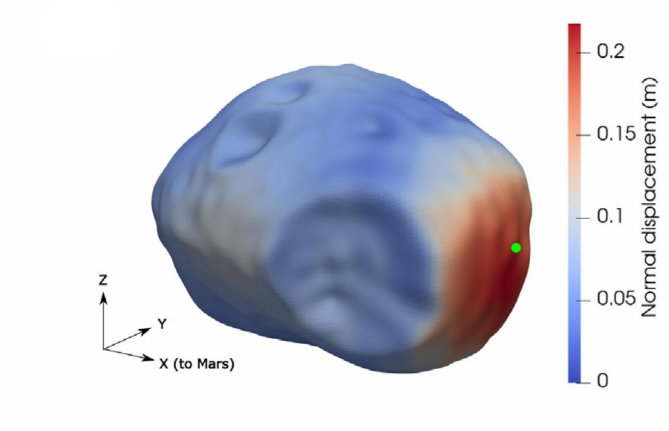}\\
    \caption{\small{Normal component of tidal deformations on the surface of the Martian moon, Phobos for an example interior structure from \citet{dmitrovskii_etal21}. The colored dot indicates the location of the sub-Mars point. As shown in this figure, the tidal displacements on the surface of such a body can vary considerably due to the irregularities in the shape. This necessitates using an appropriate 3D approach for such bodies.}}
    \label{fig:tides_phobos}
\end{figure}
\end{center}

\noindent
Although the two moons are tidally locked to Mars, their small orbital eccentricity can cause tides raised on them by Mars. Depending on the moons interior properties, such tides may be significant \citep[e.g.,][]{hurford_etal16, LeMaistre_etal13}. Recently, \citet{dmitrovskii_etal21} modeled the tidal deformations of Phobos to assess several possible interior structure models for the moon (See Figure~\ref{fig:tides_phobos}). They concluded that the currently available measurements are compatible with several models of the interior, including a rubble pile or monolithic body composed of rock or ice-rock mixtures. \citep{Campagnola_etal18,Usui_etal20}. Similar conclusions were drawn by \citet{Rambaux12, LeMaistre_etal13} and \citet{LeMaistre19} based on libration measurements. These models will be examined with the upcoming Japanese Martian Moons eXploration (MMX) mission and along with the samples returned from the moons, will be able to further constrain the origin of the Phobos and Deimos.

\clearpage

\subsection{Moons of giant planets}

\begin{figure}[ht]
\includegraphics[width=1\textwidth]{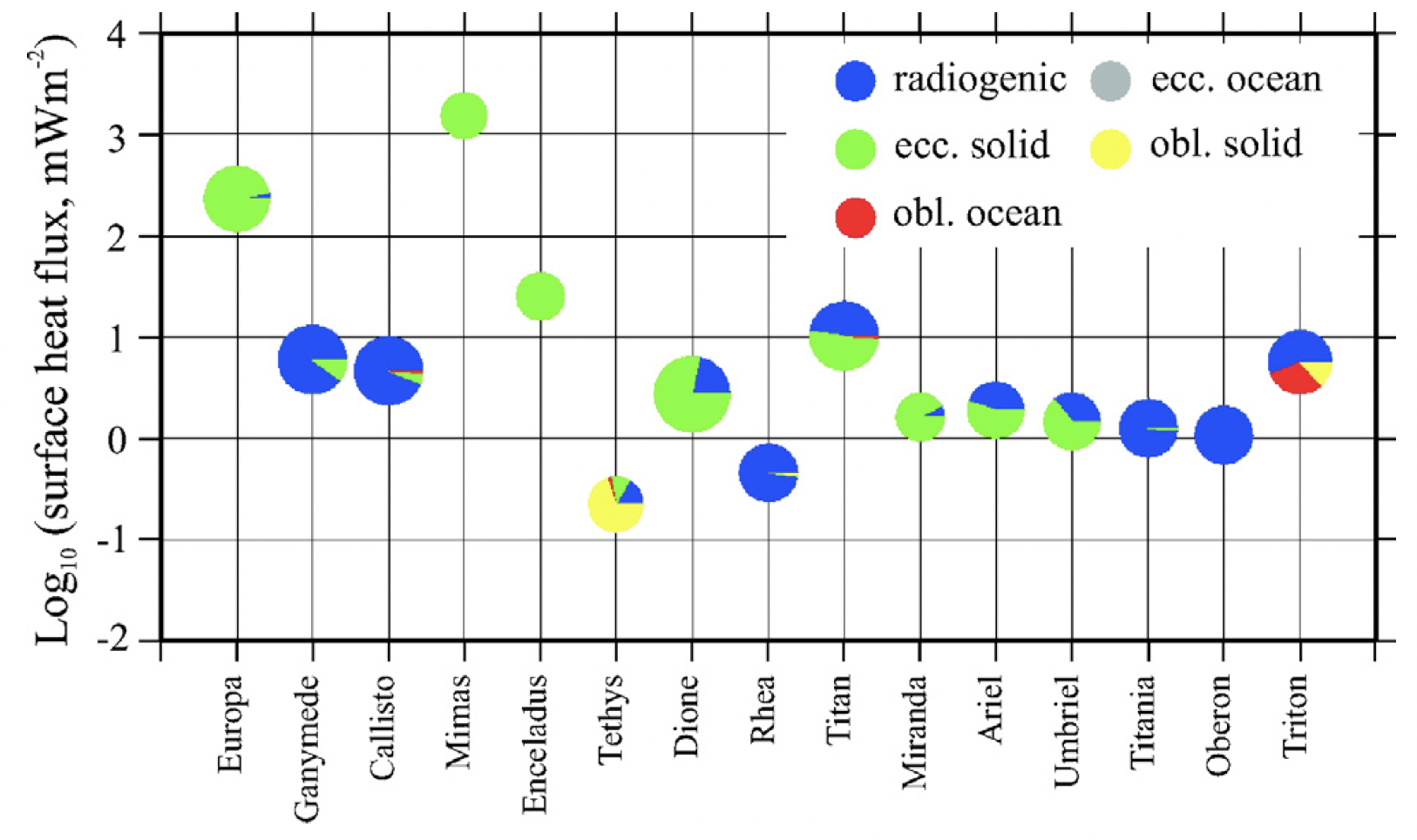}
\caption{Thermal heat budget for outer Solar System satellites, broken down in terms of radiogenic heating, and eccentricity and obliquity tides applied to the ocean and solid layers of the moons. This example assumes $k_2 = 3/2$. From \cite{chen2014}, permission pending.}
\label{fig:chen2014}
\end{figure}

The tidal response of the moons of the giant planets became a topic of great interest after it was recognised that tidal dissipation in these moons' interiors can be a significant source of heat \citep{cassen1979} (Figures~\ref{fig:chen2014} and \ref{fig:moonheatbudget}). Like for other bodies addressed in Section 5, tidal forcing also plays a role in the evolution of the moon dynamical properties (spin, orbit), and tidal stress can drive tectonic activity on some of the moons. The most famous examples are Enceladus's Tiger stripes and associated jet activity whose temporal variability is correlated with tidal forcing \citep{hedman2013}. Europa is another example, whose regional-scale ridges and cycloidal ridges have been produced from continuous forcing driving fatigue and eventual fracturing of surface material \citep{Rhoden2021}. In this section, we review the knowledge on the largest moons of giant planets with particular attention to the tidal dissipation as a heat source for different phenomenon within the interior of the moons. In Section~\ref{sec:surface_features}, we will concentrate on the surface geological features as a result of tidal activity.

\begin{figure}
\centering
\includegraphics[width=0.45\textwidth]{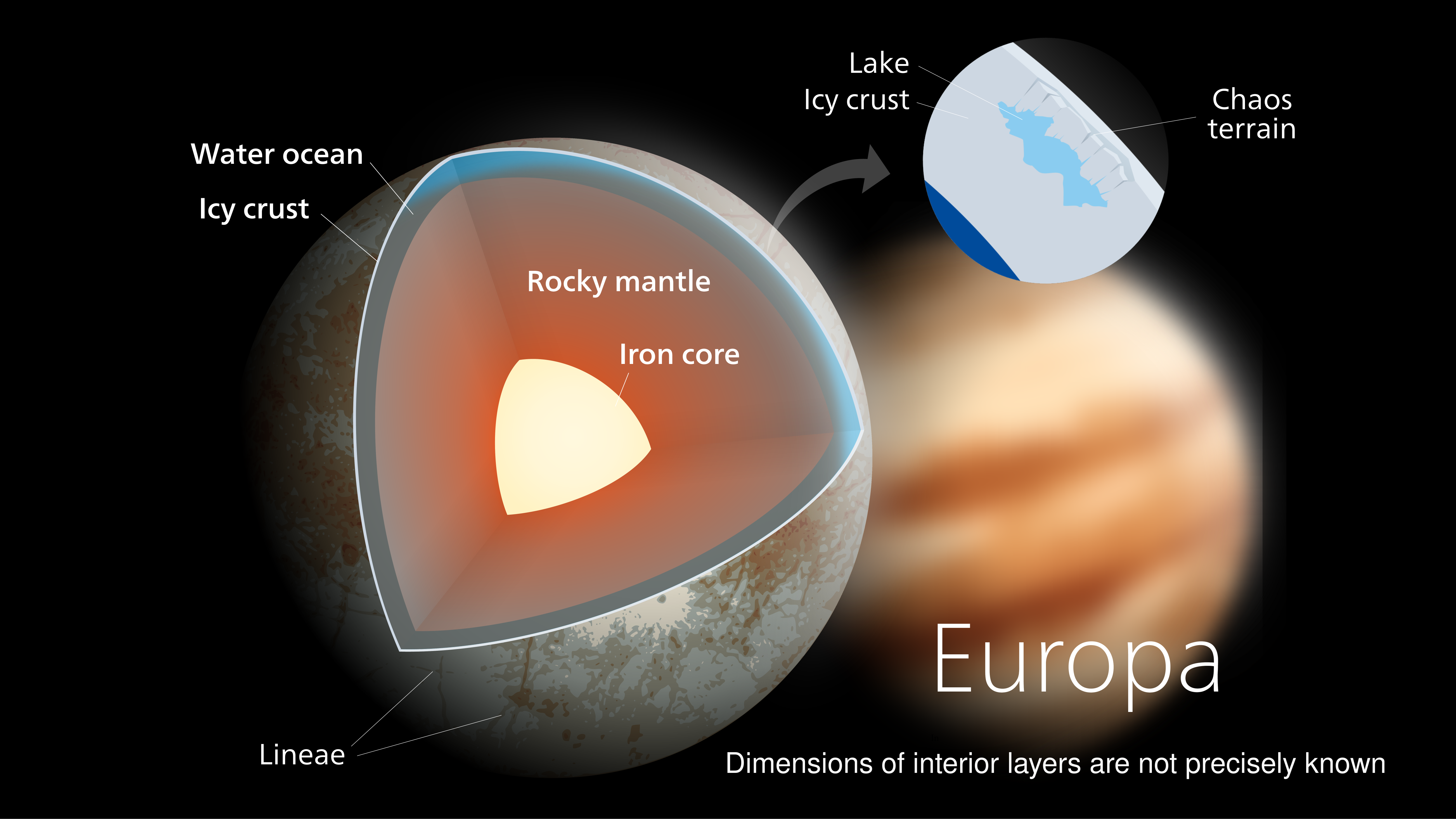}\\
\vspace{.5mm}
\includegraphics[width=0.45\textwidth]{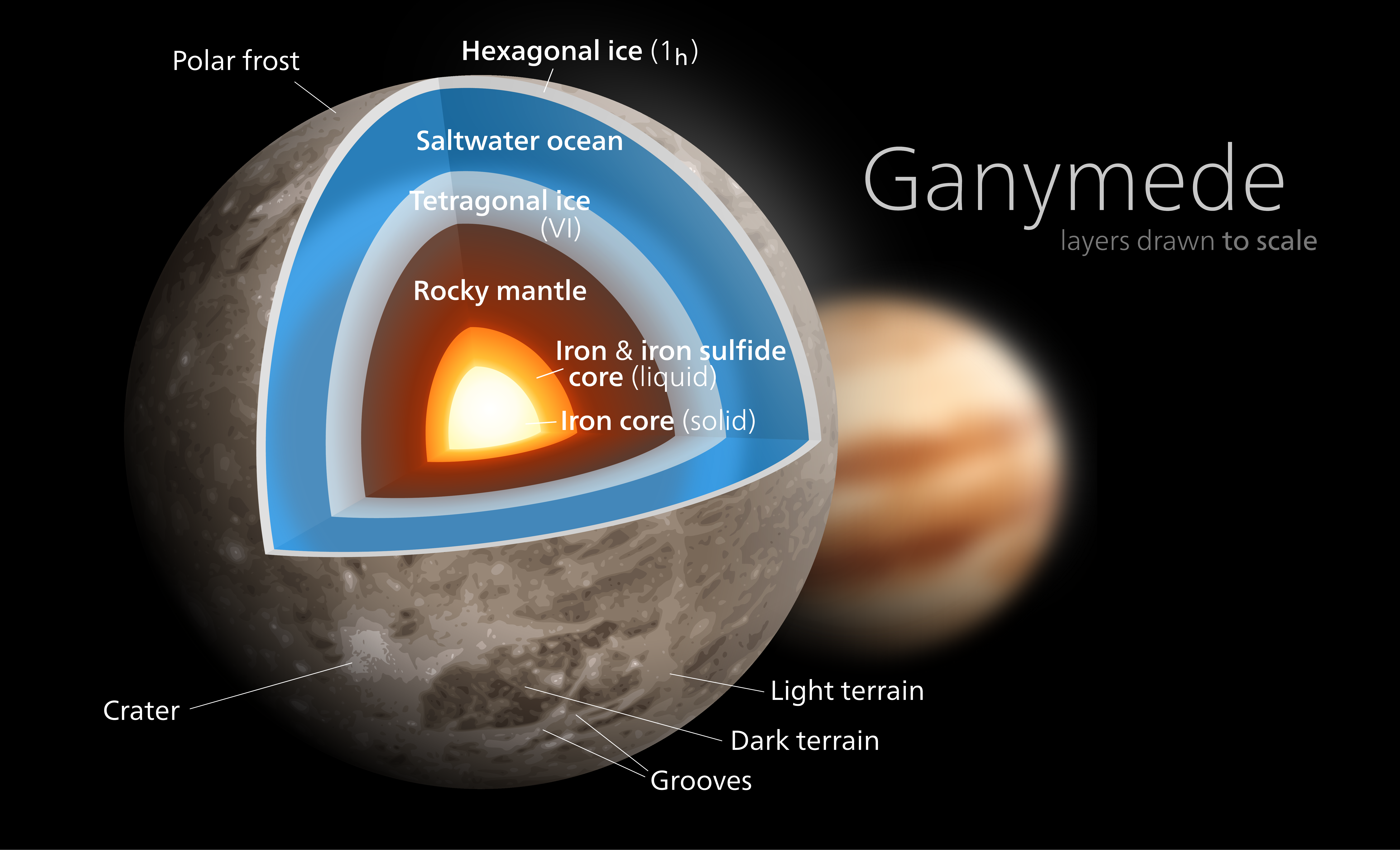}\\
\vspace{.5mm}
\includegraphics[width=0.45\textwidth]{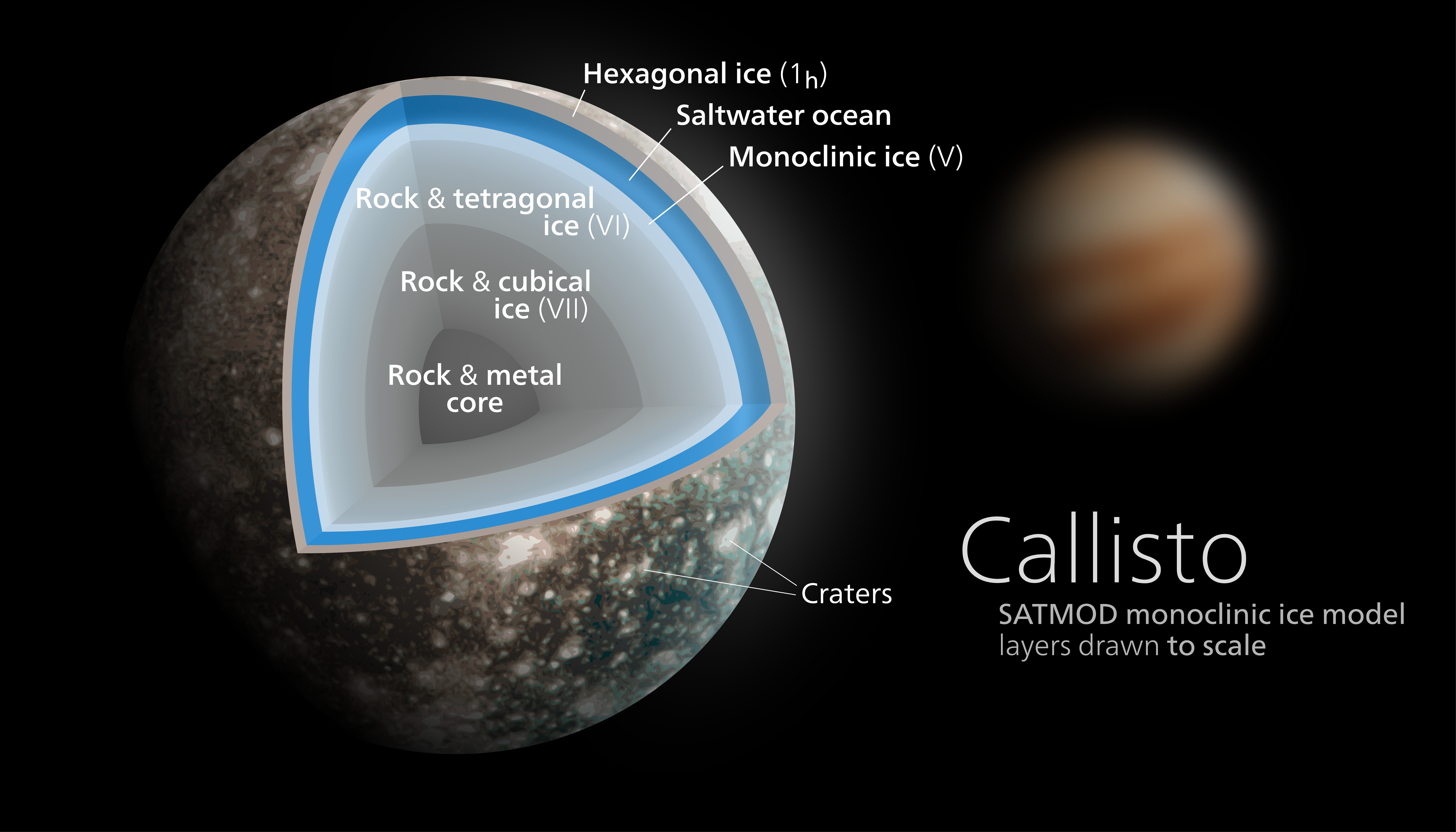}\\
\vspace{.5mm}
\includegraphics[width=0.45\textwidth]{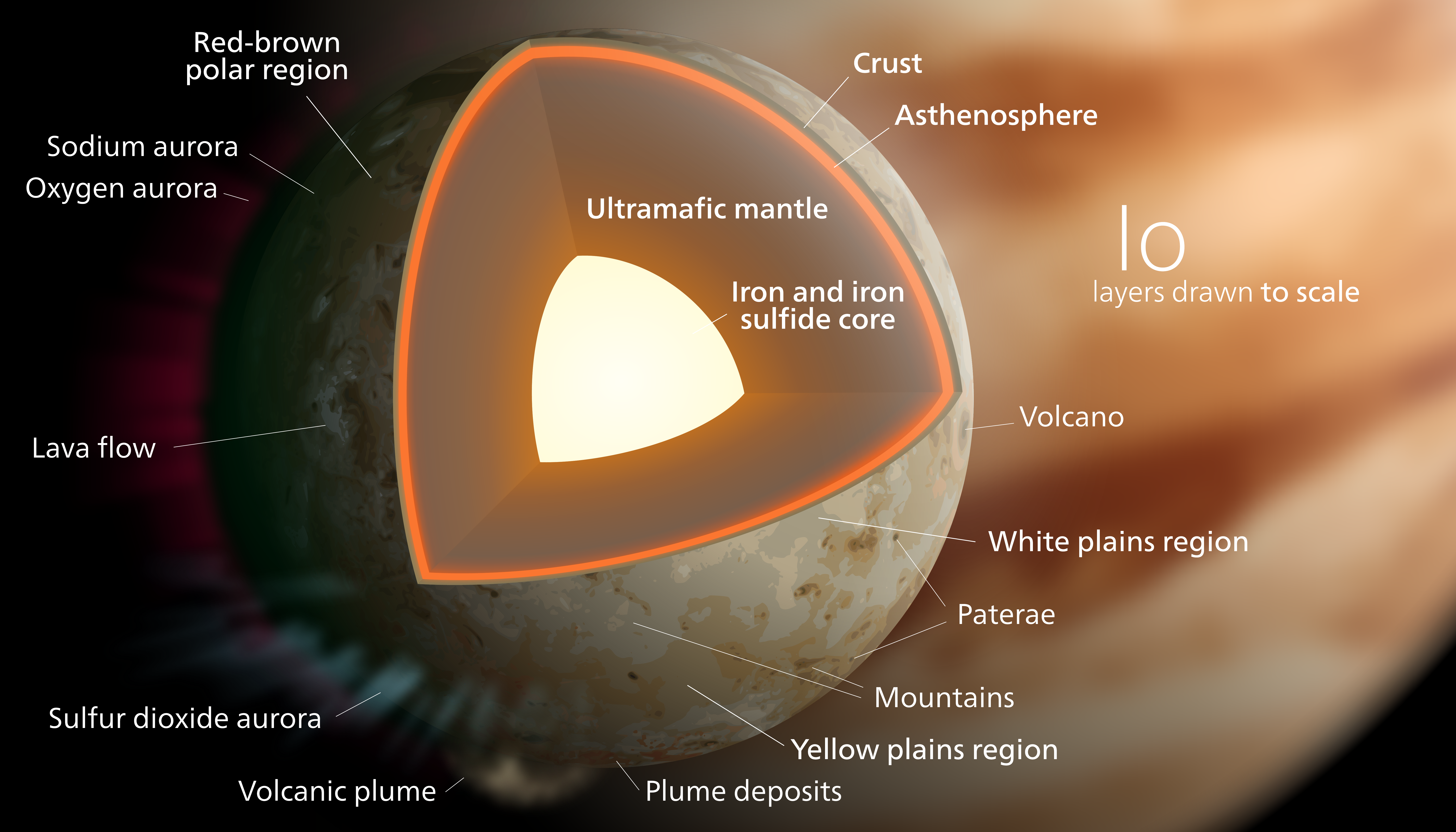}
\caption{Cross-sections of the Galilean moons' interior structure based on the gravity data from Galileo mission. Various phases of ice and ice-silicate mixtures and metallic parts are present within the moons. This necessitates appropriate interior modeling of these objects.}
\label{fig:mooninteriors}
\end{figure}

\begin{figure}[ht]
\includegraphics[width=.9\textwidth]{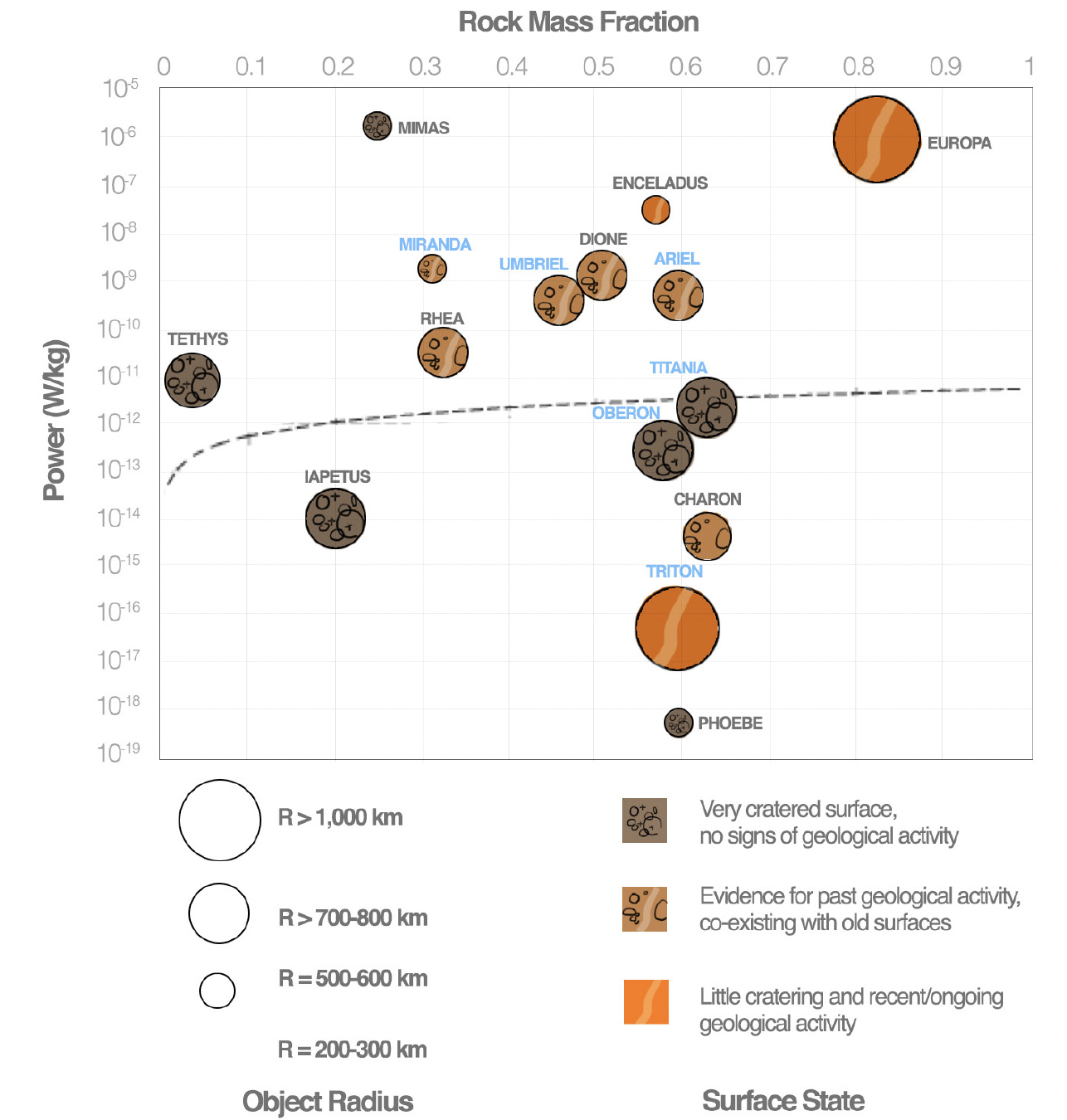}
\caption{Icy satellite rock mass fractions and first-order tidal heating power (assuming $k_2/Q= 1$) with schematic representation of
observed surface geology, which can reflect the potential for endogenic activity and the (past or current) presence of a
subsurface ocean. The dashed line indicates current-day heating due to natural decay of radioisotopes in the rock. Based on \citet{castillo2012small}, with
updated information on Charon from New Horizons data.}
\label{fig:moonheatbudget}
\end{figure}

\noindent
\textcolor{black}{Io is the most volcanically active object in the Solar system (Figure~\ref{fig:mooninteriors}).  Its tidal heating power is estimated between 65 and 125 TW \citep{lainey2009}. Io's potential for significant tidal heating was suspected \citep{peale1979} Even before Voyager 1 flew through the Jovian system. Despite several observations by missions and a sustained ground-based observation program, we still lack a full understanding of the mechanisms driving tidal heating in Io, and especially what this activity tells us about Io's deep interior. Galileo magnetic field observations hint that the moon could contain a deep magma layer, although its origin is debated (see \cite{vanhoolst2020} for discussion). Using measurements of volcanic activity as a proxy for surface heat flow, based on spacecraft datasets covering several decades, \cite{rathbun2018} found lower activity in the equatorial region at the anti-Jovian and sub-Jovian points. They also identified peak heat flow at mid-latitudes and a fourfold symmetry of upwellings distribution in longitude. The distribution of high dissipation regions inside Io is very dependent on the localization of the dissipation: in the asthenosphere or the underlying mantle (see Figure~\ref{fig:mooninteriors} for a definition of these terms). \cite{kervazo2022} demonstrated that the distribution of tidal heating between the mantle and the asthenosphere is very sensitive to the melt fraction in the asthenosphere. If this fraction is low (high), then dissipation may occur preferentially in the mantle (asthenosphere). This behavior is characterized by a drastic difference in the surface expression of tidal heating. \cite{rathbun2018} suggested that their observations were consistent with the tidal heating model by \cite{tackley2001} where large-scale convective pattern dominates the distribution of tidal heating, and smaller-scale asthenospheric instabilities  spread out the surface heat flux. On the other hand, the presence of a deep magma ocean would shift the distribution of activity peaks by 30 deg. \citep{tyler2015} in comparison to the Rathbun et al. observations. 
\cite{kervazo2022} predicted that the tidal Love numbers $h_2$ and especially $l_2$ are more sensitive than $k_2$ to the melt distribution and would allow distinguishing between a mantle-dominated and asthenosphere-dominated regimes. This difference is amplified by bulk heat production. Furthermore, \cite{vanhoolst2020} found that the diurnal libration amplitude of Io is several times larger if Io has a magma ocean compared to a partially molten asthenosphere. These authors predicted that the libration amplitude could be greater than 1 km in the former case. Upcoming missions, such as JUICE (JUpiter ICy moons Explorer, to be launched in 2023) and potential observations by Juno (two close Io flybys in 2023 and 2024) might be able to shed more light on Io's behavior.}

\noindent
The contribution of tidal heating to promoting global ice melting and then preserving a deep ocean over long-time scales is not well quantified in most bodies. Tidal forcing likely played a major role in the formation of a deep ocean in Enceladus, as the moon's small size and limited  production from radioisotope decay would prevent global scale melting. It has been suggested that Saturn's inner moons (within Titan's orbit) formed in Saturn's rings and evolved outward as a result of tidal interactions with the rings and the planet \citep{charnoz2011}. The moons could form from heterogeneous accretion of porous ice on silicate chards embedded in the rings. Then, they would exit the rings in sequence, with Rhea being the oldest moon and Mimas the youngest. This scenario could explain why the moons' densities do not follow a monotonic increase with distance to the planet, as would be expected if they formed in a circumplanetary disk.  If the moons emerged from the ring, then they could have been subject to significant heating in their early history since they formed closer to the planet and their eccentricities were potentially excited by collisions with other emerging moons \citep{charnoz2011}. However, an end-to-end evolution scenarios of these bodies has not been investigated yet.

\noindent
The feedback between tidal dissipation and orbital evolution can lead to a cycle of internal melting and freezing with geological expressions such as partial resurfacing (see for example \citet{HussmannSpohn04} for Europa). Resonance crossing, when the orbital periods of two bodies becomes near-commensurate as a result of orbital migration, can trigger significant heating. For example, Uranus's moons Miranda and Ariel have recorded in their geology evidence of high heat flow, tens of mW/m$^{2}$ vs. < 0.01 mW/m$^{2}$ for their current eccentricities (e.g., \citet{beddingfield2015}, \citet{beddingfield2021}). This high heat flow has been attributed to the crossing of resonances: a possible 5:3 resonance between Ariel and Umbriel and a 3:1 resonance between Miranda and Ariel. In resonances, the eccentricities (and sometimes the inclinations) of moons are excited.
\clearpage

\noindent
It was recently suggested that tidal forcing could also drive oceanic current (Rossby-Haurwitz  waves) with potentially important implications \citep{tyler2014}. Whether or not this mechanism could represent a long-term heat source in some of the moons is uncertain at this time.
\citet{chen2014} found that obliquity-driven tides in moon oceans could yield power of the same order as radioactive decay heat (Figure~\ref{fig:chen2014}). In the case of Neptune's moon Triton, obliquity-driven tides might be the key to the moon's geologically young surface and ongoing activity (e.g., \citet{nimmo2015}; \citet{hansen_etal21}). \textcolor{black}{Indeed, obliquity-driven tides in the ocean could generate up to 300 GW \citep{nimmo2015}, which is about five times greater than the power produced from radioisotope decay heat and six orders of magnitude greater than power generated from eccentricity tides. 
Recent work by \cite{matsuyama2018ocean} focusing on the coupling of the ocean to the ice shell showed that the outer ice shell tends to damp ocean tides and decrease their contribution to tidal heating. These authors highlighted that the obliquity phase lag is sensitive to ocean thickness and is large (18 deg. for a reference ocean thickness of 100 km below a 10-km thick shell) and may be measured by a future mission (e.g., Europa Clipper).  Furthermore, \citet{hay2022} computed that at high frequencies, tidal deformation driven by interactions among moons could be amplified by ocean dynamics. However, the amplitude of this high frequency tidal response is too small to be detectable via spacecraft.}

\noindent
In their work on the endogenic activity near the Enceladean south pole, \citet{howett2011} proposed that the outgoing energy flow associated with the tiger stripes geysers is of the order of $16$ GW. \textcolor{black}{Many explanations have been suggested for this high power, most of which have been summarized by \cite{nimmo2018}.} 
In a rough calculation, \citet{Efroimsky18Enceladus} modeled Enceladus with a homogeneous sphere composed of a Maxwell material, and calculated the amount of tidal power dissipated in it, with an input from forced libration in longitude taken into account. The so-obtained formula established a direct link between the power and the mean viscosity of Enceladus. The subsequent insertion of the measured value of heat outflow resulted in the value of the mean viscosity, which exactly coincided with the viscosity of ice near melting point. This simple estimate demonstrated that tides are sufficient to explain Enceladus's high heat power.

\noindent
Among the recent developments in our understanding of the tidal dissipation in icy moons are more elaborate models, e.g. one permitting for water circulation in porous cores (fractured but solid, or rubble piles) \citep{choblet_etal17}. Likely active in mid-sized moons (<1000 km radius, \citet{neveu2019}), this mechanism may be contributing to Enceladus's heat production, along with  the significant dissipation in the ice shell and, potentially, in the ocean. Further insights in the surface geological activity of Enceladus will be provided in Section~\ref{sec:surface_features}.

\noindent
\textcolor{black}{The implications of tidal interactions between a moon and planet depend on the $k_2/Q$ of the planet. For a long time, these parameters were poorly known. Indeed, tidal friction within Saturn causes its moons to migrate outwards, and tidal evolution determines their current position.  Based on astrometric observations obtained with the Cassini spacecraft, \citet{lainey2017} showed that dissipation in Saturn is at least one order of magnitude larger ($k_2/Q=(1.59 \pm  0.74) \times 10^{-4}$) than assumed in past theoretical studies. These authors also identified that Saturn's tidal response might be frequency dependent by comparison between the tidal evolution of the midsized moons.} 

\noindent
\textcolor{black}{Disentangling the relationships between internal evolution and dynamical evolution offers prospects to explain the origin of the Saturnian moons (and maybe the Uranian moons when more observation becomes available). If the moons formed in Saturn's rings then, the evolution to their current locations can set constraints on Saturn's $k_2/Q$. On the other hand, if the moons formed near their current locations, then Saturn's $Q$ needs to be high (about 80,000, \citet{cuk2016}), which does not appear consistent with Enceladus's high dissipation. Additional observations with Cassini lead \cite{lainey_etal20} to demonstrate that Titan is migrating away from Saturn faster than previously envisioned, leading to an estimate of Saturn's $Q$ of about 100. These authors also inferred that Titan could have formed much closer to Saturn than its current location.
\cite{fuller2016} introduced the resonance locking theory between moons and internal oscillation modes of the planet as a way to address a number of observations in the Saturnian and Galilean systems. Resonance locking opposes the constant $Q$ assumption used for decades. 
\cite{fuller2016} inferred that this effect is the dominant driver of tidal dissipation in these systems and suggested that the resonance locking theory allows the large heating rate of Enceladus to be achieved in an equilibrium eccentricity configuration. \cite{nimmo2018} pointed out that the astrometric observations at the basis of the resonance locking scenarios and follow on studies carry large uncertainties and only a narrow set of Saturn's $k_2/Q$ allows the onset of resonances. These authors further noted that the impact of resonances on the internal evolution of satellites, especially under increase dissipation peak when they cross mean motion resonances is not accounted for. Tidal dissipation within the moons could counteract Saturn's influence by decreasing eccentricity and semi-major axis.} 

\noindent
\textcolor{black}{The thermal and orbital evolution of icy moons are coupled via the thermomechanical properties of their materials (see Section~\ref{sec:coupling}).
There is evidence that Saturn's midsized moons have been very dissipative in their past, expressed in the geology (e.g., flexure relaxation, crater morphologies). This has been ascribed to resonance crossing (e.g., \citet{chen2008} for Tethys). 
Convective regions (e.g., convective plumes) can also be areas of enhanced tidal heating. \cite{sotin2002} suggested that tidal forcing can drive the formation of diapirs that may be involved in the formation of lenticulae and chaos regions at the surface of Europa. \cite{tobie_etal03} further explored this idea and predict a heat flow of 35 to 40 mW/m$^{-2}$ constant across  Europa's surface. Further details on the surface geological activity on Europa will be provided in Section~\ref{sec:surface_features}.
In the case of Titan, \cite{tobie2005} found that a few percent of ammonia in the ocean is needed to limit the dissipation in the shell (by decreasing the basal temperature of the shell) and preserve a high eccentricity (0.009), although alternative models have been suggested to explain this high eccentricity (e.g., \citet{cuk2016})
More work is needed to establish an end-to-end model of the evolution of the Saturnian moon system as a whole.}

\noindent
A recent analysis of the tidal evolution of the Uranian moons following their crossing mean motion resonances has led to constraints on their ${Q}/{k_2}$ \citep{cuk2020}. It ranges from 10$^{4}$ for Oberon (the farthest from Uranus) to 10$^{6}$ for Miranda (closest moon), suggesting a wide range of internal properties across the Uranian  moons system. Considering the moon's $k_2$ may range from 10$^{-3}$ to 10$^{-2}$ \citep{Hussmann_etal06}, we infer dissipation factors in excess of 1000 in Miranda, and potentially also Oberon, suggesting that the moons are not dissipative at present and may be near frozen. \textcolor{black}{Interestingly, the Uranian moon system is not in resonances at present.} 

\noindent
Actual constraints on the tidal parameters of icy moons are few. Multiple flybys of Saturn's moon Titan by the Cassini orbiter led to the inference of $k_2 = 0.637 \pm 0.0220$ (Iess et al. 2014).  This high value has been interpreted as evidence for a high density of the ocean, of the order of 1200 kg/m$^{3}$ (e.g., \citet{mitri2014,baland2014}). This high density points to a high content in salt \citep{baland2014} that cannot be uniquely explained by cosmochemical abundances but also involves a contribution of non-water ice volatiles (e.g., CO$_2$, NH$_3$) to oceanic solutes (e.g., bi/carbonate and ammonium ions. This high density may also reflect a large abundance of clathrate hydrate, such as CO$_2$ hydrates, whose density is about 1100 kg/m${^3}$ \citep{bostrom2021}.

%\noindent
%Other consequences of tidal heating are observed in giant planet satellite systems or are suspected to have played a major role in the past. For example, 

\noindent
Despite the critical contribution of tidal stressing to the heat budgets of large icy moons, models miss input parameters that have been obtained in conditions relevant to these objects. By lack of knowledge on the internal properties of small moons (<1000 km radius), many studies assume a dissipation factor Q = 100 for these bodies, as well as a tidal Love number based on a homogeneous interior. However, as shown above, dissipation is a function of the material frequency-dependent mechanical response, which itself depends on temperature and the history of the material. Dissipation in porous interiors (e.g., near Earth asteroids), mixtures of water ice, salts, and clathrates, that may be relevant to Europa or Titan, for example, are scarce. Tidal stresses at icy moons have a wide range of amplitudes that invoke different response mechanisms. Reproducing low stress and long forcing periods in laboratory is challenging and most cyclic forcing measurements have been obtained at frequencies greater than 10$^{-4}$ Hz. Future experimental work should aim to retire some of these knowledge gaps that prevent more high-fidelity modeling of the moon internal evolution.

\noindent
The Europa Clipper mission's gravity measurement is projected to yield Europa's tidal Love number $k_2$ with an accuracy of 0.015 (1-$\sigma$) \citep{Mazarico2021}. The combination of ground penetrating radar and imaging will yield $h_2$ \citep{steinbrugge2021} but the predicted accuracy is still a work in progress. The JUICE mission is also planning to obtain these parameters at Ganymede, in particular $k_2$ with an accuracy of 10$^{-4}$ \citep[e.g.,][]{cappuccio2020}.

\subsection{Tidal signature on planetary surfaces}\label{sec:surface_features}

\noindent
This review mainly addresses deciphering the interior structure based on measurements of tidal response presented in the form of Love numbers as well as the reciprocal effect of tidal heating and interior properties in the planetary bodies. However, it is important to mention how surface geological features as a result of tidal activities can also be used to constrain both the present-day interior and thermal-orbital evolution of the planetary bodies through their relationship to tidal stress and tidal evolution.

\noindent
Hence, we dedicate this section to overview the effect of surface features as a result of tidal activities and their interpretations for the interior in different planetary bodies, particularly the icy moons. While, a more detailed review of the tidal signatures on the surface of the planets would need a dedicated study, here we only provide a summary of the most important findings on the interior as a result of studying the surface of the icy moons.

\noindent
The extent to which the surface of a moon will be affected by tidal stress not only depends on the orbital elements such as eccentricity, obliquity, and the proximity to the parent planet, but also to  the responsiveness of the of the interior of the body to tidal deformation, which in turn is strongly affected by the presence or absence of an ocean, the rheology of the interior, and the amount of internal heat being generated within the moon \citep[e.g.,][]{greenberg_etal98,Jara-Orue_Vermeersen11, rhodenWalker22}.  Thus, geologic features on the planetary objects can provide insight as to their interior structures and thermal states. For example, the presence, absence, and style of tectonic activity can be diagnostic of a moon's thermal history \citep[e.g.,][]{Rhoden_etal20b}. Similarly, the presence of eruptions, whether in the form of surface flows or geysers, can provide insight into the cooling history of a moon and the composition of its ocean \citep[e.g.,][]{rudolph_etal22}. Lastly, indicators of past heat flows, through changes to crater morphology, can constrain the overall thermal history of a moon \citep[e.g.,][]{bland_etal12}

\noindent
Although still an area of active study, some consistent links have been identified between certain geologic features and the inferred thermal-orbital history or interior structure of a moon. We begin with tectonics. The two icy satellites with extensive canyon systems, Tethys at Saturn and Charon at Pluto, both have negligible present-day orbital eccentricities, with reason to expect that they underwent rapid despinning and circularization \citep{Peale99, Neveu_Rhoden19, cheng_etal14}. Neither moon is expected to have an ocean today due to a lack of potential heat sources, although observational constraints on Tethys's interior are quite limited \citep{castillo_etal18}. Tethys and Charon both have globally-distributed, but overall sparse, fracture systems that appear unrelated to either the canyons or tidal stress patterns \citep{castillo_etal18, Rhoden_etal20b}. Their origin is currently unknown.
\noindent
As described in \citet{Rhoden_etal20b}, the best explanation of the formation of the canyons on Tethys and Charon is that both moons began with high eccentricity orbits that generated oceans through tidal dissipation. As a result, the orbits circularized, and the lack of tidal heating led the oceans to freeze out. As oceans freeze, the thermal stresses and the increase in volume of the newly accreted ice at the base of the shell can create radial fractures within an ice shell \citep{rudolph_manga09, rudolph_etal22}. Hence, the cooling and freezing of the interior may have produced the initial fractures that developed into canyons through continued thickening of the ice shell. The lack of fractures associated with tidal stress patterns comes from the fact that circularization occurred before the ocean froze, so there were no tidal stresses available  \citet{Rhoden_etal20b}. Thus, we can hypothesize that canyon systems are indicative of a past ocean that has completely frozen out. 

\noindent
One important difference between Charon and Tethys is that only Charon shows evidence of cryovolcanism. A potential explanation for this difference is that fractures formed through ocean freezing were able to crack through the entire ice shell at Charon but not at Tethys, perhaps because the Tethys's interior did not warm up as much as Charon's, and the ice shell was correspondingly thicker. Given that Tethys is approximately 94\% ice \citep{castillo_etal18}, it may have had too little rocky material to provide sufficient radiogenic heating to reach a similar level of melting as Charon, or it may have begun with a lower eccentricity that resulted in less tidal heating. The presence of ammonia within Charon's interior, as explored by \citet{bagheri_etal22} could also facilitate eruptions and would likely be absent or less abundant within Tethys. In any case, comparing the geology of these two moons provides insight into differences in the thermal evolution of their interiors.
\noindent
On the opposite end of the tectonic spectrum are moons with oceans thicker than their overlying ice shells. Europa at Jupiter and Enceladus at Saturn \citep[e.g.,][]{nimmoPappalardo16} are the two ``confirmed" ocean worlds of this type. Both of these moons execute eccentric orbits while orbiting close to their parent planet; their subsurface oceans and warm ice shells allow for significant tidal deformation and stress in the overlying cold ice. Europa's surface is pervasively, and globally, fractured, and the fractures vary in their surface expression from linear to wavy to cycloidal \citep[e.g.,][]{kattenhornHurford09} (See Figure~\ref{fig:mooninteriors}). The shapes of cycloids, linked arcs that can span 100s of km, track the spatial and temporal changes in tidal stress throughout Europa's orbit \citep{greenberg_etal98, hoppa2001,Rhoden_etal10, Rhoden2021}. Fits to both individual cycloid shapes and the overall distribution of cycloids greatly improved when the effects of a small obliquity were added to the tidal stress model, illustrating the powerful link between the orbit, interior, and surface geology. Europa also displays strike-slip offsets along many fractures, long parallel bands formed from extension, and irregularly-shaped bands formed from compression/subsumption (see reviews by \citet{kattenhornHurford09, ProckterPatterson09}. Many strike-slip offsets have been linked to tides \citep[e.g.,][]{rhoden_etal12}, whereas band formation appears to result from large-scale plate motion \citep[e.g.,][]{kattenhorn_Prockter14}, which may be governed by different processes.

\noindent
Enceladus displays tectonic features that are morphologically-similar to those at Europa, but Enceladus's tectonism is regional, with activity concentrated near the south pole \citep{Spencer_etal06, patthoff_Kattenhorn11, CrowWillardPappalardo15}. Also, Enceladus lacks cycloids, strike-slip offsets along fractures, and bands of any kind. The four, roughly parallel fractures within the south polar terrain (dubbed ``Tiger Stripes") have orientations consistent with eccentricity-driven tidal stresses \citep{Rhoden_etal20b}, and the geyser-like eruptions emanating from the Tiger Stripes vary in output on a diurnal timescale, suggesting a link to tides \citep{Nimmo_etal07,hurford_etal07, hedman2013}. The ice shell thickness on Enceladus appears to vary considerably with location \citep[e.g.,][]{thomas_etal16, beuthe_etal16, Cadek_etal19}; the shell is thinnest at the south pole and thickest along the equator. In a heterogeneous ice shell, tidal stresses are enhanced at thin zones \citep{beuthe19}. Thus, the extent of tectonic activity may well be an indicator of ice shell thickness, with older tectonized regions representing past thin zones. This interpretation would suggest that, as at Europa, multiple generations of crisscrossing, ridge-flanked fractures are indicative of relatively thin ice shells and concentrated tidal stress.
Between these two extremes, there are a variety of heavily cratered moons with varying levels of tectonic activity, including Ganymede at Jupiter and Dione at Saturn. Both moons are thought to have thin oceans beneath much thicker ice shells, and they both orbit far enough from their parent planet to have reduced tidal heating and stress, which may be the reason for their limited geologic activity. Callisto at Jupiter and Rhea at Saturn orbit even farther away and both are heavily cratered with little other geologic activity. There is some evidence to suggest a deep, thin ocean within Callisto \citep{Khurana_etal98}, so it is possible that even heavily cratered worlds can have oceans; much less is known about the interior structure of Rhea \citep{Tortora_etal16}. Taken together, it seems that there is a correlation between ocean thickness, relative to total hydrosphere thickness, and the extent of tectonic activity. 

\noindent
One curious outlier is Saturn's smallest and innermost regular satellite, Mimas. With a higher eccentricity than its exterior neighbor, Enceladus, we might expect Mimas to be even more geologically active, akin to Io at Jupiter. However, Mimas is heavily cratered with few tectonic features and no evidence of volcanic activity. And yet, model fits to Cassini measurements of Mimas's physical librations indicate that Mimas is likely an ocean-bearing world and require that Mimas is differentiated into a rocky interior and outer hydrosphere \citep{tajeddine_etal14,Noyelles_etal17, Caudal17}. If Mimas does, indeed, have a subsurface ocean today, it may be a young ocean world, such that its surface has not had time to develop the characteristics features observed on moons such as Europa and Enceladus (see also \citet{rhodenWalker22}). 
In addition to tectonic and volcanic activity, craters can be quite diagnostic of the interior structure and heating history of a moon. Most obviously, the morphology of the initial impact can be used to constrain the thickness of the outmost layer of the moon. For example, some craters have clearly cracked through Europa's entire ice shell to the ocean beneath (see \citet{SchenkTurtle09}, for an overview), whereas several of the mid-sized moons of Saturn have large (100s of km) impact basins that show no evidence of a subsurface ocean. However, tidal dissipation within an icy satellite can also have a more subtle, but still diagnostic, effect on craters. 

\noindent
As heat is transferred from the interior of a moon to the surface, it can modify the shapes of existing craters, creating flatter floors and overall subdued morphologies. The process, called crater relaxation, records the magnitude and longevity of heat flow through the ice shell. Crater relaxation on Enceladus, Tethys, Dione, and Rhea all indicate past episodes of higher heat flows \citep{bland_etal12, white_etal13, white_etal17}. Even Enceladus, which has high present-day heat flows, would require a past epoch of considerably more dissipation to explain the observed crater shapes \citep{bland_etal12}. For the other moons, to achieve the observed levels of relaxation would imply temperatures above the melting temperature of water ice, suggesting oceans within their surface ages. In contrast, Mimas's craters show no evidence of relaxation to within the observation limit imposed by image coverage. In that case, crater shapes provide an upper limit on the extent of tidal dissipation that could have taken place within Mimas. In all of these cases, it is the link between interiors, tidal dissipation, and surface geology that allows these features to provide additional insight as to the thermal-orbital evolution of the moons and their past and present interior structures.

 \section{Summary and conclusions}
In this paper, we have attempted to provide an overview of the past, present, and future use of tides as a means of procuring information on the interior structure of planets and moons from orbiting and landed spacecraft. Since
tidal response measurements play an inordinate role in understanding the interiors of rocky and icy planetary bodies, we provided a detailed outline of the fundamental concepts connected to modeling the tidal response, including the viscoelastic behaviour of planetary materials, and how the tides couple to the thermal structure within the planetary body to determine its subsequent orbital evolution. We also provided an overview of the surface geological features on icy moons as a result of tidal activity with interpretations for their interior structure and evolution.

\noindent
This is also of fundamental importance for proper modeling and understanding of the evolution of extra-Solar planetary bodies that ultimately determines their astrobiological potential. In connection herewith are the latest extensions of modern tidal theory \citep{makarov_efroimsky13}, including tidal evolution formulae that allow for the consideration of higher-order eccentricity functions required for accurate modeling of the evolution of planetary systems in and out of e.g., spin-orbit resonances, non-synchronous rotation, and mean motion resonances \citep{Renaud_etal21,Bagheri_etal21}. For exoplanets trapped in a higher-order spin-orbit resonance for an extended amount of time, for example, important climate variations could ensue as a result of a new Solar configuration \citep{DelGenio_etal19} and thus the potential for life to develop.
\noindent
We provided an overview of the current state of investigation of the rocky bodies of our Solar system, including Mercury, Venus, the Moon, and Mars, and the icy moons of the giant planets from the point of view of tides that will invariably remain, relative to the more involved deployment of surface instrumentation, the most prominent means of inferring knowledge on the interior of the planets and satellites. The ongoing development of more refined tidal models, in combination with further experimental studies of the viscoelastic properties of planetary materials, and advances in geophysical models through continued analysis of geodetic and other large-scale measurements (e.g., InSight and LLR), promise new insights into the inner structure of terrestrial planets and moons of the Solar system.

 \section{Acknowledgments}
We acknowledge informative discussions with Ian Jackson and Ulrich Faul on material viscoelasticity, and with Valeri Makarov on tides. A.B. was supported by a grant from the Swiss National Science Foundation (project 172508). J.C.'s contribution was developed at the Jet Propulsion Laboratory, California Institute of Technology, under contract with the National Aeronautics and Space Administration. M.W. and A.C.P. gratefully acknowledge the financial support and endorsement from the DLR Management Board Young Research Group Leader Program and the Executive Board Member for Space Research and Technology.
\textcolor{black}{The artistic views of the Moon and the Saturnian moons are taken or modified from https://commons.wikimedia.org/wiki/User:Kelvin13.}
\clearpage
\section*{References}

\bibliography{references}
\bibliographystyle{abbrvnat}

 \end{document}